\documentclass[12pt]{article}
\pdfoutput=1

\usepackage[square,comma,numbers,sort&compress]{natbib}
\usepackage[left=20mm,right=20mm,top=20mm,bottom=20mm]{geometry}
\usepackage{mediabb}
\usepackage{epsf}
\usepackage{color}
\usepackage[dvipdfm]{graphicx}
\usepackage{amsmath}
\usepackage{amssymb}
\usepackage{latexsym}
\usepackage{slashed}
\usepackage{float}
\usepackage{cases}
\usepackage{multirow}
\usepackage{bm}
\usepackage{braket}

\usepackage{supertabular} 

\newcommand{\Z}{\mathcal{Z}}

\newcommand{\al}[1]{\begin{align}#1\end{align}}

\newcommand{\bp}{\begin{pmatrix}}
\newcommand{\ep}{\end{pmatrix}}
\newcommand{\bb}{\begin{bmatrix}}
\newcommand{\eb}{\end{bmatrix}}

\newcommand{\del}{\partial}

  \def\Z{Z}
 
\def\del{ \delta }  
\def\e{e}
\let\i=\relax \def\i{i}
 
\def\tpi{ 2 \pi }
\def\ldef{ := }

\def\bit{\begin{itemize}} \def\eit{\end{itemize}}
\def\ben{\begin{enumerate}} \def\een{\end{enumerate}}
\def\bec{\begin{center}} \def\eec{\end{center}}

\definecolor{check}{cmyk}{0.2,0.2,0.5,0}

\def\om{ \omega } \def\omb{ \bar{ \omega } }

\newcommand{\be}{\begin{equation}} \newcommand{\ee}{\end{equation}}
\newcommand{\bea}{\begin{eqnarray}} \newcommand{\eea}{\end{eqnarray}}

\newcommand{\fulltoday}{\number\day\space \ifcase\month\or
    January\or February\or March\or April\or May\or June\or
    July\or August\or September\or October\or November\or December\fi
    \space\number\year}

 \makeatletter
    \renewcommand{\theequation}{%
    \thesection.\arabic{equation}}
    \@addtoreset{equation}{section}
  \makeatother

\begin{document}
\allowdisplaybreaks[2]
\begin{titlepage}
\renewcommand\thefootnote{\alph{footnote}}
		\mbox{}\hfill EPHOH-15-001\\
		\mbox{}\hfill KIAS-P15002\\
		\mbox{}\hfill KOBE-TH-15-01\\
		\mbox{}\hfill KUNS-2535\\
		\mbox{}\hfill OU-HET-848\\
		\mbox{}\hfill WU-HEP-15-01\\
\vspace{2mm}
\begin{center}
{\fontsize{22pt}{0pt}\selectfont \bf{{Classification of three-generation models \\on magnetized orbifolds}}
} \\
\vspace{8mm}
	{\fontsize{14pt}{0pt}\selectfont \bf
	Tomo-hiro Abe,\,\footnote{
		E-mail: \tt t-abe@scphys.kyoto-u.ac.jp
		}}
	{}
	{\fontsize{14pt}{0pt}\selectfont \bf
	Yukihiro Fujimoto,\,\footnote{
		E-mail: \tt fujimoto@het.phys.sci.osaka-u.ac.jp
		}}
	{}
	{\fontsize{14pt}{0pt}\selectfont \bf
	Tatsuo Kobayashi,\,\footnote{
		E-mail: \tt kobayashi@particle.sci.hokudai.ac.jp
		}}
	\\[3pt]
	{\fontsize{14pt}{0pt}\selectfont \bf
	Takashi Miura,\,\footnote{
		{E-mail: \tt takashi.miura@people.kobe-u.ac.jp (miura{\_}takashi@jp.fujitsu.com)}
		}
				\footnote{Moved to \it Software Systems Laboratories, FUJITSU LABORATORIES LTD. 1-1, \\
				\textcolor{white}{spaces spaces s}Kamikodanaka 4-chome, Nakahara-ku Kawasaki, 211-8588 Japan}}
	{}
	{\fontsize{14pt}{0pt}\selectfont \bf
	Kenji Nishiwaki,\,\footnote{
		E-mail: \tt nishiken@kias.re.kr
		}}
	{}
	\\[3pt]
	{\fontsize{14pt}{0pt}\selectfont \bf
	Makoto Sakamoto,\,\,\footnote{
		E-mail: \tt dragon@kobe-u.ac.jp} }
	{\fontsize{14pt}{0pt}\selectfont \bf
	and
	Yoshiyuki Tatsuta\,\,\footnote{
		E-mail: \tt y{\_}tatsuta@akane.waseda.jp} } \\
\vspace{4mm}
	{\fontsize{13pt}{0pt}\selectfont
		${}^{\mathrm{a}}$\it Department of Physics, Kyoto University, Kyoto 606-8502, Japan \smallskip\\[3pt]
		${}^{\mathrm{b}}$\it Department of Physics, Osaka University, Toyonaka 560-0043, Japan \smallskip\\[3pt]
		${}^{\mathrm{c}}$\it Department of Physics, Hokkaido University, Sapporo 060-0810, Japan \smallskip\\[3pt]
		${}^{\mathrm{d,g}}$\it Department of Physics, Kobe University, Kobe 657-8501, Japan \smallskip\\[3pt]
		${}^{\mathrm{f}}$\it School of Physics, Korea Institute for Advanced Study,\\[2pt]
					85 Hoegiro, Dongdaemun-gu, Seoul 130-722, Republic of Korea\\[3pt]
		${}^{\mathrm{h}}$\it Department of Physics, Waseda University, Tokyo 169-8555, Japan \smallskip\\[3pt]
	}
\vspace{4mm}
{\normalsize \fulltoday}
\vspace{10mm}
\end{center}
\begin{abstract}
\fontsize{12pt}{16pt}\selectfont{
{We classify the combinations of parameters which lead {three generations} of quarks and leptons in the framework of magnetized twisted orbifolds on $T^2/Z_2$, $T^2/Z_3$, $T^2/Z_4$ and $T^2/Z_6$ with allowing nonzero discretized Wilson line phases and Scherk-Schwarz phases.
We also analyze two actual examples {with nonzero phases leading to
one-pair Higgs and five-pair Higgses} and discuss the difference from the results without nonzero phases studied previously.}
}
\end{abstract}
\end{titlepage}
\renewcommand\thefootnote{\arabic{footnote}}
\setcounter{footnote}{0}


\section{Introduction}

One of the {greatest achievements} in particle physics {is the completion} of the Standard Model~(SM) as the discovery of the SM-like Higgs boson following the works reported in~\cite{Aad:2012tfa,Chatrchyan:2012ufa}.
{As is} widely known, the SM is consistent with almost all the experimental results which {have} ever been made, while the origin of the SM configurations is still unknown, {\it e.g.}, the origin of the $SU(3)_C \times SU(2)_W \times U(1)_Y$ gauge structure and the related quantum numbers in the matter fields, or why the number of the matter generation is three accompanied with the large mass hierarchy.

Ten-dimensional (10D) super Yang-Mills~(SYM) theory on magnetized tori~\cite{Bachas:1995ik,Blumenhagen:2000wh,Angelantonj:2000hi,Blumenhagen:2000ea,Cremades:2004wa,Blumenhagen:2005mu,Blumenhagen:2006ci} possibly is a good candidate for explaining these issues simultaneously.\footnote{
{Introducing point interactions (zero-thickness branes) in the bulk space of a five-dimensional theory on $S^1$ (or a line segment) and considering various boundary conditions of fields on them~\cite{Fujimoto:2012wv,Fujimoto:2013ki,Fujimoto:2014fka} is a direction to attack the problems.}
}
Here, 4D chirality and multiple matter generations are created by the effect of magnetic fluxes in a unified gauge group.
After the introduction of the fluxes, the larger gauge group is explicitly broken down into {a} small substructure, which contains the SM counterpart.
In this direction, lots of phenomenological aspects have been {explored:} Yukawa couplings~\cite{Cremades:2004wa},
realization of quark/lepton masses and their mixing angles~{\cite{Abe:2012fj,Abe:2014vza}}, higher order couplings~\cite{Abe:2009dr}, flavor symmetries~{\cite{Abe:2009vi,Abe:2009uz,Abe:2010ii,BerasaluceGonzalez:2012vb,Honecker:2013hda,Marchesano:2013ega,Abe:2014nla}},
massive modes~\cite{Hamada:2012wj}, and others~\cite{Sakamoto:2003rh,Antoniadis:2004pp,Antoniadis:2009bg,Choi:2009pv,Kobayashi:2010an,DiVecchia:2011mf,Abe:2012ya,DeAngelis:2012jc,Abe:2013bba,Hamada:2014hpa}.\footnote{
Some related works in other stringy contexts ({\it e.g.,} intersecting D-brane model and heterotic string theory) are found in~{\cite{Cremades:2003qj,Cvetic:2003ch,Abel:2003vv,Kobayashi:2004ya,Blumenhagen:2005mu,Blumenhagen:2006ci,Kobayashi:2006wq,Ko:2007dz,Honecker:2012jd,Beye:2014nxa}}.
}

In actual model {building}, orbifolding plays {an important role}, not only {in} removing exotic particles, {but also in} deforming wave function profiles of the SM matters, which would help us to realize the observed flavor structure of the nature. 
In general under the magnetic gauge background, it is not so easy to analyze the case more than $Z_2$, namely on $T^2$, $T^2/Z_{3,4,6}$ geometries, where $Z_N$-entangled states have very complicated forms since mode functions on magnetized $T^2$ {are} described by {theta functions}.\footnote{
Note that their geometrical aspects are discussed~\cite{Katsuki:1989bf,Kobayashi:1991rp,Choi:2006qh} within the context of string theory.
In a higher-dimensional field theory, {detailed studies have been carried out}~\cite{Kawamura:2008mz,Kawamura:2007cm,Kawamura:2009sa,Kawamura:2009gr,Goto:2013jma,Goto:2014eoa}.
}
{However} recently, general structure of these geometries {was} declared in an exact and analytical way in a ``dual'' description with operator formalism~\cite{Abe:2014noa}.
{In that paper}, the authors {treated most} general cases on magnetized $T^2/Z_{2,3,4,6}$ with nonzero (discretized) Wilson line phases and/or Scherk-Schwarz phases, which {were} discussed in~\cite{Abe:2013bca} (see also~\cite{Scherk:1978ta,Scherk:1979zr,Kobayashi:1991rp,Ibanez:1986tp,Kobayashi:1990mi,Angelantonj:2005hs,Blumenhagen:2005tn,Angelantonj:2009yj,Forste:2010gw,Nibbelink:2012de}).
Note that shifted orbifold can be considered on a magnetized torus~\cite{Fujimoto:2013xha}.

Some works {toward phenomenologically realistic model building have} been done through this direction based on $U(N)$ gauge theory in~\cite{Abe:2008fi,Abe:2008sx,Abe:2012fj,Abe:2013bba,Abe:2014vza}, where only the $Z_2$ orbifolding with trivial {Wilson line phases and} Scherk-Schwarz phases is considered. See also Refs~\cite{Choi:2009pv,Kobayashi:2010an} for discussions based on $E_{6,7,8}$ groups. 
Now, because of the advent of~\cite{Abe:2014noa}, all the technical difficulties for considering the generalized case {have} been resolved and it would be phenomenologically meaningful that we start to pursue the situation with nontrivial boundary conditions under the $Z_2$ or higher $Z_{3,4,6}$.
As a first step, in this paper, we classify all the possibilities of the models with suitable three-generation matter structure of the quarks and the leptons within $U(N)$ gauge group being the original gauge structure, and {show} an example in the $Z_2$ case with nontrivial boundary conditions after discussing the effect via nonzero Wilson line {phases} and/or Scherk-Schwarz {phases}.

This paper is organized as follows.
In section~\ref{sec:T2_review}, we give {a} brief review on the description of 10D SYM theory on $(T^2)^3$ with {or} without orbifoldings $Z_{2,3,4,6}$, respectively, including magnetic fluxes, nontrivial Wilson line phases and Scherk-Schwarz phases.
In section~\ref{sec:classification}, we exhaust all the possible configurations of parameters in all the $Z_2$, $Z_3$, $Z_4$ and $Z_6$ cases with nontrivial boundary conditions.
In section~\ref{sec:examples}, after we examine how much mass hierarchy in the quark sector is deformed by introducing nonzero Wilson line phases and/or Scherk-Schwarz phases, we propose an example for realizing (semi-)realistic quark mass  patterns and the mixing structure described by the Cabibbo-Kobayashi-Maskawa~(CKM) matrix.
Section~\ref{sec:closing} is devoted to summarizing issues and discussing future prospects.
In Appendix~\ref{sec:T2ZN_information}, we show the analytical forms describing a state on the magnetized $T^2/Z_{2,3,4,6}$ orbifolds obtained in Ref.~\cite{Abe:2014noa}.
In Appendix~\ref{sec:Z2_configuration_full}, we {show our results for $T^2/Z_{2,3,4,6}$} by writing down all the independent combinations of parameters explicitly.

\section{10D SYM on generalized magnetized orbifolds \label{sec:T2_review}}

Here, we review the basics of the 10D {super} Yang-Mills theory on generalized magnetized orbifolds of $T^2/Z_2, T^2/Z_3, T^2/Z_4, T^2/Z_6$ with nonzero Wilson line {phases} and/or Scherk-Schwarz {phases}.

\subsection{$U(N)$ gauge theory on $(T^2)^3$}

Firstly, we focus on the 10D magnetized SYM theory without orbifolding~\cite{Cremades:2003qj,Hashimoto:1997gm,Abe:2008sx}, with {adopting} notations in Refs.~\cite{Abe:2013bca,Abe:2014noa},
\al{
S = \int_{M^4} d^4x \int_{(T^2)^3} d^6z \left\{ -\frac{1}{4} \text{tr} \left( F_{MN}F^{MN} \right) +
	\frac{1}{2} \text{tr} \left( \overline{\lambda} \Gamma^M i D_M \lambda \right) \right\},
	\label{10D_action}
}
which is defined on a product of 4D Minkowski space and three factorizable 2-tori, $M^4 \times (T^2)^3$.
The capital roman indices $M,N$ run over $\mu(=0,1,2,3), z_1, \overline{z_1}, z_2, \overline{z_2}, z_3, \overline{z_3}$, where the $i$-th ($i=1,2,3$) 2-torus is described by the complex {coordinates} $z_i = y_{2i+2} + i y_{2i+3}$ and its complex conjugation $\overline{z_i} = y_{2i+2} - i y_{2i+3}$ made by the two Cartesian coordinate representing the extra directions, $y_{2i+2}$ and $y_{2i+3}$.
We take each torus modulus parameter $\tau_i\ (\subset \mathbb{C})$ as $\text{Im}\tau_i > 0$ for convenience. 
We use the short-hand notation $d^6 z$ meaning $\Pi_{i=1}^3 dz_i d\overline{z_i}$.
On $T^2_i$, the coordinate $z_i$ is identified as $z_i \sim z_i+1 \sim z_i+\tau_i$.
The bulk Lagrangian holds $\mathcal{N}=1$ supersymmetry in 10D and consists of the 10D vector fields $A_M$, which {are} found in the covariant derivative $D_M$ and the field strength $F_{MN}$, and the gaugino fields $\lambda$ described by 10D Majorana-Weyl {spinors}.
The explicit forms of $D_M \lambda$ and $F_{MN}$ are given by
\al{
D_M \lambda &= \partial_M \lambda - ig \left[ A_M, \lambda \right],\\
F_{MN} &= \partial_M A_N - \partial_N A_M -ig \left[ A_M, A_N \right],
}
with the 10D gauge coupling $g$.

The gaugino fields and the 10D vector fields are Kaluza-Klein (KK) decomposed as
\al{
\lambda(x, \{z_i, \overline{z_i}\}) &=
	\sum_{l,m,n} \chi_{l,m,n}(x) \otimes \psi^{(1)}_l (z_1, \overline{z_1}) \otimes
	\psi^{(2)}_m (z_2, \overline{z_2}) \otimes \psi^{(3)}_n (z_3, \overline{z_3}), \\
A_M(x, \{z_i, \overline{z_i}\}) &=
	\sum_{l,m,n} \varphi_{l,m,n; M}(x) \otimes \phi^{(1)}_{l,M} (z_1, \overline{z_1}) \otimes
	\phi^{(2)}_{m,M} (z_2, \overline{z_2}) \otimes \phi^{(3)}_{n,M} (z_3, \overline{z_3}),
}
where $l,m,n$ are KK indices and $\psi^{(i)}_l$ is a 2D spinor describing the $l$-th KK mode on the $i$-th $T^2$, whose exact form is $\psi^{(i)}_l = \left( \psi^{(i)}_{l,+}, \psi^{(i)}_{l,-} \right)^\text{T}$ and the corresponding 2D chirality ($+$ or $-$) is denoted by $s_i$.
We {adopt} the gamma matrices $\tilde{\Gamma}^m$ (identified by the Cartesian coordinates) corresponding to the $i$-th torus as
\al{
\tilde{\Gamma}^{2i+2} = i\sigma_1,\quad
\tilde{\Gamma}^{2i+3} = i\sigma_2,
}
where {$\sigma_{1,2}$ are Pauli matrices}.
In the following part, we only focus on the zero modes $(l=m=n=0)$ and {omit} the KK indices hereafter.

We introduce factorizable Abelian magnetic fluxes on the three $T^2$ through the classical vector potential of $A_M$ in the following forms
\al{
A^{(b)}(\{z_i, \overline{z_i}\}) &= \sum_{i=1}^3 \frac{\pi}{{g} \text{Im} \tau_i}
\begin{pmatrix}
M_1^{(i)} \text{Im}\left[ (\overline{z_i} + \overline{C^{(i)}_1}) dz_i \right] \mathbf{1}_{N_1 \times N_1} & & 0 \\
& \ddots & \\
0 & & M_{n}^{(i)} \text{Im}\left[ (\overline{z_i} + \overline{C^{(i)}_n}) dz_i \right] \mathbf{1}_{N_n \times N_n}
\end{pmatrix} \notag \\
	&= \sum_{i=1}^3 \frac{1}{4 i \text{Im} \tau_i} \left( \overline{{\boldsymbol B}^{(i)}} dz_i - {\boldsymbol B}^{(i)} d\overline{z_i} \right) \notag \\
	&= \sum_{i=1}^3 \left( A_{z_i}^{(b)}(\overline{z_i}) dz_i + A_{\overline{z_i}}^{(b)}(z_i) d\overline{z_i}  \right),
	\label{1-form_potential_general}
}
where $C^{(i)}_j\ (j=1,\cdots,n)$ represent the corresponding Wilson line phases on $T^2_i$, and $M^{(i)}_j\ (j=1,\cdots,n)$ should be integers {because of} Dirac's quantization condition on $T^2_i$.
On $T^2_i$, no restriction is there on possible choices of $C^{(i)}_j$.
Under this background, the original gauge group $U(N)$ explicitly breaks down as $U(N) \to \Pi_{a=1}^n U(N_a)$ with $N = \sum_{a=1}^n N_a$.
We can derive the following relations,
\al{
A^{(b)}{(\{ z_i + \delta_{ij}, \overline{z_i} + \delta_{ij} \})} &= A^{(b)}(\{z_i, \overline{z_i}\}) + {d_j} \xi_{1}(\{z_i\}),\quad {(j=1,2,3)} \\
A^{(b)}{(\{ z_i + \delta_{ij} \tau_j, \overline{z_i} + \delta_{ij} \overline{\tau_j} \})} &= A^{(b)}(\{z_i, \overline{z_i}\}) + {d_j} \xi_{\tau}(\{z_i\}),\quad {(j=1,2,3)} \\
\xi_{1}(\{z_i\}) = \sum_{i=1}^3 \xi_{1_i}(z_i) &= \sum_{i=1}^3 \frac{1}{2\text{Im}\tau_i} \text{Im}[{\boldsymbol B}^{(i)}],\\
\xi_{\tau}(\{z_i\}) = \sum_{i=1}^3 \xi_{\tau_i}(z_i) &= \sum_{i=1}^3 \frac{1}{2\text{Im}\tau_i} \text{Im}[\overline{\tau_i}{\boldsymbol B}^{(i)}],
}
where {$d_j$} plays as an exterior derivative on {$T^2_j$}.
Here, the Lagrangian density in Eq.~(\ref{10D_action}) should be single-valued under every torus identification $z_i \sim z_i +1 \sim z_i + \tau_i\ (i=1,2,3)$, and then in the gaugino fields $\lambda(x, \{z_i, \overline{z_i}\})$, the following pseudo-periodic boundary conditions should be arranged,
\al{
\lambda{(x, \{ z_i + \delta_{ij}, \overline{z_i} + \delta_{ij} \})} &= {U_{1_j}(z_j)} \lambda(x, \{z_i, \overline{z_i}\}) {U_{1_j}(z_j)^\dagger},\quad {(j=1,2,3)} \\
\lambda{(x, \{ z_i + \delta_{ij} \tau_j, \overline{z_i} + \delta_{ij} \overline{\tau_j} \})} &= {U_{\tau_j}(z_j)} \lambda(x, \{z_i, \overline{z_i}\}) {U_{\tau_j}(z_j)^\dagger},\quad {(j=1,2,3)}
}
with
\al{
U_{1_i}(z_i) := e^{i {g} \xi_{1_i}(z_i) + 2\pi i {\boldsymbol \alpha}^{(i)}}, \quad
U_{\tau_i}(z_i) := e^{i {g} \xi_{\tau_i}(z_i) + 2\pi i {\boldsymbol \beta}^{(i)}},
	\label{boundarycondition_general}
}
\vspace{-6mm}
\al{
{\boldsymbol \alpha}^{(i)} :=
\begin{pmatrix}
\alpha^{(i)}_{1} \mathbf{1}_{N_1 \times N_1} & & 0 \\
& \ddots & \\
0 & & \alpha^{(i)}_{n}  \mathbf{1}_{N_n \times N_n}
\end{pmatrix},\quad
{\boldsymbol \beta}^{(i)} :=
\begin{pmatrix}
\beta^{(i)}_{1} \mathbf{1}_{N_1 \times N_1} & & 0 \\
& \ddots & \\
0 & & \beta^{(i)}_{n}  \mathbf{1}_{N_n \times N_n}
\end{pmatrix},
	\label{SS_matrix}
}
where $\alpha^{(i)}_{j}$ and $\beta^{(i)}_{j}$ $(j=1,\cdots,n)$ describe Scherk-Schwarz phases and can take any real numbers.

Now, we {consider} the specific case of $U(N) \to U(N_a) \times U(N_b)$ for our concrete understanding on the system.
In this case, the 1-form potential in Eq.~(\ref{1-form_potential_general}) takes the shape,
\al{
A^{(b)}(\{z_i, \overline{z_i}\}) &= \sum_{i=1}^3 \frac{\pi}{{g} \text{Im} \tau_i}
\begin{pmatrix}
M_a^{(i)} \text{Im}\left[ (\overline{z_i} + \overline{C^{(i)}_a}) dz_i \right] \mathbf{1}_{N_a \times N_a} & 0 \\
0 & M_{b}^{(i)} \text{Im}\left[ (\overline{z_i} + \overline{C^{(i)}_b}) dz_i \right] \mathbf{1}_{N_b \times N_b}
\end{pmatrix},
}
and the gaugino fields $\lambda$ and their $i$-th torus part are decomposed as follows:
\al{
\lambda(x, \{z_i, \overline{z_i}\}) =
\begin{pmatrix}
\lambda^{aa}(x, \{z_i, \overline{z_i}\}) & \lambda^{ab}(x, \{z_i, \overline{z_i}\}) \\
\lambda^{ba}(x, \{z_i, \overline{z_i}\}) & \lambda^{bb}(x, \{z_i, \overline{z_i}\})
\end{pmatrix},\quad
\psi^{(i)}(z_i, \overline{z_i}) =
\begin{pmatrix}
\psi^{(i)aa}(z_i, \overline{z_i}) & \psi^{(i)ab}(z_i, \overline{z_i}) \\
\psi^{(i)ba}(z_i, \overline{z_i}) & \psi^{(i)bb}(z_i, \overline{z_i})
\end{pmatrix}.
}
The fields $\lambda^{aa}$ and $\lambda^{bb}$ play as the gaugino fields under the unbroken gauge group $U(N_a) \times U(N_b)$, while $\lambda^{ab}$ and $\lambda^{ba}$ correspond to the bi-fundamental matter fields as $(N_a, \overline{N_b})$ and $(\overline{N_a}, N_b)$, respectively.
We obtain the zero-mode equations for these gaugino fields on the $i$-th $T^2$ with the 2D chirality ($+$ or $-$) as
\al{
\begin{pmatrix}
\partial_{\overline{z_i}} \psi^{(i)aa}_{+} &
\left[ \partial_{\overline{z_i}} + \frac{\pi}{2\text{Im} \tau_i} \left( M^{(i)}_{ab} z_i + C'^{(i)}_{ab} \right) \right] \psi^{(i)ab}_{+} \\
\left[ \partial_{\overline{z_i}} + \frac{\pi}{2\text{Im} \tau_i} \left( M^{(i)}_{ba} z_i + C'^{(i)}_{ba} \right) \right]  \psi^{(i)ba}_{+} &
\partial_{\overline{z_i}} \psi^{(i)bb}_{+}
\end{pmatrix} &= 0,
\label{2D_gauginoequation_plus} \\
\begin{pmatrix}
\partial_{z_i} \psi^{(i)aa}_{-} &
\left[ \partial_{z_i} - \frac{\pi}{2\text{Im} \tau_i} \left( M^{(i)}_{ab} \overline{z_i} + \overline{C'^{(i)}_{ab}} \right) \right] \psi^{(i)ab}_{-} \\
\left[ \partial_{z_i} - \frac{\pi}{2\text{Im} \tau_i} \left( M^{(i)}_{ba} \overline{z_i} + \overline{C'^{(i)}_{ba}} \right) \right]  \psi^{(i)ba}_{-} &
\partial_{z_i} \psi^{(i)bb}_{-}
\end{pmatrix} &= 0,
\label{2D_gauginoequation_minus}
}
with the short-hand notations $M^{(i)}_{ab} := M^{(i)}_{a} - M^{(i)}_{b}$ and $C'^{(i)}_{ab} := M^{(i)}_a C^{(i)}_a - M^{(i)}_b C^{(i)}_b$.
The block-diagonal parts of the gaugino fields, $\psi^{(i)aa}$ and $\psi^{(i)bb}$, do not feel magnetic flux and Wilson line phase, which is apparent from Eqs.~(\ref{2D_gauginoequation_plus}) and (\ref{2D_gauginoequation_minus}).
On the other hand, in the block-off-diagonal parts, $\psi^{(i)ab}$ and $\psi^{(i)ba}$, magnetic {fluxes} and Wilson line {phases} affect properties of the matter fields.
From the information in Eqs.~(\ref{boundarycondition_general}) and (\ref{SS_matrix}), the effective boundary conditions of the fields are easily written down,
\al{
\psi^{(i)ab}_{s_i}{(z_i +1, \overline{z_i} +1)} &= e^{i\frac{\pi s_i}{\text{Im} \tau_i} \text{Im} \left[ M^{(i)}_{ab} z_i + C'^{(i)}_{ab} \right] + 2 \pi i \alpha^{(i)}_{ab}} \psi^{(i)ab}_{s_i}{(z_i, \overline{z_i})}, \notag \\
\psi^{(i)ba}_{s_i}{(z_i +1, \overline{z_i} +1)} &= e^{i\frac{\pi s_i}{\text{Im} \tau_i} \text{Im} \left[ M^{(i)}_{ba} z_i + C'^{(i)}_{ba} \right] + 2 \pi i \alpha^{(i)}_{ba}} \psi^{(i)ba}_{s_i}{(z_i, \overline{z_i})}, \notag \\
\psi^{(i)aa}_{s_i}{(z_i +1, \overline{z_i} +1)} &= \psi^{(i)aa}_{s_i}{(z_i, \overline{z_i})}, \notag \\
\psi^{(i)bb}_{s_i}{(z_i +1, \overline{z_i} +1)} &= \psi^{(i)bb}_{s_i}{(z_i, \overline{z_i})},\\[4pt]
\psi^{(i)ab}_{s_i}{(z_i + \tau_i, \overline{z_i} + \overline{\tau_i})} &= e^{i\frac{\pi s_i}{\text{Im} \tau_i} \text{Im} \left[ \overline{\tau_i} \left( M^{(i)}_{ab} z_i + C'^{(i)}_{ab} \right) \right] + 2 \pi i \beta^{(i)}_{ab}} \psi^{(i)ab}_{s_i}{(z_i, \overline{z_i})}, \notag \\
\psi^{(i)ba}_{s_i}{(z_i + \tau_i, \overline{z_i} + \overline{\tau_i})} &= e^{i\frac{\pi s_i}{\text{Im} \tau_i} \text{Im} \left[ \overline{\tau_i} \left( M^{(i)}_{ba} z_i + C'^{(i)}_{ba} \right) \right] + 2 \pi i \beta^{(i)}_{ba}} \psi^{(i)ba}_{s_i}{(z_i, \overline{z_i})}, \notag \\
\psi^{(i)aa}_{s_i}{(z_i + \tau_i, \overline{z_i} + \overline{\tau_i})} &= \psi^{(i)aa}_{s_i}{(z_i, \overline{z_i})}, \notag \\
\psi^{(i)bb}_{s_i}{(z_i + \tau_i, \overline{z_i} + \overline{\tau_i})} &= \psi^{(i)bb}_{s_i}{(z_i, \overline{z_i})},
}
{with the short-hand notations, $\alpha^{(i)}_{ab} := \alpha^{(i)}_{a} - \alpha^{(i)}_{b}$ and $\beta^{(i)}_{ab} := \beta^{(i)}_{a} - \beta^{(i)}_{b}$.
We remind that $s_i$ shows the corresponding 2D chirality.}

When $M^{(i)}_{ab} > 0$, the fields $\psi^{(i)ab}_{+}$ and $\psi^{(i)ba}_{-}$ contain $|M^{(i)}_{ab}|$ normalizable zero modes, while the others $\psi^{(i)ba}_{+}$ and $\psi^{(i)ab}_{-}$ have no corresponding one.
On the other hand in $M^{(i)}_{ab} < 0$, $|M^{(i)}_{ab}|$ normalizable zero modes are generated from each of $\psi^{(i)ba}_{+}$ and $\psi^{(i)ab}_{-}$, whereas there is nothing {from} $\psi^{(i)ab}_{+}$ and $\psi^{(i)ba}_{-}$.
In the case of $M^{(i)}_{ab} = 0$, like $\psi^{(i)aa}_{s_i}$ or $\psi^{(i)bb}_{s_i}$, only one non-localized mode is generated from each of the all sectors and no phenomenological interests occur.
When $M^{(i)}_{ab} > 0${, which is equal to $M^{(i)}_{ba} < 0$}, the {wave functions} of $\psi^{(i)ab}_{+}$ and $\psi^{(i)ba}_{-}$ take the following forms:
\al{
\psi^{(i)ab}_{+}(z_i) =
\sum_{I=0}^{|M^{(i)}_{ab}|-1}
\begin{pmatrix} \Theta^{(I + \alpha^{(i)}_{ab}, \beta^{(i)}_{ab})}_{M^{(i)}_{ab},a^{(i)}_{ab}} (z_i, \tau_i) \\ 0 \end{pmatrix}, \quad
\psi^{(i)ba}_{-}(\overline{z_i}) =
\sum_{I=0}^{|M^{(i)}_{ba}|-1}
\begin{pmatrix} 0 \\ \Theta^{(I + \alpha^{(i)}_{ba}, \beta^{(i)}_{ba})}_{M^{(i)}_{ba},\overline{a^{(i)}_{ba}}} (\overline{z_i}, \overline{\tau_i})\end{pmatrix},
}
\al{
\Theta^{(I + \alpha^{(i)}_{ab}, \beta^{(i)}_{ab})}_{M^{(i)}_{ab},a^{(i)}_{ab}} (z_i, \tau_i) &=
\mathcal{N}_{|M^{(i)}_{ab}|}
e^{i\pi M^{(i)}_{ab} (z_i + a^{(i)}_{ab}){\mathrm{Im}(z_i + a^{(i)}_{ab})\over \mathrm{Im}\tau_i}}\cdot \vartheta \left[
\begin{array}{c}
{I + \alpha^{(i)}_{ab} \over M^{(i)}_{ab}} \\ -\beta^{(i)}_{ab}
\end{array}
\right] (M^{(i)}_{ab} (z_i + a^{(i)}_{ab}), M^{(i)}_{ab} \tau_i), \\
\Theta^{(I + \alpha^{(i)}_{ba}, \beta^{(i)}_{ba})}_{M^{(i)}_{ba},\overline{a^{(i)}_{ba}}} (\overline{z_i}, \overline{\tau_i}) &=
\mathcal{N}_{|M^{(i)}_{ba}|}
e^{i\pi M^{(i)}_{ba} (\overline{z_i} + \overline{a^{(i)}_{ba}}){\mathrm{Im}(\overline{z_i} + \overline{a^{(i)}_{ba}})\over \mathrm{Im} \overline{\tau_i}}}\cdot \vartheta \left[
\begin{array}{c}
{I + \alpha^{(i)}_{ba} \over M^{(i)}_{ba}} \\ -\beta^{(i)}_{ba}
\end{array}
\right] (M^{(i)}_{ba} (\overline{z_i} + \overline{a^{(i)}_{ba}}), M^{(i)}_{ba} \overline{\tau_i}),
}
with the effective Wilson line phase parameter $a^{(i)}_{ab} := C'^{(i)}_{ab}/M^{(i)}_{ab}$.
Here, $I\ (=0,\cdots,|M^{(i)}_{ab}|-1)$ discriminates the $|M^{(i)}_{ab}|$-degenerated zero-mode states.
The $\vartheta$ function is defined by 
\begin{align}
&\vartheta \left[
\begin{array}{c}
a\\ b
\end{array}
\right] (c\nu,c\tau) 
=\sum_{l=-\infty}^{\infty}e^{i\pi (a+l)^2c\tau}e^{2\pi i(a+l)(c\nu +b)} ,
\end{align}
with the properties
\begin{align}
&\vartheta \left[
\begin{array}{c}
a\\ b
\end{array}
\right] (c(\nu +n),c\tau)
=e^{2\pi i acn}\vartheta \left[
\begin{array}{c}
a\\ b
\end{array}
\right] (c\nu,c\tau), \notag \\
&\vartheta \left[
\begin{array}{c}
a\\ b
\end{array}
\right] (c(\nu +n\tau),c\tau)
=e^{-i\pi cn^2\tau -2\pi i n(c\nu +b)}\vartheta \left[
\begin{array}{c}
a\\ b
\end{array}
\right] (c\nu,c\tau), \notag \\
&\vartheta \left[
\begin{array}{c}
a+m\\ b+n
\end{array}
\right] (c\nu,c\tau)
=e^{2\pi i an}\vartheta \left[
\begin{array}{c}
a\\ b
\end{array}
\right] (c\nu,c\tau),\notag \\
&\vartheta \left[
\begin{array}{c}
a\\ b
\end{array}
\right] (c\nu,c\tau)
	=
\vartheta \left[
\begin{array}{c}
a\\ 0
\end{array}
\right] (c\nu +b,c\tau),
	\label{thetafunction_properties}
\end{align}
where $a$ and $b$ are real numbers, $c$, $m$ and $n$ are integers, and $\nu$ and $\tau$ are complex numbers with $\mathrm{Im}\tau >0$.
The following orthonormality condition determines the normalization factor $\mathcal{N}_{|M^{(i)}_{ab}|}$,
\al{
\int_{T^2_i} d^2z_i
\left( \Theta^{(I + \alpha^{(i)}_{ab}, \beta^{(i)}_{ab})}_{M^{(i)}_{ab},a^{(i)}_{ab}} (z_i, \tau_i) \right)^\ast
\left( \Theta^{(J + \alpha^{(i)}_{ab}, \beta^{(i)}_{ab})}_{M^{(i)}_{ab},a^{(i)}_{ab}} (z_i, \tau_i) \right)
&= \delta_{I,J} \quad (M^{(i)}_{ab} > 0), \notag \\
\int_{T^2_i} d^2z_i
\left( \Theta^{(I + \alpha^{(i)}_{ba}, \beta^{(i)}_{ba})}_{M^{(i)}_{ba},a^{(i)}_{ba}} (\overline{z_i}, \overline{\tau_i}) \right)^\ast
\left( \Theta^{(J + \alpha^{(i)}_{ba}, \beta^{(i)}_{ba})}_{M^{(i)}_{ba},a^{(i)}_{ba}} (\overline{z_i}, \overline{\tau_i}) \right)
&= \delta_{I,J} \quad (M^{(i)}_{ba} < 0),
}
with $d^2z_i := dz_i d\overline{z_i}$.
We note that the total number of zero modes of $\psi^{(i)ab}_{+}$ and $\psi^{(i)ba}_{-}$ (when $M^{(i)}_{ab} > 0$) {is} given as $\Pi_{i=1}^3 |M_{ab}^{(i)}|$, where we should replace $|M_{ab}^{(i)}|$ by one when $M_{ab}^{(i)} = 0$.
In the opposite case with $M^{(i)}_{ab} < 0$, situations are similar.
Here, we can derive an important relationship easily (in the case of $M^{(i)}_{ab} > 0$),
\al{
\left( \Theta^{(I + \alpha^{(i)}_{ab}, \beta^{(i)}_{ab})}_{M^{(i)}_{ab},a^{(i)}_{ab}} (z_i, \tau_i) \right)^\ast
=
\Theta^{(-I + \alpha^{(i)}_{ba}, \beta^{(i)}_{ba})}_{M^{(i)}_{ba},\overline{a^{(i)}_{ba}}} (\overline{z_i}, \overline{\tau_i}),
	\label{CC_theta}
}
where the index $I$ is identified under the condition, $\text{mod}\ |M^{(i)}_{ab}|$, and we can always redefine $-I$ as $I'(=0,\cdots, |M^{(i)}_{ab}|-1)$.

In the $U(N)$ SYM on the magnetized tori, we can realize the following gauge symmetry breaking by the flux as a typical example, $U(8) \to {U(4)_{PSC}} \times U(2)_L \times U(2)_R$, where $U(4)_{PSC}$ is the Pati-Salam color gauge group, and $U(2)_L$ and $U(2)_R$ are the left- and right-electroweak gauge groups, respectively.
Here, we can find all the {Minimal Supersymmetric Standard Model (MSSM) fields} in this case.
Precisely speaking, the requirement is that the SM gauge group $SU(3)_C \times SU(2)_L \times U(1)_Y$ should be intact under the existence of the flux. Thereby, other {possibilities} of symmetry breaking via flux, e.g., $U(8) \to U(3)_C \times U(1)_1 \times U(2)_L \times U(2)_R$ or $U(8) \to U(3)_C \times U(1)_1 \times U(2)_L \times U(1)_2 \times U(1)_3$ {are} also reasonable.

On flux background, zero-mode profiles are not only split, also localized around {points different from} each other.
Then, we can expect that hierarchical values in Yukawa couplings are created via overlap integrals in the Yukawa sector of this model.
The concrete form of the Yukawa couplings are as follows:
\al{
Y_{\mathcal{I},\mathcal{J},\mathcal{K}} &=
	c\, \lambda^{(1)}_{I_1, J_1, K_1} \lambda^{(2)}_{I_2, J_2, K_2} \lambda^{(3)}_{I_3, J_3, K_3}, \\
\lambda^{(i)}_{I_i, J_i, K_i} &= \int_{T^2_i} d^2z_i \,
	\Theta^{(I_i + \alpha^{(i)}_{I}, \beta^{(i)}_{I})}_{M^{(i)}_{I},a^{(i)}_{I}} (z_i, \tau_i)
	\Theta^{(J_i + \alpha^{(i)}_{J}, \beta^{(i)}_{J})}_{M^{(i)}_{J},a^{(i)}_{J}} (z_i, \tau_i)
	\left(
	\Theta^{(K_i + \alpha^{(i)}_{K}, \beta^{(i)}_{K})}_{M^{(i)}_{K},a^{(i)}_{K}} (z_i, \tau_i)
	\right)^\ast,
	\label{i-th_torus_Yukawa}
}
where $c$ is a constant factor via gauge structure and $\mathcal{I} = (I_1, I_2, I_3),\, \mathcal{J} = (J_1, J_2, J_3),\, \mathcal{K} = (K_1, K_2, K_3)$, respectively.
In a suitable symmetry breaking like the above examples, we can find conditions on the parameters,
\al{
M^{(i)}_{I} + M^{(i)}_{J} &= M^{(i)}_{K}, \label{selectionrune_M} \\
\alpha^{(i)}_{I} + \alpha^{(i)}_{J} &= \alpha^{(i)}_{K}, \label{selectionrune_alpha} \\
\beta^{(i)}_{I} + \beta^{(i)}_{J} &= \beta^{(i)}_{K}, \label{selectionrune_beta} \\
M^{(i)}_{I} a^{(i)}_{I} + M^{(i)}_{J} a^{(i)}_{J} &= M^{(i)}_{K} a^{(i)}_{K}, \label{selectionrune_a}
}
where we implicitly use the rule in Eq.~(\ref{selectionrune_M}) and the relation in Eq.~(\ref{CC_theta}) when we write down the actual form in Eq.~(\ref{i-th_torus_Yukawa}).
After utilizing the last property in Eq.~(\ref{thetafunction_properties}) and the following formula about the theta function,
\al{
&\vartheta \left[
\begin{array}{c}
r/N_1 \\ 0
\end{array}
\right] (z_1, N_1 \tau)
	\times
\vartheta \left[
\begin{array}{c}
s/N_2 \\ 0
\end{array}
\right] (z_2, N_2 \tau)
\quad
(r,s \in \mathbb{R}; N_1, N_2 \in \mathbb{Z}; z_1, z_2, \tau \in \mathbb{C})
\notag \\
&= \sum_{m \in Z_{N_1 + N_2}}
\vartheta \left[
\begin{array}{c}
\frac{r+s+N_1 m}{N_1 + N_2} \\ 0
\end{array}
\right] (z_1 + z_2, \tau (N_1 + N_2)) \notag \\
&\quad \times
\vartheta \left[
\begin{array}{c}
\frac{N_2 r - N_1 s + N_1 N_2 m}{N_1 N_2 (N_1 + N_2)} \\ 0
\end{array}
\right] (z_1 N_2 - z_2 N_1, \tau N_1 N_2 (N_1 + N_2)),
}
with reminding the constraints in Eqs.~(\ref{selectionrune_M})--(\ref{selectionrune_a}), we can derive the analytical result of the Yukawa coupling in Eq.~(\ref{i-th_torus_Yukawa}) as
\al{
\lambda^{(i)}_{I_i, J_i, K_i} &= \frac{\mathcal{N}_{M^{(i)}_I} \mathcal{N}_{M^{(i)}_J}}{\mathcal{N}_{M^{(i)}_K}}
	e^{\frac{i\pi}{\text{Im} \tau_i} \left[ a^{(i)}_I \text{Im}\left(M^{(i)}_I a^{(i)}_I\right) +
	a^{(i)}_J \text{Im}\left(M^{(i)}_J a^{(i)}_J\right) - a^{(i)}_K \text{Im}\left(M^{(i)}_K a^{(i)}_K\right) \right]}
	\notag \\
	&\quad \times \sum_{m \in Z_{M^{(i)}_K}}
\vartheta \left[
\begin{array}{c}
\frac{M^{(i)}_J \left(I_i + \alpha^{(i)}_I\right) - M^{(i)}_I \left(J_i + \alpha^{(i)}_J\right) + m M^{(i)}_I M^{(i)}_J}{M^{(i)}_I M^{(i)}_J M^{(i)}_K} \\ 0
\end{array}
\right] {(X, Y)} \notag \\
&\quad \times
	\delta_{I_i + \alpha^{(i)}_I + J_i + \alpha^{(i)}_J + m M^{(i)}_I, \, K_i + \alpha^{(i)}_K + \ell M^{(i)}_K},
	\label{general_Yukawa_formula}
}
with {$X := M^{(i)}_I \beta^{(i)}_J - M^{(i)}_J \beta^{(i)}_I + M^{(i)}_I M^{(i)}_J \left(a^{(i)}_I - a^{(i)}_J\right)$, $Y := \tau_i M^{(i)}_I M^{(i)}_J M^{(i)}_K$ and} possible choices of integers $\ell$.\footnote{
In another word, we consider the Kronecker's delta with the condition ``$\text{mod}\ |M^{(i)}_{K}|$''.
}

At the end of this subsection, we mention an efficient technique discussed in~\cite{Abe:2013bca,Abe:2014noa}.
In the Abelian magnetic flux, we can set either the (complex) Wilson line phase $C^{(i)}_{j}$, or the Scherk-Schwarz phases $\alpha^{(i)}_j$ and $\beta^{(i)}_j$ as zero (in each $j$ individually) without loss of generality.
This fact could make the following analysis simplified.

\subsection{$U(N)$ gauge theory on orbifolded $(T^2)^3$}

In this subsection, we examine the $U(N)$ SYM theory on magnetized $(T^2)^3$ with orbifolding.
In our configuration where the six extra dimensions are factorized as three 2-tori, apparently, we mainly focus on one of the tori $T^2_i$ for a general discussion.

On $T^2_i$, possible twisted orbifolding is to impose the covariance on the fields under the rotation with the angle $\omega$, $z_i \to \omega z_i$, where $\omega$ is $e^{2\pi i/N}$ with $N=2,3,4,6$.
In other words, $Z_2$, $Z_3$, $Z_4$ and $Z_6$ (twisted) orbifoldings are realizable on $T^2_i$.
In non-Abelian gauge theories, a nontrivial gauge structure part $P$ appears in the $Z_N$ manipulation as,
\al{
A_{\mu}{(x, \{ \omega z_i, \overline{\omega} \overline{z_i} \})} &= P A_{\mu}(x,\{z_i, \overline{z_i}\}) P^{-1}, \\
A_{z_i}{(x, \{ \omega z_i, \overline{\omega} \overline{z_i} \})} &= \overline{\omega} P A_{z_i}(x,\{z_i, \overline{z_i}\}) P^{-1}, \\
A_{\overline{z_i}}{(x, \{ \omega z_i, \overline{\omega} \overline{z_i} \})} &= \omega P A_{\overline{z_i}}(x,\{z_i, \overline{z_i}\}) P^{-1}, \\
\lambda_{s_i=+}{(x, \{ \omega z_i, \overline{\omega} \overline{z_i} \})} &= P \lambda_{s_i=+}(x,\{z_i, \overline{z_i}\}) P^{-1}, \\
\lambda_{s_i=-}{(x, \{ \omega z_i, \overline{\omega} \overline{z_i} \})} &= \omega P \lambda_{s_i=-}(x,\{z_i, \overline{z_i}\}) P^{-1} \label{ZNcondition_lambda_minus},
}
where $P$ should satisfy the conditions, $P \in U(N)$ and  $P^N = {\mathbf 1_{N \times N}}$.
Here, to prevent an additional explicit gauge symmetry breaking via the orbifoldings, we should take the following form in $P$,
\al{
P = \begin{pmatrix}
\eta_1 {\mathbf 1_{N_1 \times N_1}} & & 0 \\
& \ddots & \\
0 & & \eta_n {\mathbf 1_{N_n \times N_n}}
\end{pmatrix},
}
with $\eta_j = \{1,\omega,\cdots,\omega^{N-1}\} \ (j=1,\cdots,n)$.
Within the concrete example of $U(N) \to U(N_a) \times U(N_b)$ discussed in the previous subsection, $\psi^{(i)aa}_{+}$ and $\psi^{(i)bb}_{+}$ have trivial $Z_N$ parity {irrespective of} the values of $\eta_a$ and $\eta_b$, while $\psi^{(i)ab}_{+}$ and $\psi^{(i)ba}_{+}$ can contain nontrivial values of $\eta_a \overline{\eta_b}$, $\overline{\eta_a} \eta_b$, respectively.
The conditions for the 2D gauginos with negative chirality are evaluated with ease by use of the relation in Eq.~(\ref{ZNcondition_lambda_minus}).

Now, the gauge structure under the orbifoldings is declared, and thus we can concentrate on an actual case with the $Z_N$ parity $\eta^{(i)}$, the (2D) positive chirality and the {state-discriminating} index $I$ on $T^2_i$ as,
\al{
\psi^{(i)}_{+,\eta^{(i)}}(z_i) = \sum_{I=0}^{|M^{(i)}| - 1} \psi^{(i), I}_{+,\eta^{(i)}}(z_i),\quad
\psi^{(i), I}_{+,\eta^{(i)}}(z_i) =
\begin{pmatrix}
\widetilde{\Theta}^{(I_i+\alpha^{(i)}, \beta^{(i)})}_{M^{(i)}, a^{(i)};\eta^{(i)}} (z_i, \tau_i) \\ 0
\end{pmatrix},
}
where we assume that $M^{(i)}$ is {a} positive integer.
Note that the correspondence to the negative chirality case is basically straightforward by the replacements, $z_i \to \overline{z_i}$, $\tau_i \to \overline{\tau_i}$ and $a^{(i)} \to \overline{a^{(i)}}$.

Constructing the concrete form of $\widetilde{\Theta}^{(I_i+\alpha^{(i)}, \beta^{(i)})}_{M^{(i)}, a^{(i)}; \eta^{(i)}} (z_i, \tau_i)$ itself can be done straightforwardly just following the general recipe as
\al{
\widetilde{\Theta}^{(I_i+\alpha^{(i)}, \beta^{(i)})}_{M^{(i)}, a^{(i)}; \eta^{(i)}} (z_i, \tau_i) =
\frac{1}{N} \sum_{x=0}^{N-1} \left(\overline{\eta^{(i)}}\right)^x
\Theta^{(I_i+\alpha^{(i)}, \beta^{(i)})}_{M^{(i)}, a^{(i)}} (\omega^x z_i, \tau_i),
	\label{ZN_rotation}
}
where, apparently, this state has the eigenvalue $\eta^{(i)}$ under $z_i \to \omega z_i$.
Since the function $\Theta(z_i)$ forms a complete set, then in principle, we can express the $Z_N$-rotated one as a linear combination of $\Theta(z_i)$.
If we derive the coefficients of such a linear combination analytically, we can construct the $T^2_i/Z_N$ mode functions in an exact manner.\footnote{
Within the direct calculations with {theta functions}, it is very {difficult} to fix analytical forms of the coefficients.
Fortunately, the method developed in Ref.~\cite{Abe:2014noa} based on operator formalism resolves the technical difficulty.
By considering a ``dual'' two-dimensional quantum mechanical system, we can estimate all the coefficients in an analytical way.
}
Here, following the discussions in~\cite{Abe:2014noa}, we take the following {basis},
\al{
a^{(i)} = 0,
\label{our_gauge_choice}
}
for simplicity.
Within taking the Abelian magnetic flux, we can always select this choice without any loss of generality~\cite{Abe:2013bca,Abe:2014noa}.

By use of the results in~\cite{Abe:2014noa}, the $Z_N$-transformed states (in the {basis}~(\ref{our_gauge_choice})) can be expressed as
\al{
\Theta^{(I_i+\alpha^{(i)}, \beta^{(i)})}_{M^{(i)}, 0} (\omega^x z_i, \tau_i)
=
\sum_{J_i = 0}^{|M^{(i)}|-1} C_{I_i J_i}^{(\omega^x)}
\Theta^{(J_i+\alpha^{(i)}, \beta^{(i)})}_{M^{(i)}, 0} (z_i, \tau_i),
}
with the expansion coefficients
\al{
C_{I_i J_i}^{(\omega^x)} = \int_{T^2_i} d^2z_i
\left( \Theta^{(J_i+\alpha^{(i)}, \beta^{(i)})}_{M^{(i)}, 0} (z_i, \tau_i) \right)^\ast
\Theta^{(I_i+\alpha^{(i)}, \beta^{(i)})}_{M^{(i)}, 0} (\omega^x z_i, \tau_i),
}
{where explicit forms of $C_{I_i J_i}^{(\omega^x)}$ are summarized in Appendix~\ref{sec:T2ZN_information}.}
Using this information, the relation in~(\ref{ZN_rotation}) is written,
\al{
\widetilde{\Theta}^{(I_i+\alpha^{(i)}, \beta^{(i)})}_{M^{(i)}, 0; \eta^{(i)}} (z_i, \tau_i) &=
\sum_{J_i = 0}^{|M^{(i)}|-1}
\left( \frac{1}{N} \sum_{x=0}^{N-1} \left(\overline{\eta^{(i)}}\right)^x C_{I_i J_i}^{(\omega^x)} \right)
\Theta^{(J_i+\alpha^{(i)}, \beta^{(i)})}_{M^{(i)}, 0} (z_i, \tau_i) \notag \\
&=:
\sum_{J_i = 0}^{|M^{(i)}|-1}
M_{I_i J_i}^{(Z_N; \eta^{(i)})}
\Theta^{(J_i+\alpha^{(i)}, \beta^{(i)})}_{M^{(i)}, 0} (z_i, \tau_i),
	\label{defining_M}
}
where the number of independent physical states can be calculated as
\al{
(\text{$\#$ of physical states}) = \text{Rank}\left[ M_{I_i J_i}^{(Z_N; \eta^{(i)})} \right].
}
In general after orbifolding, the number of independent physical states should be reduced as $\text{Rank}\left[ M_{I_i J_i}^{(Z_N; \eta^{(i)})} \right] < |M^{(i)}|$, but exact numbers are only available after concrete calculation.
This procedure is possible since operator formalism tells us the analytical forms of $C_{I_i J_i}^{(\omega^x)}$~\cite{Abe:2014noa}.
The {detailed} information on the numbers of the survived physical states and the forms of $C_{I_i J_i}^{(\omega^x)}$ are summarized in Appendix~\ref{sec:T2ZN_information}.

Also, operator formalism helps us when we {calculate} the kinetic terms on $T^2_i/Z_N$ described as
\al{
\mathcal{K}_{I_i J_i}^{(Z_N; \eta^{(i)})} =
\int_{T^2_i} d^2z_i
\left( \widetilde{\Theta}^{(I_i+\alpha^{(i)}, \beta^{(i)})}_{M^{(i)}, 0; \eta^{(i)}} (z_i, \tau_i) \right)^\ast
\widetilde{\Theta}^{(J_i+\alpha^{(i)}, \beta^{(i)})}_{M^{(i)}, 0; \eta^{(i)}} (z_i, \tau_i).
}
After some calculations in operators and states, we can reach the simple correspondence,
\al{
\mathcal{K}_{I_i J_i}^{(Z_N; \eta^{(i)})} = M_{J_i I_i}^{(\eta^{(i)})}.
	\label{KandM_correspondence}
}
Here, we should mention that the kinetic terms $\mathcal{K}_{I_i J_i}^{(Z_N; \eta^{(i)})}$ are in general non-diagonal, and thereby we should diagonalize them to know the correct forms with suitable normalization.
This process itself is, at least in principle, doable for any larger $|M^{(i)}|$ through the Gram-Schmidt process for orthonormalization in linear algebra.
After the unitary transformation with the corresponding diagonalizing matrix $U^{(Z_N; \eta^{(i)})}$, $\mathcal{K}_{I_i J_i}^{(Z_N; \eta^{(i)})}$ should be performed as
\al{
\mathcal{K}^{(Z_N; \eta^{(i)})} \to \left(U^{(Z_N; \eta^{(i)})}\right)^\dagger \mathcal{K}^{(Z_N; \eta^{(i)})} U^{(Z_N; \eta^{(i)})}
=
\text{diag}(\underbrace{1,\cdots,1}_{{\text{Rank}\left[ \mathcal{K}^{(Z_N; \eta^{(i)})} \right]}}, 0,\cdots,0).
	\label{diagonalizing_K}
}
The row index of the $|M^{(i)}|$-by-$|M^{(i)}|$ diagonalizing matrix $U^{(Z_N; \eta^{(i)})}$ means a discriminator for the states on the original geometry $T^2_i$, whereas the column index identifies the physical eigenstates, where this index is meaningful only up to {$\text{Rank}\left[ \mathcal{K}^{(Z_N; \eta^{(i)})} \right] - 1$ from zero}.
Therefore in the physical eigenstates after considering the correct normalization in the kinetic terms,
$T^2_i/Z_N$ mode function should be
\al{
\widetilde{\Theta}^{(I_i+\alpha^{(i)}, \beta^{(i)})}_{M^{(i)}, 0; \eta^{(i)}} (z_i, \tau_i) \to
\sum_{I_i = 0}^{|M^{(i)}|-1}
\widetilde{\Theta}^{(I_i+\alpha^{(i)}, \beta^{(i)})}_{M^{(i)}, 0; \eta^{(i)}} (z_i, \tau_i)
\left(U^{(Z_N; \eta^{(i)})}\right)_{I_i I'_i},
	\label{to_physicalbasis}
}
where $I'_i$ {is} the index of physical eigenstates.

Based on the knowledge, we briefly discuss {the way of calculating Yukawa couplings} under the $Z_N$ symmetry.
The $T^2_i$ part is formulated (with the {basis} fixing in Eq.~(\ref{our_gauge_choice})) in the ``\,$T^2_i$\,'' eigenbasis as
\al{
\widetilde{\lambda}^{(i)}_{I_i, J_i, K_i} &= \int_{T^2_i} d^2z_i \,
	\widetilde{\Theta}^{(I_i + \alpha^{(i)}_{I}, \beta^{(i)}_{I})}_{M^{(i)}_{I}, 0; \eta_{I}^{(i)}} (z_i, \tau_i)
	\widetilde{\Theta}^{(J_i + \alpha^{(i)}_{J}, \beta^{(i)}_{J})}_{M^{(i)}_{J}, 0; \eta_{J}^{(i)}} (z_i, \tau_i)
	\left(
	\widetilde{\Theta}^{(K_i + \alpha^{(i)}_{K}, \beta^{(i)}_{K})}_{M^{(i)}_{K}, 0; \eta_{K}^{(i)}} (z_i, \tau_i)
	\right)^\ast,
	\label{T2ZN_Yukawa}
}
where we find the condition on the $Z_N$ parities (via the invariance of the system),
\al{
\eta_{I}^{(i)} \eta_{J}^{(i)} \overline{\eta_{K}^{(i)}} = 1.
}
Note that a complex-conjugated state {holds} the corresponding complex-conjugated $Z_N$ parity, which is clearly understandable from Eq.~(\ref{ZN_rotation}).
Now, evaluating $\widetilde{\lambda}^{(i)}_{I_i, J_i, K_i}$ turns out to be straightforward with the help of Eqs.~(\ref{defining_M}), (\ref{KandM_correspondence}), (\ref{diagonalizing_K}) and (\ref{to_physicalbasis}).

\section{Possibilities of {three generations} in $U(N)$ models \label{sec:classification}}

In this section, we classify all the reasonable possibilities of {three-generation} models under the $Z_2$, $Z_3$, $Z_4$ and $Z_6$ orbifoldings.
When we start from 10D $U(N)$ SYM theory, the basic pattern of the gauge symmetry breaking under the magnetic flux is $U(N) \to U(N_a) \times U(N_b) \times U(N_c)$ with $N = N_a + N_b + N_c$, where the corresponding 1-form potential is
\al{
&A^{(b)}(\{z_i, \overline{z_i}\}) = \sum_{i=1}^3 \frac{\pi}{q \text{Im} \tau_i} \times \notag \\
&\quad\text{diag}\left(
M_a^{(i)} \text{Im}\left[ ( \overline{z_i} + \overline{C^{(i)}_a} ) dz_i \right] \mathbf{1}_{N_a \times N_a},
M_b^{(i)} \text{Im}\left[ ( \overline{z_i} + \overline{C^{(i)}_b} ) dz_i \right] \mathbf{1}_{N_b \times N_b},
M_{c}^{(i)} \text{Im}\left[ ( \overline{z_i} + \overline{C^{(i)}_c} ) dz_i \right] \mathbf{1}_{N_c \times N_c}\right).
}
Following the notation in the previous {example} of $U(N) \to U(N_a) \times U(N_b)$, we find six types of bi-fundamental matter fields, $\lambda^{ab}, \lambda^{bc}, \lambda^{ca}, \lambda^{ba}, \lambda^{cb}, \lambda^{ac}$, whose gauge properties are $(N_a, \overline{N_b})$, $(N_b, \overline{N_c})$, $(\overline{N_a}, N_c)$, $(\overline{N_a}, N_b)$, $(\overline{N_b}, N_c)$, $(N_a, \overline{N_c})$, respectively.
When we {adopt} the choice, $N_a = 4$, $N_b = 2$, $N_c = 2$, $U(4)_{PSC} \times U(2)_L \times U(2)_R$ gauge groups are realized from the $U(8)$ group up to $U(1)$ factors, where the subscripts $PSC$, $L$ and $R$ denote the Pati-Salam color, left- and right-electroweak gauge groups, respectively.\footnote{
Some of the combinations of the {$U(1)$} part would be {anomalous. Then they could} be massive and decoupled via the Green-Schwarz mechanism.
}
In such a situation, when the actual chirality of the gaugino is left (negative), {$\lambda^{ab}$ corresponds to the left-handed quarks and leptons, and $\lambda^{ca}$ accords with (charge-conjugated) right-handed quarks and leptons, respectively}.
When the magnetic fluxes are suitably assigned, the situation with three-generations is materialized.
Besides, $\lambda^{bc}$ plays as up-type and down-type Higgsinos.
After we assume that (4D $\mathcal{N}=1$) supersymmetry is preserved at least locally at the $ab$, $bc$ and $ca$ sectors, the corresponding Higgses via extra-dimensional components of the 10D vector fields {are} still massless under the fluxes and the number of the fields are the same with Higgsino fields.
Also, no tachyonic mode is expected at the tree level.
Here, in general, multiple Higgs fields appear from the $bc$ sector.
Interestingly, when $\lambda^{ab}$, $\lambda^{bc}$ and $\lambda^{ca}$ have zero modes, $\lambda^{ba}$, $\lambda^{cb}$ and $\lambda^{ac}$ cannot contain any zero mode and thus no exotic particle {arises} from these fermionic sectors.
In the case of the actual chirality being right (positive), we should flip the roles of the two categories.

As a phenomenological point of view, we can consider the following additional breakdowns originating from flux, $U(4)_{PSC} \to U(3)_C \times U(1)_1$ and $U(2)_R \to U(1)_2 \times U(1)_3$ (up to U(1) factors), where $U(3)_C$ is the color gauge group (up to {a $U(1)$ factor}).
Under the latter breaking, the up-type and down-type Higgsino/Higgs sectors can feel different magnetic fluxes {individually.
Consequently, the numbers} of the two types of the fields become not the same.

\subsection{Strategy in classification}

In the $U(N)$ model, the effective Yukawa couplings after the integration among the six extra dimensions are symbolically written {as}
\al{
Y_{ijk} = {c} \int_{(T^2)^3} d^6z \psi_{L_i}(z) \psi_{R_j}(z) \phi_{H_k}(z),
}
where {$c$ is a constant factor.} $\psi_{L_i}(z)$, $\psi_{R_j}(z)$ and $\phi_{H_k}(z)$ denote zero mode {wave functions} of the {left-, right-handed} matter fields and the Higgs fields, and their generations are indicated by the indices $i$, $j$ and $k$, respectively.
When the left- and right-handed matters are localized among different tori, the Yukawa couplings are factorized like
\al{
Y_{ijk} = a_{ik} b_{jk}.
}
{The rank of this type of {matrices} (about $i$ and $j$) is one and no room would be there for generating realistic flavor structure.}
We can reach the same conclusion about the Higgs fields.
Thereby, realizing the Yukawa matrix holding rank three requires the situation that all the three types of the fields are localized on {the same} torus like
\al{
Y_{ijk} = a^{(1)}_{ijk} a^{(2)} a^{(3)}.
	\label{Yukawa_structure}
}
In this case, the contributions from $a^{(2)}$ and $a^{(3)}$ only affect the overall normalization, {and the structure of the Yukawa matrix is} governed only by $a^{(1)}$.
{Thus, we can} mainly focus on the torus determining the matter generations.

Based on the knowledge which we have obtained through the discussions before, we can easily understand the constraints on the parameters describing each sector in the $U(N)$ theory,
\al{
M^{(i)}_{ab} + M^{(i)}_{bc} + M^{(i)}_{ca} &= 0, \notag \\
\alpha^{(i)}_{ab} + \alpha^{(i)}_{bc} + \alpha^{(i)}_{ca} &= 0, \notag \\
\beta^{(i)}_{ab} + \beta^{(i)}_{bc} + \beta^{(i)}_{ca} &= 0, \notag \\
M^{(i)}_{ab} a^{(i)}_{ab} + M^{(i)}_{bc} a^{(i)}_{bc} + M^{(i)}_{ca} a^{(i)}_{ca} &= 0, \notag \\
\eta^{(i)}_{ab} \eta^{(i)}_{bc} \eta^{(i)}_{ca} &= 1,
	\label{constraints_on_parameters}
}
{where the above parameters are defined by the fundamental ones like $M_{ab}^{(i)} = M^{(i)}_{a} - M^{(i)}_{b}$ except the $Z_N$ parities}.
The $Z_N$ parities are described as
\al{
\eta_{ab}^{(i)} = \eta_{a}^{(i)} \overline{\eta_{b}^{(i)}}, \quad
\eta_{bc}^{(i)} = \eta_{b}^{(i)} \overline{\eta^{(i)}_{c}}, \quad
\eta_{ca}^{(i)} = \eta_{c}^{(i)} \overline{\eta_{a}^{(i)}}.
}
When we remember the property under complex conjugation in Eq.~(\ref{CC_theta}), we can reach the form in Eq.~(\ref{i-th_torus_Yukawa}).
As discussed in~\cite{Abe:2013bca,Abe:2014noa}, it is convenient to choose a specific {basis} where either the Wilson line {phases} or the Scherk-Schwarz phases {are} zero.
In the case of Abelian magnetic flux, this is possible without loss of generality.
The way of the correspondence in the parameters in between the two {bases} where the (complex) Wilson line {phases are} zero $\{\alpha, \beta, Ma = 0\}$ and the Scherk-Schwarz phases are zero $\{ \tilde{\alpha} = \tilde{\beta} = 0, M \tilde{a} \}$ {is} as follows with the torus modulus parameter $\tau$~\cite{Abe:2013bca}:
\al{
M \tilde{a} = \alpha \tau - \beta.
	\label{interprelation_rule}
}
{In the following analysis, like the discussions in the previous section, we {adopt} the basis where the Scherk-Schwarz phases are nonzero, and correspondingly the Wilson line phases are zero.
The interpretation into the basis with nonzero Wilson line phases and Scherk-Schwarz phases being zero is straightforward, where we just obey the relation in Eq.~(\ref{interprelation_rule}).}

In the following part, we search for all the phenomenological possibilities with three generations in the quarks and the leptons.
Here, we {adopt} the strategy that we scan over all the combinations of the effective parameters describing the $ab$, $bc$ and $ca$ sectors under the constraints in Eq.~(\ref{constraints_on_parameters}).
Note that the $ba$, $cb$ and $ac$ sectors are completely fixed after we determine the configurations of the above three parts.
As we {discussed} before, the number of generations is exactly calculable with the help of operator formalism and actual numbers are summarized in Appendix~\ref{sec:T2ZN_information}.
We use this information for the selection cut in realized matter generations.
Since our attention is on the Yukawa structure schematically depicted in Eq.~(\ref{Yukawa_structure}), it is enough to focus on a 2-torus of the geometry.
Thus hereafter in this section, we drop the index $i$ for identifying the tori.

Before going to scanning, we {focus on} how many configurations we should consider to exhaust possibilities under the requirements in Eq.~(\ref{constraints_on_parameters}).
Apparently from~(\ref{constraints_on_parameters}), the information on the {$bc$} sector is totally determined after we set the parameters of the other two sectors.
Here, we look at the condition on the magnetic fluxes.
The first line of Eq.~(\ref{constraints_on_parameters}) tells us that at least one of the signs of the three fluxes should be different from the others
{and it is enough that we focus on the two possibilities in the signs of the fluxes of the matter sectors}, $M_{ab}$ and {$M_{ca}$} as,\footnote{
{There is another possibility of $M_{ab} > 0,\ M_{bc} < 0$, but this case is physically the same as $M_{ab} < 0,\ M_{bc} > 0$.}
}
\al{
M_{ab} < 0,\ {M_{ca}} < 0; \quad
M_{ab} < 0,\ {M_{ca}} > 0.
	\label{scanningcondition_M_1}
}
Besides, after ignoring the difference coming from the combinatorics, we can introduce the additional condition,
\al{
{|M_{ab}| \leq |{M_{ca}}|}.
	\label{scanningcondition_M_2}
}
Consequently, we only take into account of the magnetic fluxes being possible under the conditions in (\ref{scanningcondition_M_1}) and (\ref{scanningcondition_M_2}).

We also comment on the other parameters.
After the orbifolding, the Wilson line {phases} and the Scherk-Schwarz phases no longer take arbitrary values~\cite{Abe:2013bca,Abe:2014noa}.
At this moment in time, {combinatorics} degrees of freedom {are} fixed by our preconditions for analysis in Eqs.~(\ref{scanningcondition_M_1}) and (\ref{scanningcondition_M_2}), and then we should scan over all the possibilities in the other parameters.
The possible values of the Wilson line {phases} and the Scherk-Schwarz phases on $T^2/Z_N$ {in the specific basis} are depicted in Appendix~\ref{sec:T2ZN_information}.

\subsection{Results}

Now, {we are ready to evaluate} the numbers of allowed possibilities in effective parameter configurations of the $Z_2$, $Z_3$, $Z_4$ and $Z_6$ cases.
Summaries of our analysis are found in Tables~\ref{tbl:Z2_classification_result}\,($Z_2$), \ref{tbl:Z3_classification_result}\,($Z_3$), \ref{tbl:Z4_classification_result}\,($Z_4$) and \ref{tbl:Z6_classification_result}\,($Z_6$), where ``General cases'' and ``Trivial BC's only'' means the cases with and without nontrivial {Scherk-Schwarz} phases, respectively.
Here, we only show the numbers of allowed parameter configurations in $M_{ab},M_{ca} < 0$ and $M_{ab} < 0,\, M_{ca} > 0$ separately.
Corresponding numbers of the Higgs pairs ($N_H$) are also shown.
If an allowed configuration says $M_{bc} = 0$, where one non-localized Higgs pair appears without magnetic background and this situation {would not be} interesting in the phenomenological point of view, we discriminate this case as ``$1_\text{trivial}$'' from the one Higgs cases with magnetic flux ``$1$''.

We mention about the $Z_2$ result.
The ``Trivial BC's only'' case in $Z_2$ was already analyzed in Ref.~\cite{Abe:2008sx} and our result is totally consistent with them.
The complete information on $Z_{2,3,4,6}$ configurations is summarized in Appendix~\ref{sec:Z2_configuration_full}.
Brief comments are also given here.
After allowing nonzero Wilson line {phases}, the configurations with seven Higgs pairs arise in the $Z_2$ case as shown in Table~\ref{tbl:Z2_classification_result}.
In the cases of $Z_3$, $Z_4$ and $Z_6$ summarized in Tables~\ref{tbl:Z3_classification_result}, \ref{tbl:Z4_classification_result} and \ref{tbl:Z6_classification_result}, we find the situations where two or four Higgs pairs are realized, while those with nine pairs never occur.

\begin{table}[H]
\begin{center}
\begin{tabular}{|l|l||l|l|} \hline
\multicolumn{2}{|c||}{General cases} & \multicolumn{2}{|c|}{Trivial BC's only} \\ \hline
$M_{ab},M_{ca} < 0$ & $M_{ab} < 0,\, M_{ca} > 0$ & $M_{ab},M_{ca} < 0$ & $M_{ab} < 0,\, M_{ca} > 0$ \\ \hline \hline
$41\,(N_{H} = 5)$ & $16\,(N_{H} = 1_{\text{trivial}})$ & $5\,(N_{H} = 5)$ & $4\,(N_{H} = 1_{\text{trivial}})$ \\
$56\,(N_{H} = 6)$ & $65\,(N_{H} = 1)$ & $2\,(N_{H} = 6)$ & $5\,(N_{H} = 1)$ \\
$30\,(N_{H} = 7)$ &    &     &    \\
$8\,(N_{H} = 8)$ &     & $2\,(N_{H} = 8)$ &    \\
$1\,(N_{H} = 9)$ &     & $1\,(N_{H} = 9)$ &    \\ \hline
\multicolumn{2}{|c||}{$136+81=217$ in total} & \multicolumn{2}{|c|}{$10+9=19$ in total} \\ \hline
\end{tabular}
\caption{Result in $Z_2$ case.
``General cases'' and ``Trivial BC's only'' means the cases with and without nontrivial {Scherk-Schwarz} phases, respectively.
Corresponding numbers of the Higgs pairs ($N_H$) are also shown.
The case indicated by $1_{\text{trivial}}$ means the one Higgs pair appears under non-magnetized background in $bc$ sector.
}
\label{tbl:Z2_classification_result}
\end{center}
\end{table}

\begin{table}[H]
\begin{center}
\begin{tabular}{|l|l||l|l|} \hline
\multicolumn{2}{|c||}{General cases} & \multicolumn{2}{|c|}{Trivial BC's only} \\ \hline
$M_{ab},M_{ca} < 0$ & $M_{ab} < 0,\, M_{ca} > 0$ & $M_{ab},M_{ca} < 0$ & $M_{ab} < 0,\, M_{ca} > 0$ \\ \hline \hline
$11\,(N_{H} = 4)$ & $17\,(N_{H} = 1_{\text{trivial}})$ & $1\,(N_{H} = 4)$ & $9\,(N_{H} = 1_{\text{trivial}})$ \\
$83\,(N_{H} = 5)$ & $142\,(N_{H} = 1)$ & $6\,(N_{H} = 5)$ & $27\,(N_{H} = 1)$ \\
$190\,(N_{H} = 6)$ & $21\,(N_{H} = 2)$ & $7\,(N_{H} = 6)$ &    \\
$83\,(N_{H} = 7)$ &     & $6\,(N_{H} = 7)$ &    \\
$11\,(N_{H} = 8)$ &     & $1\,(N_{H} = 8)$ &    \\ \hline
\multicolumn{2}{|c||}{$378+180=558$ in total} & \multicolumn{2}{|c|}{$21+36=57$ in total} \\ \hline
\end{tabular}
\caption{Result in $Z_3$ case. The convention is the same with in Table~\ref{tbl:Z2_classification_result}.}
\label{tbl:Z3_classification_result}
\end{center}
\end{table}

\begin{table}[H]
\begin{center}
\begin{tabular}{|l|l||l|l|} \hline
\multicolumn{2}{|c||}{General cases} & \multicolumn{2}{|c|}{Trivial BC's only} \\ \hline
$M_{ab},M_{ca} < 0$ & $M_{ab} < 0,\, M_{ca} > 0$ & $M_{ab},M_{ca} < 0$ & $M_{ab} < 0,\, M_{ca} > 0$ \\ \hline \hline
$9\,(N_{H} = 4)$ & $24\,(N_{H} = 1_{\text{trivial}})$ & $3\,(N_{H} = 4)$ & $12\,(N_{H} = 1_{\text{trivial}})$ \\
$128\,(N_{H} = 5)$ & $228\,(N_{H} = 1)$ & $37\,(N_{H} = 5)$ & $60\,(N_{H} = 1)$ \\
$254\,(N_{H} = 6)$ & $18\,(N_{H} = 2)$ & $59\,(N_{H} = 6)$ & $6\,(N_{H} = 2)$ \\
$120\,(N_{H} = 7)$ &     & $27\,(N_{H} = 7)$ &    \\
$17\,(N_{H} = 8)$ &     & $10\,(N_{H} = 8)$ &    \\ \hline
\multicolumn{2}{|c||}{$528+270=798$ in total} & \multicolumn{2}{|c|}{$136+78=214$ in total} \\ \hline
\end{tabular}
\caption{Result in $Z_4$ case. The convention is the same with in Table~\ref{tbl:Z2_classification_result}.}
\label{tbl:Z4_classification_result}
\end{center}
\end{table}

\begin{table}[H]
\begin{center}
\begin{tabular}{|l|l||l|l|} \hline
\multicolumn{2}{|c||}{General cases} & \multicolumn{2}{|c|}{Trivial BC's only} \\ \hline
$M_{ab},M_{ca} < 0$ & $M_{ab} < 0,\, M_{ca} > 0$ & $M_{ab},M_{ca} < 0$ & $M_{ab} < 0,\, M_{ca} > 0$ \\ \hline \hline
$14\,(N_{H} = 4)$ & $24\,(N_{H} = 1_{\text{trivial}})$ & $4\,(N_{H} = 4)$ & $12\,(N_{H} = 1_{\text{trivial}})$ \\
$156\,(N_{H} = 5)$ & $282\,(N_{H} = 1)$ & $45\,(N_{H} = 5)$ & $73\,(N_{H} = 1)$ \\
$326\,(N_{H} = 6)$ & $27\,(N_{H} = 2)$ & $76\,(N_{H} = 6)$ & $8\,(N_{H} = 2)$ \\
$150\,(N_{H} = 7)$ &     & $36\,(N_{H} = 7)$ &    \\
$20\,(N_{H} = 8)$ &     & $10\,(N_{H} = 8)$ &    \\ \hline
\multicolumn{2}{|c||}{$666+333=999$ in total} & \multicolumn{2}{|c|}{$171+93=264$ in total} \\ \hline
\end{tabular}
\caption{Result in $Z_6$ case. The convention is the same with in Table~\ref{tbl:Z2_classification_result}.}
\label{tbl:Z6_classification_result}
\end{center}
\end{table}

\section{Examples in $Z_2$ with nontrivial twisting \label{sec:examples}}

Based on the results which we have achieved in the previous section, we {can} examine how much {nontrivial} boundary conditions by the Scherk-Schwarz phases (or the Wilson line phases) affect the flavor structure in some specific models with the $Z_2$ orbifolding.
Here, we {adopt} the {basis} with the Wilson line phases being zero and only debate with the torus where all the matter fields are assumed to be localized.
As we already discussed, the above assumption is justified by the phenomenological requirement for flavors.
Here, we rewrite the expressions of the Yukawa couplings {$\lambda_{I, J, K}$ on $T^2$, and $\widetilde{\lambda}_{I, J, K}$ on $T^2/Z_N$}, in Eqs.~(\ref{i-th_torus_Yukawa}) and (\ref{T2ZN_Yukawa}), respectively,
\al{
\lambda_{I, J, K} &= \int_{T^2} d^2z \,
	\Theta^{(I + \alpha_{I}, \beta_{I})}_{M_{I}, 0} (z, \tau)
	\Theta^{(J + \alpha_{J}, \beta_{J})}_{M_{J}, 0} (z, \tau)
	\left(
	\Theta^{(K + \alpha_{K}, \beta_{K})}_{M_{K}, 0} (z, \tau)
	\right)^\ast, \\
\widetilde{\lambda}_{I, J, K} &= \int_{T^2} d^2z \,
	\widetilde{\Theta}^{(I + \alpha_{I}, \beta_{I})}_{M_{I}, 0; \eta_{I}} (z, \tau)
	\widetilde{\Theta}^{(J + \alpha_{J}, \beta_{J})}_{M_{J}, 0; \eta_{J}} (z, \tau)
	\left(
	\widetilde{\Theta}^{(K + \alpha_{K}, \beta_{K})}_{M_{K}, 0; \eta_{K}} (z, \tau)
	\right)^\ast,
}
with dropping the torus index $i$ for convenience.
From Eq.~(\ref{general_Yukawa_formula}), the analytical form of the Yukawa on $T^2$ without nonzero Wilson line {phases} is represented as
\al{
\lambda_{I, J, K} &= \frac{\mathcal{N}_{M_I} \mathcal{N}_{M_J}}{\mathcal{N}_{M_K}}
\sum_{m \in Z_{M_K}}
\vartheta \left[
\begin{array}{c}
\frac{M_J \left(I + \alpha_I\right) - M_I \left(J + \alpha_J\right) + m M_I M_J}{M_I M_J M_K} \\ 0
\end{array}
\right] (M_I \beta_J - M_J \beta_I, \tau M_I M_J M_K) \notag \\
&\quad \times \delta_{I + \alpha_I + J + \alpha_J + m M_I, \, K + \alpha_K + \ell M_K}.
	\label{general_Yukawa_formula_2}
}
In the ``kinetic'' eigenbasis, the Yukawa {couplings are} expressed as
\al{
\widetilde{\lambda}'_{I',J',K'} = \sum_{I=0}^{|M_I|-1} \sum_{J=0}^{|M_J|-1} \sum_{K=0}^{|M_K|-1}
\widetilde{\lambda}_{I, J, K} \left(U^{Z_2; \eta_I}\right)_{I,I'} \left(U^{Z_2; {\eta_J}}\right)_{J,J'}
\left({U^{Z_2; {\eta_K}}}\right)_{K,K'}^\ast
}
where the indices for identifying kinetic eigenstates, $I'$, $J'$, $K'$, have $\text{Rank}\left[ M^{(Z_2; \eta_I)} \right]$, $\text{Rank}\left[ M^{(Z_2; \eta_J)} \right]$, $\text{Rank}\left[ M^{(Z_2; \eta_K)} \right]$ numbers of nonzero configurations, respectively. 
In general, the mixing effect through $U^{Z_N; \eta_I}$ contributes to the physics.
{However} in the $Z_2$ case, the kinetic terms are diagonal in the $Z_2$-orbifolded basis (irrespective of values of the Scherk-Schwarz phases) and we can set the diagonalizing matrix when we calculate Yukawa couplings as
\al{
\left(U^{Z_N; \eta_I}\right)_{J,J'} \to  \delta_{J,J'}.
}
The reason behind this simplicity is that the structure of orbifolded functions is uninvolved, where we estimate the number of survived physical modes analytically and explicitly as in Appendix~\ref{sec:T2ZN_information}.\footnote{
When $T^2/Z_{3,4,6}$, we can also analytically evaluate the numbers in operator analysis, but an explicit formula is not obtained yet.
}
In the $Z_{3,4,6}$ orbifoldings, nontrivial effects via $U^{Z_N; \eta_I}$ might be expected.

\subsection{Numerical analysis in one-pair Higgs model}

From the result in the previous section on the $Z_2$ general case, we see the number of the possibilities with one-pair Higgs is strikingly enlarged as $65$ from $5$.
Note that the modulus parameter $\tau$ of the torus does not receive any restriction from the $Z_2$ orbifolding, in contrast to the other three cases.
As discussed in~\cite{Abe:2008sx}, when the greater value {the modulus parameter} $\tau$ takes, the more significant hierarchy could be realized.
{However,} in the simple five cases without nonzero Scherk-Schwarz {phases}, we can check that no sizable hierarchy is generated even when we take a large value.
{For example, even when we {adopt} $\tau = 10 \, i$, only up to $\mathcal{O}(10^2)$-order hierarchy can be generated.}

Through the following example, we will see how the nonzero phases improve the magnitude in the hierarchy.
Here, we consider the configuration,
\al{
&(M_{bc}, M_{ca}, M_{ab}) = (-2, -4, +6), \notag \\
&(\alpha_{bc}, \alpha_{ca}, \alpha_{ab}) = (0, 0, 0), \notag \\
&(\beta_{bc}, \beta_{ca}, \beta_{ab}) = (1/2, 0, 1/2), \notag \\
&(\eta_{bc}, \eta_{ca}, \eta_{ab}) = (1, 1, 1), \notag \\[6pt]
&(M_{bc'}, M_{c'a}, M_{ab}) = (-1, -5, +6), \notag \\
&(\alpha_{bc'}, \alpha_{c'a}, \alpha_{ab}) = (0, 0, 0), \notag \\
&(\beta_{bc'}, \beta_{c'a}, \beta_{ab}) = (1/2, 0, 1/2), \notag \\
&(\eta_{bc'}, \eta_{c'a}, \eta_{ab}) = (1, 1, 1),
	\label{oneHiggs_parameters}
}
where $U(2)_{R}$ breaks down {to} $U(1)_{R1}$ and $U(1)_{R2}$ identified by the indices $c$ and $c'$.
Now, {the left-handed quarks, right-handed up/down-type quarks} and {up}/down-type Higgs bosons live in the sectors, $ab$, $bc/bc'$ and $ca/c'a$, respectively.
This case belongs to the category ``$M_{ab} < 0, M_{ca} > 0$'' in the previous classification and the correspondence to Eq.~(\ref{general_Yukawa_formula_2}) is straightforward, {\it e.g.}, $|M_{bc}| \to M_{I}, \ |M_{ca}| \to M_{J}, \ |M_{ab}| \to M_{K}$.
By following the formula in Eq.~(\ref{general_Yukawa_formula}), each element of the up/down Yukawa couplings are calculated analytically by use of the forms $\eta_N^{(u)}/\eta_N^{(d)}$,
\al{
\eta_N^{(u)} := \vartheta
\left[
\begin{array}{c}
N/M_u \\ 0
\end{array}
\right] (-2, M_u \tau),\quad
\eta_N^{(d)} := \vartheta
\left[
\begin{array}{c}
N/M_d \\ 0
\end{array}
\right] (-5/2, M_d \tau),
}
where they are functions of $N$ and $M_u$ and $M_d$ are defined as $M_u := |M_{ab} M_{bc} M_{ca}| = 48$ and $M_d := |M_{ab} M_{bc'} M_{c'a}| = 30$.
{$\eta_N^{(u)}$ and $\eta_N^{(d)}$ are equivalent to $\eta_{N+ M_u}^{(u)}$ and $\eta_{N+ M_d}^{(d)}$, respectively.}
When we change $N$, the values of the functions {alter} exponentially, whose maxima and minima are around $N=0$ and $N=M_{u,d}/2$, respectively.
Apart {from} the trivial Scherk-Schwarz case, the ``$c\nu$'' index of the theta function in $\eta_N^{(u,d)}$ can take nonzero values.
The correspondence to the $c\nu = 0$ expression is easily obtainable when we utilize the first relation in Eq.~(\ref{thetafunction_properties})
as
\al{
\eta_N^{(u)} = e^{2\pi i \frac{N}{M_u} (-2)}
\vartheta
\left[
\begin{array}{c}
N/M_u \\ 0
\end{array}
\right] (0, M_u \tau), \quad
\eta_N^{(d)} = e^{2\pi i \frac{N}{M_d} \left(-\frac{5}{2}\right)}
\vartheta
\left[
\begin{array}{c}
N/M_d \\ 0
\end{array}
\right] (0, M_d \tau),
	\label{eta_relation}
}
where there are only $\mathcal{O}(1)$ differences between {$\eta^{(u,d)}_N$} with nonzero $c\nu$ and with $c\nu = 0$.

On the other hand, allowed values of $N$ with nonzero Scherk-Schwarz phases are changed since {such phases modify} the selection rule described with Kronecker's delta in Eq.~(\ref{general_Yukawa_formula_2}).
When we pick up only the most significant term in each matrix element with ignoring $\mathcal{O}(1)$ factor,
the results are as follows,
\al{
\widetilde{\lambda'}^{(u)}_{I',J'} \sim
\begin{pmatrix}
\eta_0^{(u)} & 0 & \eta_{12}^{(u)} \\
0 & \eta_{2}^{(u)} & 0 \\
\eta_{8}^{(u)} & 0 & \eta_{4}^{(u)}
\end{pmatrix},\quad
\widetilde{\lambda'}^{(d)}_{I',J'} \sim
\begin{pmatrix}
\eta_{0}^{(d)} & \eta_{6}^{(d)} & \eta_{12}^{(d)} \\
\eta_{5}^{(d)} & \eta_{1}^{(d)} & \eta_{23}^{(d)} \\
\eta_{10}^{(d)} & \eta_{4}^{(d)} & \eta_{2}^{(d)}
\end{pmatrix},
	\label{oneHiggs_Yukawas}
}
where we omit to show the index about the Higgs because only one Higgs pair {is} there.
Now, we show the resultant values in Table~\ref{tbl:oneHiggs_result},\footnote{
In the numerical calculation, we include all the sub-leading terms which we omit to represent in Eq.~(\ref{oneHiggs_Yukawas}).
{In this calculation, note that the ratio of top and bottom masses is determined only by the ratio of VEVs, so-called $\tan{\beta}$ in the MSSM since we assume a MSSM-like Higgs sector.}
}
where we notice that the desirable amount of hierarchy is suitably realized in the mass values.\footnote{
{Here, we assume that the Yukawa couplings are proportional to the mass values, like in the SM.}
}
However, if we try to generate the realistic CKM mixing, a more complicated mixing pattern would be required.

\begin{table}[t]
\begin{center}
\begin{tabular}{|c||c|c|} \hline
 & {Obtained} values & {Actual} values \\ \hline
$(m_u, m_c, m_t)/m_t$ & $(9.9 \times 10^{-6}, 2.8 \times 10^{-2}, 1)$ & $(1.5 \times 10^{-5}, 7.5 \times 10^{-3}, 1)$ \\ \hline
$(m_d, m_s, m_b)/m_b$ & $(5.0 \times 10^{-3}, 1.6 \times 10^{-2}, 1)$ & $(1.2 \times 10^{-3}, 2.3 \times 10^{-2}, 1)$ \\ \hline
$|V_{\text{CKM}}|$ &
$
\begin{pmatrix}
0.99 & 3.1 \times 10^{-8} & 0 \\
3.1 \times 10^{-8} & 0.99 & 2.6 \times 10^{-13} \\
0 & 2.6 \times 10^{-13} & 0.99
\end{pmatrix}
$
&
$
\begin{pmatrix}
0.97 & 0.23 & 0.0035 \\
0.23 & 0.97 & 0.041 \\
0.0087 & 0.040 & 1.0
\end{pmatrix}
$ \\ \hline
\end{tabular}
\caption{The mass ratios of the quarks and the absolute values of the CKM matrix elements when we {adopt} the parameters in Eq.~(\ref{oneHiggs_parameters}) with one Higgs pair.
We use the experimental values in Ref.~\cite{Agashe:2014kda}.}
\label{tbl:oneHiggs_result}
\end{center}
\end{table}

\subsection{Gaussian Froggatt-Nielsen mechanism with discrete Scherk-Schwarz phases}

Here, we would like to search for a configuration with realistic observed values by following the strategy discussed in~\cite{Abe:2014vza}, named Gaussian Froggatt-Nielsen mechanism.
As pointed out in~\cite{Abe:2014vza}, especially when $M_{u,d}$ are as large as $\mathcal{O}(10^2)$ or more, the forms of $\eta_N^{(u,d)}$ are approximately described by the following simple forms with good precision,
\al{
\eta_N^{(u,d)} \sim e^{- \frac{\text{Im}[\tau]\pi}{M_{u,d}} N^2},
}
where the real part of $\tau$ only contributes to this as {an} $\mathcal{O}(1)$ phase factor.
Also, we would remember that the nonzeroness in ``$c\nu$'' of the theta function does not lead to a sizable effect, only up to $\mathcal{O}(1)$ as in Eq.~(\ref{eta_relation}).
{Under some assumptions in multiple Higgs VEVs}, the Yukawa couplings would be symbolically expressed as
\al{
Y_{I',J'} \sim e^{-c(a_{I'} + b_{J'})^2},
} 
where $c$ is a common constant factor and {$a_{I'}$ and $b_{J'}$ are determined only by the magnitudes of the magnetic fluxes on the left- and right-handed fermion sectors, respectively.}
It seems to be similar to another familiar one named Froggatt-Nielsen form~\cite{Froggatt:1978nt} as
\al{
Y_{I',J'} \sim e^{-c(a_{I'} + b_{J'})},
}
where in this case, a linear form appears and $a_{I'}$ and $b_{J'}$ correspond to quantum numbers named Froggatt-Nielsen charges.
When we realize the following shape,
\al{
(Y)_{I',J'} \ll (Y)_{M',N'},
}
for $I' \leq M'$ and $J' \leq N'$, we would regenerate a desirable form seen in the CKM matrix.

In the following part, we analyze the flavor structure through Gaussian Froggatt-Nielsen mechanism with discrete Scherk-Schwarz phases in the {parameters},
\al{
&(M_{ab}, M_{ca}, M_{bc}) = (-5, -4, +9), \notag \\
&(\alpha_{ab}, \alpha_{ca}, \alpha_{bc}) = (0, 0, 0), \notag \\
&(\beta_{ab}, \beta_{ca}, \beta_{bc}) = (1/2, 0, 1/2), \notag \\
&(\eta_{ab}, \eta_{ca}, \eta_{bc}) = (1, 1, 1), \notag \\[6pt]
&(M_{ab}, M_{c'a}, M_{bc'}) = (-5, -7, +12), \notag \\
&(\alpha_{ab}, \alpha_{c'a}, \alpha_{bc'}) = (0, 0, 0), \notag \\
&(\beta_{ab}, \beta_{c'a}, \beta_{bc'}) = (1/2, 1/2, 0), \notag \\
&(\eta_{ab}, \eta_{c'a}, \eta_{bc'}) = (1, -1, -1),
	\label{fiveHiggs_parameters}
}
where the notation is the same with that in the previous analysis.
This configuration is categorized in the case, ``$M_{ab}, M_{ca} < 0$'', where {multiple Higgs fields tend to be obtained.}
The correspondence to Eq.~(\ref{general_Yukawa_formula_2}) is straightforward, {\it e.g.}, $|M_{ab}| \to M_{I}, \ |M_{ca}| \to M_{J}, \ |M_{bc}| \to M_{K}$.

Our choice predicts five pairs of up- and down-type Higgs bosons, where enough such possibilities are found in the $Z_2$ parameter landscape.
With neglecting $\mathcal{O}(1)$ phase factors and dropping sub-leading terms,
we can focus on the following part,
\al{
\eta_N^{(u)} := \vartheta
\left[
\begin{array}{c}
N/M_u \\ 0
\end{array}
\right] (-2, M_u \tau),\quad
\eta_N^{(d)} := \vartheta
\left[
\begin{array}{c}
N/M_d \\ 0
\end{array}
\right] (-1, M_d \tau), \\[6pt]
\widetilde{\lambda'}^{(u)}_{I',J',K'=3} \sim
\begin{pmatrix}
\eta_{60}^{(u)} & \eta_{15}^{(u)} & \eta_{30}^{(u)} \\
\eta_{24}^{(u)} & \eta_{21}^{(u)} & \eta_{6}^{(u)} \\
\eta_{12}^{(u)} & \eta_{3}^{(u)} & \eta_{42}^{(u)}
\end{pmatrix},\quad
\widetilde{\lambda'}^{(u)}_{I',J',K'=4} \sim
\begin{pmatrix}
\eta_{20}^{(u)} & \eta_{25}^{(u)} & \eta_{70}^{(u)} \\
\eta_{16}^{(u)} & \eta_{11}^{(u)} & \eta_{34}^{(u)} \\
\eta_{52}^{(u)} & \eta_{7}^{(u)} & \eta_{2}^{(u)}
\end{pmatrix}, \\
\widetilde{\lambda'}^{(d)}_{I',J',K'=3} \sim
\begin{pmatrix}
\eta_{40}^{(d)} & \eta_{80}^{(d)} & \eta_{20}^{(d)} \\
\eta_{68}^{(d)} & \eta_{32}^{(d)} & \eta_{8}^{(d)} \\
\eta_{16}^{(d)} & \eta_{4}^{(d)} & \eta_{64}^{(d)}
\end{pmatrix},\quad
\widetilde{\lambda'}^{(d)}_{I',J',K'=4} \sim
\begin{pmatrix}
\eta_{145}^{(d)} & \eta_{25}^{(d)} & \eta_{85}^{(d)} \\
\eta_{23}^{(d)} & \eta_{73}^{(d)} & \eta_{13}^{(d)} \\
\eta_{61}^{(d)} & \eta_{11}^{(d)} & \eta_{1}^{(d)}
\end{pmatrix},
}
with $M_u = 180$ and $M_d = 420$.
When we assume that the two Higgs pairs{, $H_{u3(d3)}$ and $H_{u4(d4)}$,} contain VEVs and the other do not, the effective Yukawa couplings $Y^{(u,d)}$ are described with the following parameters $\rho_{u,d}$,
\al{
\rho_u = \frac{\langle H_{u3} \rangle}{\langle H_{u4} \rangle},\quad
\rho_d = \frac{\langle H_{d3} \rangle}{\langle H_{d4} \rangle},
}
\al{
Y^{(u)} \sim
\begin{pmatrix}
\eta_{20}^{(u)} & \rho_u \eta_{15}^{(u)} & \rho_u \eta_{30}^{(u)} \\
\eta_{16}^{(u)} & \eta_{11}^{(u)} & \rho_u \eta_{6}^{(u)} \\
\rho_u \eta_{12}^{(u)} & \rho_u \eta_{3}^{(u)} & \eta_{2}^{(u)}
\end{pmatrix}, \quad
Y^{(d)} \sim
\begin{pmatrix}
\rho_d \eta_{40}^{(d)} & \eta_{25}^{(d)} & \rho_d \eta_{20}^{(d)} \\
\eta_{23}^{(d)} & \rho_d \eta_{32}^{(d)} & \rho_d \eta_{8}^{(d)} \\
\rho_d \eta_{16}^{(d)} & \rho_d \eta_{4}^{(d)} & \eta_{1}^{(d)}
\end{pmatrix}.
	\label{effectiveYukawas_5Higgs}
}
The choice in the two ratios in the Higgs VEVs, $\rho_u = 0.03$ and $\rho_d = 0.49$, leads to the \mbox{{(semi-)realistic}} observed values summarized in Table~\ref{tbl:fiveHiggs_result}.\footnote{
In the numerical calculation, we include all the sub-leading terms which we omit to represent in Eq.~(\ref{effectiveYukawas_5Higgs}).
{Also like the previous calculation, we do not touch the ratio between $m_t$ and $m_b$ described by the ratio of the VEVs.}
}

\begin{table}[t]
\begin{center}
\begin{tabular}{|c||c|c|} \hline
 & {Obtained} values & {Actual} values \\ \hline
$(m_u, m_c, m_t)/m_t$ & $(3.0 \times 10^{-5}, 3.6 \times 10^{-2}, 1)$ & $(1.5 \times 10^{-5}, 7.5 \times 10^{-3}, 1)$ \\ \hline
$(m_d, m_s, m_b)/m_b$ & $(2.9 \times 10^{-4}, 8.9 \times 10^{-3}, 1)$ & $(1.2 \times 10^{-3}, 2.3 \times 10^{-2}, 1)$ \\ \hline
$|V_{\text{CKM}}|$ &
$
\begin{pmatrix}
0.97 & 0.23 & 0.011 \\
0.23 & 0.97 & 0.026 \\
0.005 & 0.027 & 1.00
\end{pmatrix}
$
&
$
\begin{pmatrix}
0.97 & 0.23 & 0.0035 \\
0.23 & 0.97 & 0.041 \\
0.0087 & 0.040 & 1.0
\end{pmatrix}
$ \\ \hline
\end{tabular}
\caption{The mass ratios of the quarks and the absolute values of the CKM matrix elements when we {adopt} the parameters in Eq.~(\ref{fiveHiggs_parameters}) with five Higgs pairs.
We use the experimental values in Ref.~\cite{Agashe:2014kda}.}
\label{tbl:fiveHiggs_result}
\end{center}
\end{table}

\section{Summary and Discussion \label{sec:closing}}

{In this paper, based on the analytical formulation discussed in Ref.~\cite{Abe:2014noa}, we have classified all the individual possibilities of flavor models generating suitable three generations in the quarks and the leptons on generalized $T^2$-based magnetized orbifolds, namely $T^2/Z_2$, $T^2/Z_3$, $T^2/Z_4$, $T^2/Z_6$, with nontrivial Wilson line phases and/or Scherk-Schwarz phases.
Also, we examined the effects caused by the nontrivial phases in mass hierarchy in a configuration with one Higgs pair, and {constructed} a concrete model with the phases being nonzero values in the case of five Higgs pairs, accompanied by a realization of the SM quark structure, that is the observed mass values and mixing angles, with rather good precision.

Now, there are lots of possible suitable candidates with three generations in the choice of fluxes and related parameters, and thus to search for suitable configurations to describe the nature on $T^2/Z_{3,4,6}$ could be an interesting direction.
Here, nontrivial contributions via diagonalizing matrices of kinetic terms would be nontrivial and situations might be different from in the $Z_2$. Revealing this point is also an important work in the model building on magnetized background.
Also, like in~\cite{Abe:2012fj,Abe:2013bba}, full model embedding to 10D SYM theory is a meaningful task.}


\section*{Acknowledgments}

{Y.T. would like to thank Hiroyuki Abe for useful comments.}
This work was supported in part by scientific grants from the {Ministry of Education, Culture,
Sports, Science, and Technology} under Grants No.~25$\cdot$3825 (Y.F.), No.~25400252 (T.K.) and No.~20540274 (M.S.).



\appendix
\section*{Appendix}
\section{Information on $T^2/Z_N$ orbifold}\label{sec:T2ZN_information}

In this Appendix, we summarize the information on the numbers of the survived physical states and the forms of $C_{I_i J_i}^{(\omega^x)}$ in $T^2/Z_N$ orbifolds $(N=2,3,4,6)$ discussed in~\cite{Abe:2014noa}.
Hereafter in this section, we omit the subscript $i$ and use small roman letters for showing the indices identifying the states on $T^2$ (without orbifolding) for simplicity.
Also, we replace $\alpha$ and $\beta$ as $\alpha_1$ and $\alpha_2$, respectively.
Note that we evaluate the numbers of generations in the $Z_{3,4,6}$ cases within the range required in the discussion in section~\ref{sec:classification}.
{Note that all the following formulas are for the situation where the value of the magnetic flux is positive.
The correspondence to the negative case is understandable, {\it e.g.,} through Eq.~(\ref{CC_theta}).}

\subsection{$T^2/\Z_2$}

On the $T^2 / \Z_2$ ($\omega = e^{\pi i} = -1$), allowed values of Scherk-Schwarz phases are following four patterns,
\be
	( \alpha_1 , \alpha_2 ) = \left( 0 , 0 \right) , \left( \frac{ 1 }{ 2 } , 0 \right) , \left( 0 , \frac{ 1 }{ 2 } \right) , \left( \frac{ 1 }{ 2 } , \frac{ 1 }{ 2 } \right) \, .
\ee
The analytic form of the expansion coefficient is
\begin{align}
	C_{ j k }^{ ( \om^1 ) } = \e^{ - \tpi \i \cdot \frac{ 2 \alpha_2 }{ M } ( j + \alpha_1 ) } {\del_{ - 2 \alpha_1 - j , k }} \, .
\end{align}
We can find easily the formula of the number of the physical states by analyzing the matrix
\be
	M_{ j k }^{ ( \Z_2 ; \eta ) } = \frac{ 1 }{ 2 } \left( \del_{ j , k } + \eta \, C_{ j k }^{ ( \om^1 ) } \right) \quad ( \eta = \pm 1 ) \, .
\ee

\paragraph{$( \alpha_1 , \alpha_2 ) = ( 0 , 0 )$ case}

In this pattern, the number of independent physical states is given by
\be
\begin{split}
	( \textrm{{\#} of identical physical states for $\eta = + 1$} ) &= \left[ \frac{ M }{ 2 } \right] + 1 , \\
	( \textrm{{\#} of identical physical states for $\eta = - 1$} ) &= \left[ \frac{ M - 1 }{ 2 } \right] ,
\end{split}
\ee
where $\left[ x \right]$ is floor function.

This case has been already studied.
The above equations obtained by the operator formalism analysis reproduce the same result as the earlier study, {Table} 1 in~\cite{Abe:2008sx}.

\paragraph{$( \alpha_1 , \alpha_2 ) = ( \frac{ 1 }{ 2 } , 0 )$ case}

In this pattern, the number of independent physical states is given by
\be
\begin{split}
	( \textrm{{\#} of identical physical states for $\eta = + 1$} ) &= \left[ \frac{ M + 1 }{ 2 } \right] , \\
	( \textrm{{\#} of identical physical states for $\eta = - 1$} ) &= \left[ \frac{ M }{ 2 } \right] .
\end{split}
\ee

\paragraph{$( \alpha_1 , \alpha_2 ) = ( 0 , \frac{ 1 }{ 2 } )$ case}

In this pattern, the number of independent physical states is given by
\be
\begin{split}
	( \textrm{{\#} of identical physical states for $\eta = + 1$} ) &= \left[ \frac{ M + 1 }{ 2 } \right] , \\
	( \textrm{{\#} of identical physical states for $\eta = - 1$} ) &= \left[ \frac{ M }{ 2 } \right] .
\end{split}
\ee

\paragraph{$( \alpha_1 , \alpha_2 ) = ( \frac{ 1 }{ 2 } , \frac{ 1 }{ 2 } )$ case}

In this pattern, the number of independent physical states is given by
\be
\begin{split}
	( \textrm{{\#} of identical physical states for $\eta = + 1$} ) &= \left[ \frac{ M }{ 2 } \right] , \\
	( \textrm{{\#} of identical physical states for $\eta = - 1$} ) &= \left[ \frac{ M + 1 }{ 2 } \right] .
\end{split}
\ee

\subsection{$T^2/\Z_3$}

On the $T^2 / \Z_3$ ($\omega = e^{2\pi i/3}$), allowed values of Scherk-Schwarz phases are
\be
	\alpha \ldef \alpha_1 = \alpha_2 = \left\{ \begin{aligned}
		& 0 , \frac{ 1 }{ 3 } , \frac{ 2 }{ 3 } \quad\, ( {\rm for} ~ M = {\rm even} ) \, , \\
		& \frac{ 1 }{ 6 } , \frac{ 3 }{ 6 } , \frac{ 5 }{ 6 } \quad ( {\rm for} ~ M = {\rm odd} ) \, 
. \end{aligned} \right.
\ee
After analyzing the matrix
\be
	M_{ j k }^{ ( \Z_3 ; \eta ) } = \frac{ 1 }{ 3 } \sum_{ x = 0 }^2 \bar{ \eta }^x C_{ j k }^{ ( \om^x ) } \quad ( \eta = 1 , \om , \omb ) \, ,
\ee
described by the elements
	\begin{align}
	&C^{(\omega)}_{jk}=\frac{1}{\sqrt{M}} e^{-i\frac{\pi}{12} + i\frac{3\pi\alpha^2}{M} }e^{i\frac{\pi}{M}k(k+6\alpha)+2\pi i \frac{j\cdot k}{M}},\nonumber\\
	&C^{(\omega^2)}_{jk}=\frac{1}{\sqrt{M}} e^{i\frac{\pi}{12} - i\frac{3\pi\alpha^2}{M}-i\frac{\pi}{M}j(j+6\alpha)}e^{-2\pi i \frac{j\cdot k}{M}},
	\end{align}
we obtain the results shown in {Tables}~\ref{gen.table-Z3_00}, \ref{gen.table-Z3_24}, \ref{gen.table-Z3_15}, \ref{gen.table-Z3_33}.
\begin{table}[H]
\bec
\begin{tabular}{|c|c|ccccccccccc|} \hline
	\multicolumn{2}{|c|}{$| M |$} & 2 & 4 & 6 & 8 & 10 & 12 & 14 & 16 & 18 & 20 & 22 \\ \hline
	\multirow{3}{*}{$\eta$}	& 1 & 1 & 1 & 3 & 3 & 3 & 5 & 5 & 5 & 7 & 7 & 7 \\
		& $\omega$				& 0 & 2 & 2 & 2 & 4 & 4 & 4 & 6 & 6 & 6 & 8 \\
		& $\bar{ \omega }$		& 1 & 1 & 1 & 3 & 3 & 3 & 5 & 5 & 5 & 7 & 7 \\ \hline
\end{tabular}
\eec
\vspace{-3mm}
\caption{The numbers of linearly independent zero-mode eigenstates with $Z_3$ eigenvalue $\eta$ for $M = {\rm even}$ and $( \alpha_1 , \alpha_2 ) = ( 0 , 0 )$ on $T^2 / \Z_3$.}
\label{gen.table-Z3_00}
\end{table}
\begin{table}[H]
\bec
\begin{tabular}{|c|c|ccccccccccc|} \hline
	\multicolumn{2}{|c|}{$| M |$} & 2 & 4 & 6 & 8 & 10 & 12 & 14 & 16 & 18 & 20 & 22 \\ \hline
	\multirow{3}{*}{$\eta$}	& 1 & 1 & 2 & 2 & 3 & 4 & 4 & 5 & 6 & 6 & 7 & 8 \\
		& $\omega$				& 1 & 1 & 2 & 3 & 3 & 4 & 5 & 5 & 6 & 7 & 7 \\
		& $\bar{ \omega }$		& 0 & 1 & 2 & 2 & 3 & 4 & 4 & 5 & 6 & 6 & 7 \\ \hline
\end{tabular}
\eec
\vspace{-3mm}
\caption{The numbers of linearly independent zero-mode eigenstates with $Z_3$ eigenvalue $\eta$ for $M = {\rm even}$ and $( \alpha_1 , \alpha_2 ) = ( \frac{ 1 }{ 3 } , \frac{ 1 }{ 3 } ) , ( \frac{ 2 }{ 3 } , \frac{ 2 }{ 3 } ) $ on $T^2 / \Z_3$.}
\label{gen.table-Z3_24}
\end{table}
\begin{table}[H]
\bec
\begin{tabular}{|c|c|ccccccccccc|} \hline
	\multicolumn{2}{|c|}{$| M |$} & 1 & 3 & 5 & 7 & 9 & 11 & 13 & 15 & 17 & 19 & 21 \\ \hline
	\multirow{3}{*}{$\eta$}	& 1 & 1 & 1 & 2 & 3 & 3 & 4 & 5 & 5 & 6 & 7 & 7 \\
		& $\omega$				& 0 & 1 & 2 & 2 & 3 & 4 & 4 & 5 & 6 & 6 & 7 \\
		& $\bar{ \omega }$		& 0 & 1 & 1 & 2 & 3 & 3 & 4 & 5 & 5 & 6 & 7 \\ \hline
\end{tabular}
\eec
\vspace{-3mm}
\caption{The numbers of linearly independent zero-mode eigenstates with $Z_3$ eigenvalue $\eta$ for $M = {\rm odd}$ and $( \alpha_1 , \alpha_2 ) = ( \frac{ 1 }{ 6 } , \frac{ 1 }{ 6 } ) , ( \frac{ 5 }{ 6 } , \frac{ 5 }{ 6 } ) $ on $T^2 / \Z_3$.}
\label{gen.table-Z3_15}
\end{table}
\begin{table}[H]
\bec
\begin{tabular}{|c|c|ccccccccccc|} \hline
	\multicolumn{2}{|c|}{$| M |$} & 1 & 3 & 5 & 7 & 9 & 11 & 13 & 15 & 17 & 19 & 21 \\ \hline
	\multirow{3}{*}{$\eta$}	& 1 & 0 & 2 & 2 & 2 & 4 & 4 & 4 & 6 & 6 & 6 & 8 \\
		& $\omega$				& 1 & 1 & 1 & 3 & 3 & 3 & 5 & 5 & 5 & 7 & 7 \\
		& $\bar{ \omega }$		& 0 & 0 & 2 & 2 & 2 & 4 & 4 & 4 & 6 & 6 & 6 \\ \hline
\end{tabular}
\eec
\vspace{-3mm}
\caption{The numbers of linearly independent zero-mode eigenstates with $Z_3$ eigenvalue $\eta$ for $M = {\rm odd}$ and $( \alpha_1 , \alpha_2 ) = ( \frac{ 3 }{ 6 } , \frac{ 3 }{ 6 } ) $ on $T^2 / \Z_3$.}
\label{gen.table-Z3_33}
\end{table}

\subsection{$T^2/\Z_4$}

On the $T^2 / \Z_4$ ($\omega = e^{\pi i/2} = i$), allowed values of Scherk-Schwarz phases are
\be
	\alpha \ldef \alpha_1 = \alpha_2 = 0 , \frac{ 1 }{ 2 } \, .
\ee
After analyzing the matrix
\be
	M_{ j k }^{ ( \Z_4 ; \eta ) } = \frac{ 1 }{ 4 } \sum_{ x = 0 }^3 \bar{ \eta }^x C_{ j k }^{ ( \om^x ) } \quad ( \eta = \pm 1 , \pm \i ) \, ,
\ee
described by the elements
	\begin{align}
	&C_{jk}^{(\omega)}=\frac{1}{\sqrt{M}} e^{2\pi i \frac{\alpha^2}{M}} e^{2\pi i \frac{j\cdot k}{M}+2\pi i\frac{2\alpha}{M}k },\nonumber\\
	&C_{jk}^{(\omega^2)}=e^{-2\pi i \frac{2\alpha}{M}(\alpha+j)} \delta_{{-2\alpha-j,k}},\nonumber\\
	&C_{jk}^{(\omega^3)}=\frac{1}{\sqrt{M}} e^{-2\pi i \frac{\alpha^2}{M}-2\pi i \frac{2\alpha}{M}j }e^{-2\pi i \frac{j\cdot k}{M}},
	\end{align}
we obtain the results summarized in {Tables}~\ref{gen.table-Z4_0} and \ref{gen.table-Z4_1}.
\begin{table}[H]
\bec
\begin{tabular}{|c|c|ccccccccccccccc|} \hline
	\multicolumn{2}{|c|}{$| M |$}	& 1 & 2 & 3 & 4 & 5 & 6 & 7 & 8 & 9 & 10 & 11 & 12 & 13 & 14 & 15 \\ \hline
	\multirow{4}{*}{$\eta$}	& $+ 1$	& 1 & 1 & 1 & 2 & 2 & 2 & 2 & 3 & 3 & 3 & 3 & 4 & 4 & 4 & 4 \\
		& $+ \i$					& 0 & 0 & 1 & 1 & 1 & 1 & 2 & 2 & 2 & 2 & 3 & 3 & 3 & 3 & 4 \\
		& $- 1$						& 0 & 1 & 1 & 1 & 1 & 2 & 2 & 2 & 2 & 3 & 3 & 3 & 3 & 4 & 4 \\
		& $- \i$					& 0 & 0 & 0 & 0 & 1 & 1 & 1 & 1 & 2 & 2 & 2 & 2 & 3 & 3 & 3 \\ \hline \hline
	\multicolumn{2}{|c|}{$| M |$}	& 16 & 17 & 18 & 19 & 20 & 21 & 22 & 23 & 24 & 25 & 26 & 27 & 28 & 29 & 30 \\ \hline
	\multirow{4}{*}{$\eta$}	& $+ 1$	& 5 & 5 & 5 & 5 & 6 & 6 & 6 & 6 & 7 & 7 & 7 & 7 & 8 & 8 & 8 \\
		& $+ \i$					& 4 & 4 & 4 & 5 & 5 & 5 & 5 & 6 & 6 & 6 & 6 & 7 & 7 & 7 & 7 \\
		& $- 1$						& 4 & 4 & 5 & 5 & 5 & 5 & 6 & 6 & 6 & 6 & 7 & 7 & 7 & 7 & 8 \\
		& $- \i$					& 3 & 4 & 4 & 4 & 4 & 5 & 5 & 5 & 5 & 6 & 6 & 6 & 6 & 7 & 7 \\ \hline
\end{tabular}
\eec
\vspace{-3mm}
\caption{The numbers of linearly independent zero-mode eigenstates with $Z_4$ eigenvalue $\eta$ for $( \alpha_1 , \alpha_2 ) = ( 0 , 0 ) $ on $T^2 / \Z_4$.}
\label{gen.table-Z4_0}
\end{table}

\begin{table}[H]
\bec
\begin{tabular}{|c|c|ccccccccccccccc|} \hline
	\multicolumn{2}{|c|}{$| M |$}	& 1 & 2 & 3 & 4 & 5 & 6 & 7 & 8 & 9 & 10 & 11 & 12 & 13 & 14 & 15 \\ \hline
	\multirow{4}{*}{$\eta$}	& $+ 1$	& 0 & 1 & 1 & 1 & 1 & 2 & 2 & 2 & 2 & 3 & 3 & 3 & 3 & 4 & 4 \\
		& $+ \i$					& 1 & 1 & 1 & 1 & 2 & 2 & 2 & 2 & 3 & 3 & 3 & 3 & 4 & 4 & 4 \\
		& $- 1$						& 0 & 0 & 0 & 1 & 1 & 1 & 1 & 2 & 2 & 2 & 2 & 3 & 3 & 3 & 3 \\
		& $- \i$					& 0 & 0 & 1 & 1 & 1 & 1 & 2 & 2 & 2 & 2 & 3 & 3 & 3 & 3 & 4 \\ \hline \hline
	\multicolumn{2}{|c|}{$| M |$}	& 16 & 17 & 18 & 19 & 20 & 21 & 22 & 23 & 24 & 25 & 26 & 27 & 28 & 29 & 30 \\ \hline
	\multirow{4}{*}{$\eta$}	& $+ 1$	& 4 & 4 & 5 & 5 & 5 & 5 & 6 & 6 & 6 & 6 & 7 & 7 & 7 & 7 & 8 \\
		& $+ \i$					& 4 & 5 & 5 & 5 & 5 & 6 & 6 & 6 & 6 & 7 & 7 & 7 & 7 & 8 & 8 \\
		& $- 1$						& 4 & 4 & 4 & 4 & 5 & 5 & 5 & 5 & 6 & 6 & 6 & 6 & 7 & 7 & 7 \\
		& $- \i$					& 4 & 4 & 4 & 5 & 5 & 5 & 5 & 6 & 6 & 6 & 6 & 7 & 7 & 7 & 7 \\ \hline
\end{tabular}
\eec
\vspace{-3mm}
\caption{The numbers of linearly independent zero-mode eigenstates with $Z_4$ eigenvalue $\eta$ for $( \alpha_1 , \alpha_2 ) = ( \frac{ 1 }{ 2 } , \frac{ 1 }{ 2 } ) $ on $T^2 / \Z_3$.}
\label{gen.table-Z4_1}
\end{table}

\subsection{$T^2/\Z_6$}

On the $T^2 / \Z_6$ ($\omega = e^{\pi i/3}$), allowed values of Scherk-Schwarz phases are
\be
	\alpha \ldef \alpha_1 = \alpha_2
	= \left\{ \begin{aligned} & 0 \quad\, ( {\rm for} ~ M = {\rm even} ) \\ & \frac{ 1 }{ 2 } \quad ( {\rm for} ~ M = {\rm odd} ) \, 
. \end{aligned} \right.
\ee
After analyzing the matrix
\be
	M_{ j k }^{ ( \Z_6 ; \eta ) } = \frac{ 1 }{ 6 } \sum_{ x = 0 }^5 \bar{ \eta }^x C_{ j k }^{ ( \om^x ) } \quad ( \eta = 1 , \om , \om^2 , \om^3 , \om^4 , \om^5 ) \, ,
\ee
described by the elements
	\begin{align}
	&C_{jk}^{(\omega)}=\frac{1}{\sqrt{M}} e^{i\frac{\pi}{12} + i\frac{\pi}{M}\alpha^2  }e^{-i\frac{\pi}{M}k^2 +2\pi i\frac{\alpha}{M}k +2\pi i\frac{j\cdot k}{M} },\nonumber\\
	&C_{jk}^{(\omega^2)}=\frac{1}{\sqrt{M}} e^{-i\frac{\pi}{12} + i\frac{3\pi\alpha^2}{M} +i\frac{\pi}{M}j^2 +2\pi i \frac{\alpha}{M}j}e^{i\frac{4 \pi\alpha}{M}k +2\pi i \frac{j\cdot k}{M}},\nonumber\\
	&C_{jk}^{(\omega^3)}=e^{-i\frac{4\pi\alpha^2}{M} - i\frac{4\pi\alpha}{M} j}\delta_{{-2\alpha-j,k}},\nonumber\\
	&C_{jk}^{(\omega^4)}=\frac{1}{\sqrt{M}} e^{i\frac{\pi}{12} - i\frac{3\pi\alpha^2}{M} - i\frac{4\pi\alpha}{M}j}e^{-i\frac{\pi}{M}k^2 -2\pi i\frac{\alpha}{M}k-2\pi i \frac{j\cdot k}{M}},\nonumber\\
	&C_{jk}^{(\omega^5)}=\frac{1}{\sqrt{M}} e^{-i\frac{\pi}{12} -i \frac{\pi}{M}\alpha^2{+}i\frac{\pi}{M}j^2 -2\pi i \frac{\alpha}{M}j}e^{-2\pi i \frac{j\cdot k}{M}},
	\end{align}
we obtain the results in Tables~\ref{gen.table-Z6_0} and \ref{gen.table-Z6_1}.
\begin{table}[H]
\bec
\begin{tabular}{|c|c|ccccccccccc|} \hline
	\multicolumn{2}{|c|}{$| M |$} & 2 & 4 & 6 & 8 & 10 & 12 & 14 & 16 & 18 & 20 & 22 \\ \hline
	\multirow{6}{*}{$\eta$}	& 1 & 1 & 1 & 2 & 2 & 2 & 3 & 3 & 3 & 4 & 4 & 4 \\
		& $\omega$				& 0 & 1 & 1 & 1 & 2 & 2 & 2 & 3 & 3 & 3 & 4 \\
		& $\omega^2$			& 1 & 1 & 1 & 2 & 2 & 2 & 3 & 3 & 3 & 4 & 4 \\
		& $\omega^3$			& 0 & 0 & 1 & 1 & 1 & 2 & 2 & 2 & 3 & 3 & 3 \\
		& $\omega^4$			& 0 & 1 & 1 & 1 & 2 & 2 & 2 & 3 & 3 & 3 & 4 \\
		& $\omega^5$			& 0 & 0 & 0 & 1 & 1 & 1 & 2 & 2 & 2 & 3 & 3 \\ \hline \hline
	\multicolumn{2}{|c|}{$| M |$} & 24 & 26 & 28 & 30 & 32 & 34 & 36 & 38 & 40 & 42 & 44 \\ \hline
	\multirow{6}{*}{$\eta$}	& 1 & 5 & 5 & 5 & 6 & 6 & 6 & 7 & 7 & 7 & 8 & 8 \\
		& $\omega$				& 4 & 4 & 5 & 5 & 5 & 6 & 6 & 6 & 7 & 7 & 7 \\
		& $\omega^2$			& 4 & 5 & 5 & 5 & 6 & 6 & 6 & 7 & 7 & 7 & 8 \\
		& $\omega^3$			& 4 & 4 & 4 & 5 & 5 & 5 & 6 & 6 & 6 & 7 & 7 \\
		& $\omega^4$			& 4 & 4 & 5 & 5 & 5 & 6 & 6 & 6 & 7 & 7 & 7 \\
		& $\omega^5$			& 3 & 4 & 4 & 4 & 5 & 5 & 5 & 6 & 6 & 6 & 7 \\ \hline
\end{tabular}
\eec
\vspace{-3mm}
\caption{The numbers of linearly independent zero-mode eigenstates with $Z_6$ eigenvalue $\eta$ for $M = {\rm even}$ and $( \alpha_1 , \alpha_2 ) = ( 0 , 0 ) $ on $T^2 / \Z_6$.}
\label{gen.table-Z6_0}
\end{table}

\begin{table}[H]
\bec
\begin{tabular}{|c|c|ccccccccccc|} \hline
	\multicolumn{2}{|c|}{$| M |$} & 1 & 3 & 5 & 7 & 9 & 11 & 13 & 15 & 17 & 19 & 21 \\ \hline
	\multirow{6}{*}{$\eta$}	& 1 & 0 & 1 & 1 & 1 & 2 & 2 & 2 & 3 & 3 & 3 & 4 \\
		& $\omega$				& 1 & 1 & 1 & 2 & 2 & 2 & 3 & 3 & 3 & 4 & 4 \\
		& $\omega^2$			& 0 & 0 & 1 & 1 & 1 & 2 & 2 & 2 & 3 & 3 & 3 \\
		& $\omega^3$			& 0 & 1 & 1 & 1 & 2 & 2 & 2 & 3 & 3 & 3 & 4 \\
		& $\omega^4$			& 0 & 0 & 0 & 1 & 1 & 1 & 2 & 2 & 2 & 3 & 3 \\
		& $\omega^5$			& 0 & 0 & 1 & 1 & 1 & 2 & 2 & 2 & 3 & 3 & 3 \\ \hline \hline
	\multicolumn{2}{|c|}{$| M |$} & 23 & 25 & 27 & 29 & 31 & 33 & 35 & 37 & 39 & 41 & 43 \\ \hline
	\multirow{6}{*}{$\eta$}	& 1 & 4 & 4 & 5 & 5 & 5 & 6 & 6 & 6 & 7 & 7 & 7 \\
		& $\omega$				& 4 & 5 & 5 & 5 & 6 & 6 & 6 & 7 & 7 & 7 & 8 \\
		& $\omega^2$			& 4 & 4 & 4 & 5 & 5 & 5 & 6 & 6 & 6 & 7 & 7 \\
		& $\omega^3$			& 4 & 4 & 5 & 5 & 5 & 6 & 6 & 6 & 7 & 7 & 7 \\
		& $\omega^4$			& 3 & 4 & 4 & 4 & 5 & 5 & 5 & 6 & 6 & 6 & 7 \\
		& $\omega^5$			& 4 & 4 & 4 & 5 & 5 & 5 & 6 & 6 & 6 & 7 & 7 \\ \hline
\end{tabular}
\eec
\vspace{-3mm}
\caption{The numbers of linearly independent zero-mode eigenstates with $Z_6$ eigenvalue $\eta$ for $M = {\rm odd}$ and $( \alpha_1 , \alpha_2 ) = ( \frac{ 1 }{ 2 } , \frac{ 1 }{ 2 } ) $ on $T^2 / \Z_6$.}
\label{gen.table-Z6_1}
\end{table}


\section{Possible parameter configurations in $Z_{2,3,4,6}$}\label{sec:Z2_configuration_full}

In this part, we show the explicit information on the possible configurations in {the cases of $Z_{2,3,4,6}$} with nontrivial boundary conditions.
{In the analysis, we {adopt} the basis where the Scherk-Schwarz phases are nonzero, and correspondingly the Wilson line phases are zero.}
The interpretation into the {basis} {with nonzero Wilson line phases and Scherk-Schwarz phases being zero} is straightforward, where we just obey the relation in Eq.~(\ref{interprelation_rule}).

{Results of the two cases mentioned in Eq.~(\ref{scanningcondition_M_1}), $M_{ab} < 0,\ M_{bc} < 0$, and, $M_{ab} < 0,\ M_{bc} > 0$, are separately stored in tables~\ref{tbl:T2Z2_negative} and \ref{tbl:T2Z2_positive} ($Z_2$), \ref{tbl:T2Z3_negative} and \ref{tbl:T2Z3_positive} ($Z_3$), \ref{tbl:T2Z4_negative} and \ref{tbl:T2Z4_positive} ($Z_4$) and \ref{tbl:T2Z6_negative} and \ref{tbl:T2Z6_positive} ($Z_6$), respectively.}
Here, we skip to represent the parameters of $bc$ ({Higgsino}/Higgs) sector since they are not independent and automatically fixed by the configurations of the other two sectors through the constraints in Eq.~(\ref{constraints_on_parameters}).
If an allowed configuration says $M_{bc} = 0$, where one non-localized Higgs pair appears without magnetic background and this situation would be not interesting in the phenomenological point of view, we discriminate this case as ``$\fbox{1}$'' from the one Higgs cases with magnetic flux ``$1$''.
{A $Z_N$ parity $\eta$ is represented as a function of $i$ in this form, $\eta = \omega^i$ ($\omega = e^{2 \pi i/N}$).}

Note that the ``Trivial BC's only'' case in $Z_2$ was already analyzed in Ref.~\cite{Abe:2008sx} and our result is completely consistent with the previous ones.
{In the $Z_{3,4,6}$ orbifoldings, the two values of the Scherk-Schwarz phases should be the same as $\alpha = \beta$.}

\newpage

\subsection{$T^2/Z_2$}

\begin{center}
\tablehead{
 \hline
 \multicolumn{9}{|c|}{$T^2/Z_2$ with $M_{ab} < 0,\, M_{ca} < 0$} \\ \hline \hline
 \multicolumn{4}{|c||}{$ab$-sector} & \multicolumn{4}{c||}{$ca$-sector} & $bc$-sector \\ \hline
 $M_{ab}$ & $\omega^i_{ab}$ & $\alpha_{ab}$ & $\beta_{ab}$ & $M_{ca}$ & $\omega^i_{ca}$ & $\alpha_{ca}$ & $\beta_{ca}$ & \# of Higgs \\ \hline\hline
}
\tabletail{\hline}
\tablecaption{Possible parameter configurations on $T^2/Z_2$ with $M_{ab} < 0,\, M_{ca} < 0$.} \label{tbl:T2Z2_negative}
\begin{supertabular}{|c|c|c|c||c|c|c|c||c|}
$-4$ & $0$ & $0$ & $0$  &  $-4$ & $0$ & $0$ & $0$  &  $5$\\\hline
$-4$ & $0$ & $0$ & $0$  &  $-5$ & $0$ & $0$ & $0$  &  $5$\\
$-4$ & $0$ & $0$ & $0$  &  $-5$ & $0$ & $0$ & $1/2$  &  $5$\\
$-4$ & $0$ & $0$ & $0$  &  $-5$ & $0$ & $1/2$ & $0$  &  $5$\\
$-4$ & $0$ & $0$ & $0$  &  $-5$ & $1$ & $1/2$ & $1/2$  &  $5$\\\hline
$-5$ & $0$ & $0$ & $0$  &  $-5$ & $0$ & $0$ & $0$  &  $6$\\
$-5$ & $0$ & $0$ & $0$  &  $-5$ & $0$ & $0$ & $1/2$  &  $5$\\
$-5$ & $0$ & $0$ & $0$  &  $-5$ & $0$ & $1/2$ & $0$  &  $5$\\
$-5$ & $0$ & $0$ & $1/2$  &  $-5$ & $0$ & $0$ & $1/2$  &  $6$\\
$-5$ & $0$ & $0$ & $1/2$  &  $-5$ & $0$ & $1/2$ & $0$  &  $5$\\
$-5$ & $0$ & $1/2$ & $0$  &  $-5$ & $0$ & $1/2$ & $0$  &  $6$\\
$-5$ & $0$ & $0$ & $0$  &  $-5$ & $1$ & $1/2$ & $1/2$  &  $5$\\
$-5$ & $0$ & $0$ & $1/2$  &  $-5$ & $1$ & $1/2$ & $1/2$  &  $5$\\
$-5$ & $0$ & $1/2$ & $0$  &  $-5$ & $1$ & $1/2$ & $1/2$  &  $5$\\
$-5$ & $1$ & $1/2$ & $1/2$  &  $-5$ & $1$ & $1/2$ & $1/2$  &  $6$\\\hline
$-4$ & $0$ & $0$ & $0$  &  $-6$ & $0$ & $0$ & $1/2$  &  $5$\\
$-4$ & $0$ & $0$ & $0$  &  $-6$ & $0$ & $1/2$ & $0$  &  $5$\\
$-4$ & $0$ & $0$ & $0$  &  $-6$ & $0$ & $1/2$ & $1/2$  &  $5$\\
$-4$ & $0$ & $0$ & $0$  &  $-6$ & $1$ & $0$ & $1/2$  &  $5$\\
$-4$ & $0$ & $0$ & $0$  &  $-6$ & $1$ & $1/2$ & $0$  &  $5$\\
$-4$ & $0$ & $0$ & $0$  &  $-6$ & $1$ & $1/2$ & $1/2$  &  $5$\\\hline
$-5$ & $0$ & $0$ & $0$  &  $-6$ & $0$ & $0$ & $1/2$  &  $6$\\
$-5$ & $0$ & $0$ & $0$  &  $-6$ & $0$ & $1/2$ & $0$  &  $6$\\
$-5$ & $0$ & $0$ & $0$  &  $-6$ & $0$ & $1/2$ & $1/2$  &  $5$\\
$-5$ & $0$ & $0$ & $1/2$  &  $-6$ & $0$ & $0$ & $1/2$  &  $6$\\
$-5$ & $0$ & $0$ & $1/2$  &  $-6$ & $0$ & $1/2$ & $0$  &  $5$\\
$-5$ & $0$ & $0$ & $1/2$  &  $-6$ & $0$ & $1/2$ & $1/2$  &  $6$\\
$-5$ & $0$ & $1/2$ & $0$  &  $-6$ & $0$ & $0$ & $1/2$  &  $5$\\
$-5$ & $0$ & $1/2$ & $0$  &  $-6$ & $0$ & $1/2$ & $0$  &  $6$\\
$-5$ & $0$ & $1/2$ & $0$  &  $-6$ & $0$ & $1/2$ & $1/2$  &  $6$\\
$-5$ & $0$ & $0$ & $0$  &  $-6$ & $1$ & $0$ & $1/2$  &  $5$\\
$-5$ & $0$ & $0$ & $0$  &  $-6$ & $1$ & $1/2$ & $0$  &  $5$\\
$-5$ & $0$ & $0$ & $0$  &  $-6$ & $1$ & $1/2$ & $1/2$  &  $6$\\
$-5$ & $0$ & $0$ & $1/2$  &  $-6$ & $1$ & $0$ & $1/2$  &  $5$\\
$-5$ & $0$ & $0$ & $1/2$  &  $-6$ & $1$ & $1/2$ & $0$  &  $6$\\
$-5$ & $0$ & $0$ & $1/2$  &  $-6$ & $1$ & $1/2$ & $1/2$  &  $5$\\
$-5$ & $0$ & $1/2$ & $0$  &  $-6$ & $1$ & $0$ & $1/2$  &  $6$\\
$-5$ & $0$ & $1/2$ & $0$  &  $-6$ & $1$ & $1/2$ & $0$  &  $5$\\
$-5$ & $0$ & $1/2$ & $0$  &  $-6$ & $1$ & $1/2$ & $1/2$  &  $5$\\
$-5$ & $1$ & $1/2$ & $1/2$  &  $-6$ & $0$ & $0$ & $1/2$  &  $5$\\
$-5$ & $1$ & $1/2$ & $1/2$  &  $-6$ & $0$ & $1/2$ & $0$  &  $5$\\
$-5$ & $1$ & $1/2$ & $1/2$  &  $-6$ & $0$ & $1/2$ & $1/2$  &  $5$\\
$-5$ & $1$ & $1/2$ & $1/2$  &  $-6$ & $1$ & $0$ & $1/2$  &  $6$\\
$-5$ & $1$ & $1/2$ & $1/2$  &  $-6$ & $1$ & $1/2$ & $0$  &  $6$\\
$-5$ & $1$ & $1/2$ & $1/2$  &  $-6$ & $1$ & $1/2$ & $1/2$  &  $6$\\\hline
$-6$ & $0$ & $0$ & $1/2$  &  $-6$ & $0$ & $0$ & $1/2$  &  $7$\\
$-6$ & $0$ & $0$ & $1/2$  &  $-6$ & $0$ & $1/2$ & $0$  &  $6$\\
$-6$ & $0$ & $0$ & $1/2$  &  $-6$ & $0$ & $1/2$ & $1/2$  &  $6$\\
$-6$ & $0$ & $1/2$ & $0$  &  $-6$ & $0$ & $1/2$ & $0$  &  $7$\\
$-6$ & $0$ & $1/2$ & $0$  &  $-6$ & $0$ & $1/2$ & $1/2$  &  $6$\\
$-6$ & $0$ & $1/2$ & $1/2$  &  $-6$ & $0$ & $1/2$ & $1/2$  &  $7$\\
$-6$ & $0$ & $0$ & $1/2$  &  $-6$ & $1$ & $0$ & $1/2$  &  $5$\\
$-6$ & $0$ & $0$ & $1/2$  &  $-6$ & $1$ & $1/2$ & $0$  &  $6$\\
$-6$ & $0$ & $0$ & $1/2$  &  $-6$ & $1$ & $1/2$ & $1/2$  &  $6$\\
$-6$ & $0$ & $1/2$ & $0$  &  $-6$ & $1$ & $0$ & $1/2$  &  $6$\\
$-6$ & $0$ & $1/2$ & $0$  &  $-6$ & $1$ & $1/2$ & $0$  &  $5$\\
$-6$ & $0$ & $1/2$ & $0$  &  $-6$ & $1$ & $1/2$ & $1/2$  &  $6$\\
$-6$ & $0$ & $1/2$ & $1/2$  &  $-6$ & $1$ & $0$ & $1/2$  &  $6$\\
$-6$ & $0$ & $1/2$ & $1/2$  &  $-6$ & $1$ & $1/2$ & $0$  &  $6$\\
$-6$ & $0$ & $1/2$ & $1/2$  &  $-6$ & $1$ & $1/2$ & $1/2$  &  $5$\\
$-6$ & $1$ & $0$ & $1/2$  &  $-6$ & $1$ & $0$ & $1/2$  &  $7$\\
$-6$ & $1$ & $0$ & $1/2$  &  $-6$ & $1$ & $1/2$ & $0$  &  $6$\\
$-6$ & $1$ & $0$ & $1/2$  &  $-6$ & $1$ & $1/2$ & $1/2$  &  $6$\\
$-6$ & $1$ & $1/2$ & $0$  &  $-6$ & $1$ & $1/2$ & $0$  &  $7$\\
$-6$ & $1$ & $1/2$ & $0$  &  $-6$ & $1$ & $1/2$ & $1/2$  &  $6$\\
$-6$ & $1$ & $1/2$ & $1/2$  &  $-6$ & $1$ & $1/2$ & $1/2$  &  $7$\\\hline
$-4$ & $0$ & $0$ & $0$  &  $-7$ & $0$ & $1/2$ & $1/2$  &  $5$\\
$-4$ & $0$ & $0$ & $0$  &  $-7$ & $1$ & $0$ & $0$  &  $5$\\
$-4$ & $0$ & $0$ & $0$  &  $-7$ & $1$ & $0$ & $1/2$  &  $5$\\
$-4$ & $0$ & $0$ & $0$  &  $-7$ & $1$ & $1/2$ & $0$  &  $5$\\\hline
$-5$ & $0$ & $0$ & $0$  &  $-7$ & $0$ & $1/2$ & $1/2$  &  $6$\\
$-5$ & $0$ & $0$ & $1/2$  &  $-7$ & $0$ & $1/2$ & $1/2$  &  $6$\\
$-5$ & $0$ & $1/2$ & $0$  &  $-7$ & $0$ & $1/2$ & $1/2$  &  $6$\\
$-5$ & $0$ & $0$ & $0$  &  $-7$ & $1$ & $0$ & $0$  &  $5$\\
$-5$ & $0$ & $0$ & $0$  &  $-7$ & $1$ & $0$ & $1/2$  &  $6$\\
$-5$ & $0$ & $0$ & $0$  &  $-7$ & $1$ & $1/2$ & $0$  &  $6$\\
$-5$ & $0$ & $0$ & $1/2$  &  $-7$ & $1$ & $0$ & $0$  &  $6$\\
$-5$ & $0$ & $0$ & $1/2$  &  $-7$ & $1$ & $0$ & $1/2$  &  $5$\\
$-5$ & $0$ & $0$ & $1/2$  &  $-7$ & $1$ & $1/2$ & $0$  &  $6$\\
$-5$ & $0$ & $1/2$ & $0$  &  $-7$ & $1$ & $0$ & $0$  &  $6$\\
$-5$ & $0$ & $1/2$ & $0$  &  $-7$ & $1$ & $0$ & $1/2$  &  $6$\\
$-5$ & $0$ & $1/2$ & $0$  &  $-7$ & $1$ & $1/2$ & $0$  &  $5$\\
$-5$ & $1$ & $1/2$ & $1/2$  &  $-7$ & $0$ & $1/2$ & $1/2$  &  $5$\\
$-5$ & $1$ & $1/2$ & $1/2$  &  $-7$ & $1$ & $0$ & $0$  &  $6$\\
$-5$ & $1$ & $1/2$ & $1/2$  &  $-7$ & $1$ & $0$ & $1/2$  &  $6$\\
$-5$ & $1$ & $1/2$ & $1/2$  &  $-7$ & $1$ & $1/2$ & $0$  &  $6$\\\hline
$-6$ & $0$ & $0$ & $1/2$  &  $-7$ & $0$ & $1/2$ & $1/2$  &  $7$\\
$-6$ & $0$ & $1/2$ & $0$  &  $-7$ & $0$ & $1/2$ & $1/2$  &  $7$\\
$-6$ & $0$ & $1/2$ & $1/2$  &  $-7$ & $0$ & $1/2$ & $1/2$  &  $7$\\
$-6$ & $0$ & $0$ & $1/2$  &  $-7$ & $1$ & $0$ & $0$  &  $6$\\
$-6$ & $0$ & $0$ & $1/2$  &  $-7$ & $1$ & $0$ & $1/2$  &  $6$\\
$-6$ & $0$ & $0$ & $1/2$  &  $-7$ & $1$ & $1/2$ & $0$  &  $7$\\
$-6$ & $0$ & $1/2$ & $0$  &  $-7$ & $1$ & $0$ & $0$  &  $6$\\
$-6$ & $0$ & $1/2$ & $0$  &  $-7$ & $1$ & $0$ & $1/2$  &  $7$\\
$-6$ & $0$ & $1/2$ & $0$  &  $-7$ & $1$ & $1/2$ & $0$  &  $6$\\
$-6$ & $0$ & $1/2$ & $1/2$  &  $-7$ & $1$ & $0$ & $0$  &  $7$\\
$-6$ & $0$ & $1/2$ & $1/2$  &  $-7$ & $1$ & $0$ & $1/2$  &  $6$\\
$-6$ & $0$ & $1/2$ & $1/2$  &  $-7$ & $1$ & $1/2$ & $0$  &  $6$\\
$-6$ & $1$ & $0$ & $1/2$  &  $-7$ & $0$ & $1/2$ & $1/2$  &  $6$\\
$-6$ & $1$ & $1/2$ & $0$  &  $-7$ & $0$ & $1/2$ & $1/2$  &  $6$\\
$-6$ & $1$ & $1/2$ & $1/2$  &  $-7$ & $0$ & $1/2$ & $1/2$  &  $6$\\
$-6$ & $1$ & $0$ & $1/2$  &  $-7$ & $1$ & $0$ & $0$  &  $7$\\
$-6$ & $1$ & $0$ & $1/2$  &  $-7$ & $1$ & $0$ & $1/2$  &  $7$\\
$-6$ & $1$ & $0$ & $1/2$  &  $-7$ & $1$ & $1/2$ & $0$  &  $6$\\
$-6$ & $1$ & $1/2$ & $0$  &  $-7$ & $1$ & $0$ & $0$  &  $7$\\
$-6$ & $1$ & $1/2$ & $0$  &  $-7$ & $1$ & $0$ & $1/2$  &  $6$\\
$-6$ & $1$ & $1/2$ & $0$  &  $-7$ & $1$ & $1/2$ & $0$  &  $7$\\
$-6$ & $1$ & $1/2$ & $1/2$  &  $-7$ & $1$ & $0$ & $0$  &  $6$\\
$-6$ & $1$ & $1/2$ & $1/2$  &  $-7$ & $1$ & $0$ & $1/2$  &  $7$\\
$-6$ & $1$ & $1/2$ & $1/2$  &  $-7$ & $1$ & $1/2$ & $0$  &  $7$\\\hline
$-7$ & $0$ & $1/2$ & $1/2$  &  $-7$ & $0$ & $1/2$ & $1/2$  &  $8$\\
$-7$ & $0$ & $1/2$ & $1/2$  &  $-7$ & $1$ & $0$ & $0$  &  $7$\\
$-7$ & $0$ & $1/2$ & $1/2$  &  $-7$ & $1$ & $0$ & $1/2$  &  $7$\\
$-7$ & $0$ & $1/2$ & $1/2$  &  $-7$ & $1$ & $1/2$ & $0$  &  $7$\\
$-7$ & $1$ & $0$ & $0$  &  $-7$ & $1$ & $0$ & $0$  &  $8$\\
$-7$ & $1$ & $0$ & $0$  &  $-7$ & $1$ & $0$ & $1/2$  &  $7$\\
$-7$ & $1$ & $0$ & $0$  &  $-7$ & $1$ & $1/2$ & $0$  &  $7$\\
$-7$ & $1$ & $0$ & $1/2$  &  $-7$ & $1$ & $0$ & $1/2$  &  $8$\\
$-7$ & $1$ & $0$ & $1/2$  &  $-7$ & $1$ & $1/2$ & $0$  &  $7$\\
$-7$ & $1$ & $1/2$ & $0$  &  $-7$ & $1$ & $1/2$ & $0$  &  $8$\\\hline
$-4$ & $0$ & $0$ & $0$  &  $-8$ & $1$ & $0$ & $0$  &  $5$\\\hline
$-5$ & $0$ & $0$ & $0$  &  $-8$ & $1$ & $0$ & $0$  &  $6$\\
$-5$ & $0$ & $0$ & $1/2$  &  $-8$ & $1$ & $0$ & $0$  &  $6$\\
$-5$ & $0$ & $1/2$ & $0$  &  $-8$ & $1$ & $0$ & $0$  &  $6$\\
$-5$ & $1$ & $1/2$ & $1/2$  &  $-8$ & $1$ & $0$ & $0$  &  $6$\\\hline
$-6$ & $0$ & $0$ & $1/2$  &  $-8$ & $1$ & $0$ & $0$  &  $7$\\
$-6$ & $0$ & $1/2$ & $0$  &  $-8$ & $1$ & $0$ & $0$  &  $7$\\
$-6$ & $0$ & $1/2$ & $1/2$  &  $-8$ & $1$ & $0$ & $0$  &  $7$\\
$-6$ & $1$ & $0$ & $1/2$  &  $-8$ & $1$ & $0$ & $0$  &  $7$\\
$-6$ & $1$ & $1/2$ & $0$  &  $-8$ & $1$ & $0$ & $0$  &  $7$\\
$-6$ & $1$ & $1/2$ & $1/2$  &  $-8$ & $1$ & $0$ & $0$  &  $7$\\\hline
$-7$ & $0$ & $1/2$ & $1/2$  &  $-8$ & $1$ & $0$ & $0$  &  $8$\\
$-7$ & $1$ & $0$ & $0$  &  $-8$ & $1$ & $0$ & $0$  &  $8$\\
$-7$ & $1$ & $0$ & $1/2$  &  $-8$ & $1$ & $0$ & $0$  &  $8$\\
$-7$ & $1$ & $1/2$ & $0$  &  $-8$ & $1$ & $0$ & $0$  &  $8$\\\hline
$-8$ & $1$ & $0$ & $0$  &  $-8$ & $1$ & $0$ & $0$  &  $9$\\\hline
\end{supertabular}
\end{center}

\newpage

\begin{center}
\tablehead{
 \hline
 \multicolumn{9}{|c|}{$T^2/Z_2$ with $M_{ab} < 0,\, M_{ca} > 0$} \\ \hline \hline
 \multicolumn{4}{|c||}{$ab$-sector} & \multicolumn{4}{c||}{$ca$-sector} & $bc$-sector \\ \hline
 $M_{ab}$ & $\omega^i_{ab}$ & $\alpha_{ab}$ & $\beta_{ab}$ & $M_{ca}$ & $\omega^i_{ca}$ & $\alpha_{ca}$ & $\beta_{ca}$ & \# of Higgs \\ \hline\hline
}
\tabletail{\hline}
\tablecaption{Possible parameter configurations on $T^2/Z_2$ with $M_{ab} < 0,\, M_{ca} > 0$.} \label{tbl:T2Z2_positive}
\begin{supertabular}{|c|c|c|c||c|c|c|c||c|}
$-4$ & $0$ & $0$ & $0$  &  $4$ & $0$ & $0$ & $0$  &  $\fbox{1}$\\\hline
$-4$ & $0$ & $0$ & $0$  &  $5$ & $0$ & $0$ & $0$  &  $1$\\
$-4$ & $0$ & $0$ & $0$  &  $5$ & $0$ & $0$ & $1/2$  &  $1$\\
$-4$ & $0$ & $0$ & $0$  &  $5$ & $0$ & $1/2$ & $0$  &  $1$\\
$-4$ & $0$ & $0$ & $0$  &  $5$ & $1$ & $1/2$ & $1/2$  &  $1$\\\hline
$-5$ & $0$ & $0$ & $0$  &  $5$ & $0$ & $0$ & $0$  &  $\fbox{1}$\\
$-5$ & $0$ & $0$ & $1/2$  &  $5$ & $0$ & $0$ & $1/2$  &  $\fbox{1}$\\
$-5$ & $0$ & $1/2$ & $0$  &  $5$ & $0$ & $1/2$ & $0$  &  $\fbox{1}$\\
$-5$ & $1$ & $1/2$ & $1/2$  &  $5$ & $1$ & $1/2$ & $1/2$  &  $\fbox{1}$\\\hline
$-4$ & $0$ & $0$ & $0$  &  $6$ & $0$ & $0$ & $1/2$  &  $1$\\
$-4$ & $0$ & $0$ & $0$  &  $6$ & $0$ & $1/2$ & $0$  &  $1$\\
$-4$ & $0$ & $0$ & $0$  &  $6$ & $0$ & $1/2$ & $1/2$  &  $1$\\
$-4$ & $0$ & $0$ & $0$  &  $6$ & $1$ & $0$ & $1/2$  &  $1$\\
$-4$ & $0$ & $0$ & $0$  &  $6$ & $1$ & $1/2$ & $0$  &  $1$\\
$-4$ & $0$ & $0$ & $0$  &  $6$ & $1$ & $1/2$ & $1/2$  &  $1$\\\hline
$-5$ & $0$ & $0$ & $0$  &  $6$ & $0$ & $0$ & $1/2$  &  $1$\\
$-5$ & $0$ & $0$ & $0$  &  $6$ & $0$ & $1/2$ & $0$  &  $1$\\
$-5$ & $0$ & $0$ & $1/2$  &  $6$ & $0$ & $0$ & $1/2$  &  $1$\\
$-5$ & $0$ & $0$ & $1/2$  &  $6$ & $0$ & $1/2$ & $1/2$  &  $1$\\
$-5$ & $0$ & $1/2$ & $0$  &  $6$ & $0$ & $1/2$ & $0$  &  $1$\\
$-5$ & $0$ & $1/2$ & $0$  &  $6$ & $0$ & $1/2$ & $1/2$  &  $1$\\
$-5$ & $0$ & $0$ & $0$  &  $6$ & $1$ & $1/2$ & $1/2$  &  $1$\\
$-5$ & $0$ & $0$ & $1/2$  &  $6$ & $1$ & $1/2$ & $0$  &  $1$\\
$-5$ & $0$ & $1/2$ & $0$  &  $6$ & $1$ & $0$ & $1/2$  &  $1$\\
$-5$ & $1$ & $1/2$ & $1/2$  &  $6$ & $1$ & $0$ & $1/2$  &  $1$\\
$-5$ & $1$ & $1/2$ & $1/2$  &  $6$ & $1$ & $1/2$ & $0$  &  $1$\\
$-5$ & $1$ & $1/2$ & $1/2$  &  $6$ & $1$ & $1/2$ & $1/2$  &  $1$\\\hline
$-6$ & $0$ & $0$ & $1/2$  &  $6$ & $0$ & $0$ & $1/2$  &  $\fbox{1}$\\
$-6$ & $0$ & $1/2$ & $0$  &  $6$ & $0$ & $1/2$ & $0$  &  $\fbox{1}$\\
$-6$ & $0$ & $1/2$ & $1/2$  &  $6$ & $0$ & $1/2$ & $1/2$  &  $\fbox{1}$\\
$-6$ & $1$ & $0$ & $1/2$  &  $6$ & $1$ & $0$ & $1/2$  &  $\fbox{1}$\\
$-6$ & $1$ & $1/2$ & $0$  &  $6$ & $1$ & $1/2$ & $0$  &  $\fbox{1}$\\
$-6$ & $1$ & $1/2$ & $1/2$  &  $6$ & $1$ & $1/2$ & $1/2$  &  $\fbox{1}$\\\hline
$-4$ & $0$ & $0$ & $0$  &  $7$ & $0$ & $1/2$ & $1/2$  &  $1$\\
$-4$ & $0$ & $0$ & $0$  &  $7$ & $1$ & $0$ & $0$  &  $1$\\
$-4$ & $0$ & $0$ & $0$  &  $7$ & $1$ & $0$ & $1/2$  &  $1$\\
$-4$ & $0$ & $0$ & $0$  &  $7$ & $1$ & $1/2$ & $0$  &  $1$\\\hline
$-5$ & $0$ & $0$ & $0$  &  $7$ & $0$ & $1/2$ & $1/2$  &  $1$\\
$-5$ & $0$ & $0$ & $1/2$  &  $7$ & $0$ & $1/2$ & $1/2$  &  $1$\\
$-5$ & $0$ & $1/2$ & $0$  &  $7$ & $0$ & $1/2$ & $1/2$  &  $1$\\
$-5$ & $0$ & $0$ & $0$  &  $7$ & $1$ & $0$ & $1/2$  &  $1$\\
$-5$ & $0$ & $0$ & $0$  &  $7$ & $1$ & $1/2$ & $0$  &  $1$\\
$-5$ & $0$ & $0$ & $1/2$  &  $7$ & $1$ & $0$ & $0$  &  $1$\\
$-5$ & $0$ & $0$ & $1/2$  &  $7$ & $1$ & $1/2$ & $0$  &  $1$\\
$-5$ & $0$ & $1/2$ & $0$  &  $7$ & $1$ & $0$ & $0$  &  $1$\\
$-5$ & $0$ & $1/2$ & $0$  &  $7$ & $1$ & $0$ & $1/2$  &  $1$\\
$-5$ & $1$ & $1/2$ & $1/2$  &  $7$ & $1$ & $0$ & $0$  &  $1$\\
$-5$ & $1$ & $1/2$ & $1/2$  &  $7$ & $1$ & $0$ & $1/2$  &  $1$\\
$-5$ & $1$ & $1/2$ & $1/2$  &  $7$ & $1$ & $1/2$ & $0$  &  $1$\\\hline
$-6$ & $0$ & $0$ & $1/2$  &  $7$ & $0$ & $1/2$ & $1/2$  &  $1$\\
$-6$ & $0$ & $1/2$ & $0$  &  $7$ & $0$ & $1/2$ & $1/2$  &  $1$\\
$-6$ & $0$ & $1/2$ & $1/2$  &  $7$ & $0$ & $1/2$ & $1/2$  &  $1$\\
$-6$ & $0$ & $0$ & $1/2$  &  $7$ & $1$ & $1/2$ & $0$  &  $1$\\
$-6$ & $0$ & $1/2$ & $0$  &  $7$ & $1$ & $0$ & $1/2$  &  $1$\\
$-6$ & $0$ & $1/2$ & $1/2$  &  $7$ & $1$ & $0$ & $0$  &  $1$\\
$-6$ & $1$ & $0$ & $1/2$  &  $7$ & $1$ & $0$ & $0$  &  $1$\\
$-6$ & $1$ & $0$ & $1/2$  &  $7$ & $1$ & $0$ & $1/2$  &  $1$\\
$-6$ & $1$ & $1/2$ & $0$  &  $7$ & $1$ & $0$ & $0$  &  $1$\\
$-6$ & $1$ & $1/2$ & $0$  &  $7$ & $1$ & $1/2$ & $0$  &  $1$\\
$-6$ & $1$ & $1/2$ & $1/2$  &  $7$ & $1$ & $0$ & $1/2$  &  $1$\\
$-6$ & $1$ & $1/2$ & $1/2$  &  $7$ & $1$ & $1/2$ & $0$  &  $1$\\\hline
$-7$ & $0$ & $1/2$ & $1/2$  &  $7$ & $0$ & $1/2$ & $1/2$  &  $\fbox{1}$\\
$-7$ & $1$ & $0$ & $0$  &  $7$ & $1$ & $0$ & $0$  &  $\fbox{1}$\\
$-7$ & $1$ & $0$ & $1/2$  &  $7$ & $1$ & $0$ & $1/2$  &  $\fbox{1}$\\
$-7$ & $1$ & $1/2$ & $0$  &  $7$ & $1$ & $1/2$ & $0$  &  $\fbox{1}$\\\hline
$-4$ & $0$ & $0$ & $0$  &  $8$ & $1$ & $0$ & $0$  &  $1$\\\hline
$-5$ & $0$ & $0$ & $0$  &  $8$ & $1$ & $0$ & $0$  &  $1$\\
$-5$ & $0$ & $0$ & $1/2$  &  $8$ & $1$ & $0$ & $0$  &  $1$\\
$-5$ & $0$ & $1/2$ & $0$  &  $8$ & $1$ & $0$ & $0$  &  $1$\\
$-5$ & $1$ & $1/2$ & $1/2$  &  $8$ & $1$ & $0$ & $0$  &  $1$\\\hline
$-6$ & $0$ & $0$ & $1/2$  &  $8$ & $1$ & $0$ & $0$  &  $1$\\
$-6$ & $0$ & $1/2$ & $0$  &  $8$ & $1$ & $0$ & $0$  &  $1$\\
$-6$ & $0$ & $1/2$ & $1/2$  &  $8$ & $1$ & $0$ & $0$  &  $1$\\
$-6$ & $1$ & $0$ & $1/2$  &  $8$ & $1$ & $0$ & $0$  &  $1$\\
$-6$ & $1$ & $1/2$ & $0$  &  $8$ & $1$ & $0$ & $0$  &  $1$\\
$-6$ & $1$ & $1/2$ & $1/2$  &  $8$ & $1$ & $0$ & $0$  &  $1$\\\hline
$-7$ & $0$ & $1/2$ & $1/2$  &  $8$ & $1$ & $0$ & $0$  &  $1$\\
$-7$ & $1$ & $0$ & $0$  &  $8$ & $1$ & $0$ & $0$  &  $1$\\
$-7$ & $1$ & $0$ & $1/2$  &  $8$ & $1$ & $0$ & $0$  &  $1$\\
$-7$ & $1$ & $1/2$ & $0$  &  $8$ & $1$ & $0$ & $0$  &  $1$\\\hline
$-8$ & $1$ & $0$ & $0$  &  $8$ & $1$ & $0$ & $0$  &  $\fbox{1}$\\\hline
\end{supertabular}
\end{center}


\newpage

\subsection{$T^2/Z_3$}

\begin{center}
\tablehead{
 \hline
 \multicolumn{7}{|c|}{$T^2/Z_3$ with $M_{ab} < 0,\, M_{ca} < 0$} \\ \hline \hline
 \multicolumn{3}{|c||}{$ab$-sector} & \multicolumn{3}{c||}{$ca$-sector} & $bc$-sector \\ \hline
 $M_{ab}$ & $\omega^i_{ab}$ & $\alpha_{ab}$  &  $M_{ca}$ & $\omega^i_{ca}$ & $\alpha_{ca}$  &  \# of Higgs \\ \hline\hline
}
\tabletail{\hline}
\tablecaption{Possible parameter configurations on $T^2/Z_3$ with $M_{ab} < 0,\, M_{ca} < 0$.} \label{tbl:T2Z3_negative}
\begin{supertabular}{|c|c|c||c|c|c||c|}
$-6$ & $0$ & $0$  &  $-6$ & $0$ & $0$  &  $5$\\\hline
$-6$ & $0$ & $0$  &  $-7$ & $0$ & $1/6$  &  $5$\\
$-6$ & $0$ & $0$  &  $-7$ & $0$ & $5/6$  &  $5$\\
$-6$ & $0$ & $0$  &  $-7$ & $1$ & $1/2$  &  $4$\\\hline
$-7$ & $0$ & $1/6$  &  $-7$ & $0$ & $1/6$  &  $5$\\
$-7$ & $0$ & $1/6$  &  $-7$ & $0$ & $5/6$  &  $5$\\
$-7$ & $0$ & $5/6$  &  $-7$ & $0$ & $5/6$  &  $5$\\
$-7$ & $0$ & $1/6$  &  $-7$ & $1$ & $1/2$  &  $4$\\
$-7$ & $0$ & $5/6$  &  $-7$ & $1$ & $1/2$  &  $4$\\
$-7$ & $1$ & $1/2$  &  $-7$ & $1$ & $1/2$  &  $4$\\\hline
$-6$ & $0$ & $0$  &  $-8$ & $0$ & $0$  &  $5$\\
$-6$ & $0$ & $0$  &  $-8$ & $0$ & $1/3$  &  $5$\\
$-6$ & $0$ & $0$  &  $-8$ & $0$ & $2/3$  &  $5$\\
$-6$ & $0$ & $0$  &  $-8$ & $1$ & $1/3$  &  $4$\\
$-6$ & $0$ & $0$  &  $-8$ & $1$ & $2/3$  &  $4$\\
$-6$ & $0$ & $0$  &  $-8$ & $2$ & $0$  &  $4$\\\hline
$-7$ & $0$ & $1/6$  &  $-8$ & $0$ & $0$  &  $5$\\
$-7$ & $0$ & $1/6$  &  $-8$ & $0$ & $1/3$  &  $6$\\
$-7$ & $0$ & $1/6$  &  $-8$ & $0$ & $2/3$  &  $5$\\
$-7$ & $0$ & $5/6$  &  $-8$ & $0$ & $0$  &  $5$\\
$-7$ & $0$ & $5/6$  &  $-8$ & $0$ & $1/3$  &  $5$\\
$-7$ & $0$ & $5/6$  &  $-8$ & $0$ & $2/3$  &  $6$\\
$-7$ & $0$ & $1/6$  &  $-8$ & $1$ & $1/3$  &  $4$\\
$-7$ & $0$ & $1/6$  &  $-8$ & $1$ & $2/3$  &  $5$\\
$-7$ & $0$ & $5/6$  &  $-8$ & $1$ & $1/3$  &  $5$\\
$-7$ & $0$ & $5/6$  &  $-8$ & $1$ & $2/3$  &  $4$\\
$-7$ & $0$ & $1/6$  &  $-8$ & $2$ & $0$  &  $5$\\
$-7$ & $0$ & $5/6$  &  $-8$ & $2$ & $0$  &  $5$\\
$-7$ & $1$ & $1/2$  &  $-8$ & $0$ & $0$  &  $4$\\
$-7$ & $1$ & $1/2$  &  $-8$ & $0$ & $1/3$  &  $5$\\
$-7$ & $1$ & $1/2$  &  $-8$ & $0$ & $2/3$  &  $5$\\
$-7$ & $1$ & $1/2$  &  $-8$ & $1$ & $1/3$  &  $5$\\
$-7$ & $1$ & $1/2$  &  $-8$ & $1$ & $2/3$  &  $5$\\
$-7$ & $1$ & $1/2$  &  $-8$ & $2$ & $0$  &  $6$\\\hline
$-8$ & $0$ & $0$  &  $-8$ & $0$ & $0$  &  $5$\\
$-8$ & $0$ & $0$  &  $-8$ & $0$ & $1/3$  &  $6$\\
$-8$ & $0$ & $0$  &  $-8$ & $0$ & $2/3$  &  $6$\\
$-8$ & $0$ & $1/3$  &  $-8$ & $0$ & $1/3$  &  $6$\\
$-8$ & $0$ & $1/3$  &  $-8$ & $0$ & $2/3$  &  $5$\\
$-8$ & $0$ & $2/3$  &  $-8$ & $0$ & $2/3$  &  $6$\\
$-8$ & $0$ & $0$  &  $-8$ & $1$ & $1/3$  &  $5$\\
$-8$ & $0$ & $0$  &  $-8$ & $1$ & $2/3$  &  $5$\\
$-8$ & $0$ & $1/3$  &  $-8$ & $1$ & $1/3$  &  $5$\\
$-8$ & $0$ & $1/3$  &  $-8$ & $1$ & $2/3$  &  $5$\\
$-8$ & $0$ & $2/3$  &  $-8$ & $1$ & $1/3$  &  $5$\\
$-8$ & $0$ & $2/3$  &  $-8$ & $1$ & $2/3$  &  $5$\\
$-8$ & $0$ & $0$  &  $-8$ & $2$ & $0$  &  $6$\\
$-8$ & $0$ & $1/3$  &  $-8$ & $2$ & $0$  &  $5$\\
$-8$ & $0$ & $2/3$  &  $-8$ & $2$ & $0$  &  $5$\\
$-8$ & $1$ & $1/3$  &  $-8$ & $1$ & $1/3$  &  $5$\\
$-8$ & $1$ & $1/3$  &  $-8$ & $1$ & $2/3$  &  $6$\\
$-8$ & $1$ & $2/3$  &  $-8$ & $1$ & $2/3$  &  $5$\\
$-8$ & $1$ & $1/3$  &  $-8$ & $2$ & $0$  &  $6$\\
$-8$ & $1$ & $2/3$  &  $-8$ & $2$ & $0$  &  $6$\\
$-8$ & $2$ & $0$  &  $-8$ & $2$ & $0$  &  $5$\\\hline
$-6$ & $0$ & $0$  &  $-9$ & $0$ & $1/6$  &  $5$\\
$-6$ & $0$ & $0$  &  $-9$ & $0$ & $5/6$  &  $5$\\
$-6$ & $0$ & $0$  &  $-9$ & $1$ & $1/6$  &  $5$\\
$-6$ & $0$ & $0$  &  $-9$ & $1$ & $1/2$  &  $4$\\
$-6$ & $0$ & $0$  &  $-9$ & $1$ & $5/6$  &  $5$\\
$-6$ & $0$ & $0$  &  $-9$ & $2$ & $1/6$  &  $5$\\
$-6$ & $0$ & $0$  &  $-9$ & $2$ & $5/6$  &  $5$\\\hline
$-7$ & $0$ & $1/6$  &  $-9$ & $0$ & $1/6$  &  $6$\\
$-7$ & $0$ & $1/6$  &  $-9$ & $0$ & $5/6$  &  $5$\\
$-7$ & $0$ & $5/6$  &  $-9$ & $0$ & $1/6$  &  $5$\\
$-7$ & $0$ & $5/6$  &  $-9$ & $0$ & $5/6$  &  $6$\\
$-7$ & $0$ & $1/6$  &  $-9$ & $1$ & $1/6$  &  $5$\\
$-7$ & $0$ & $1/6$  &  $-9$ & $1$ & $1/2$  &  $5$\\
$-7$ & $0$ & $1/6$  &  $-9$ & $1$ & $5/6$  &  $5$\\
$-7$ & $0$ & $5/6$  &  $-9$ & $1$ & $1/6$  &  $5$\\
$-7$ & $0$ & $5/6$  &  $-9$ & $1$ & $1/2$  &  $5$\\
$-7$ & $0$ & $5/6$  &  $-9$ & $1$ & $5/6$  &  $5$\\
$-7$ & $0$ & $1/6$  &  $-9$ & $2$ & $1/6$  &  $5$\\
$-7$ & $0$ & $1/6$  &  $-9$ & $2$ & $5/6$  &  $6$\\
$-7$ & $0$ & $5/6$  &  $-9$ & $2$ & $1/6$  &  $6$\\
$-7$ & $0$ & $5/6$  &  $-9$ & $2$ & $5/6$  &  $5$\\
$-7$ & $1$ & $1/2$  &  $-9$ & $0$ & $1/6$  &  $5$\\
$-7$ & $1$ & $1/2$  &  $-9$ & $0$ & $5/6$  &  $5$\\
$-7$ & $1$ & $1/2$  &  $-9$ & $1$ & $1/6$  &  $5$\\
$-7$ & $1$ & $1/2$  &  $-9$ & $1$ & $1/2$  &  $6$\\
$-7$ & $1$ & $1/2$  &  $-9$ & $1$ & $5/6$  &  $5$\\
$-7$ & $1$ & $1/2$  &  $-9$ & $2$ & $1/6$  &  $6$\\
$-7$ & $1$ & $1/2$  &  $-9$ & $2$ & $5/6$  &  $6$\\\hline
$-8$ & $0$ & $0$  &  $-9$ & $0$ & $1/6$  &  $6$\\
$-8$ & $0$ & $0$  &  $-9$ & $0$ & $5/6$  &  $6$\\
$-8$ & $0$ & $1/3$  &  $-9$ & $0$ & $1/6$  &  $6$\\
$-8$ & $0$ & $1/3$  &  $-9$ & $0$ & $5/6$  &  $6$\\
$-8$ & $0$ & $2/3$  &  $-9$ & $0$ & $1/6$  &  $6$\\
$-8$ & $0$ & $2/3$  &  $-9$ & $0$ & $5/6$  &  $6$\\
$-8$ & $0$ & $0$  &  $-9$ & $1$ & $1/6$  &  $5$\\
$-8$ & $0$ & $0$  &  $-9$ & $1$ & $1/2$  &  $6$\\
$-8$ & $0$ & $0$  &  $-9$ & $1$ & $5/6$  &  $5$\\
$-8$ & $0$ & $1/3$  &  $-9$ & $1$ & $1/6$  &  $6$\\
$-8$ & $0$ & $1/3$  &  $-9$ & $1$ & $1/2$  &  $5$\\
$-8$ & $0$ & $1/3$  &  $-9$ & $1$ & $5/6$  &  $5$\\
$-8$ & $0$ & $2/3$  &  $-9$ & $1$ & $1/6$  &  $5$\\
$-8$ & $0$ & $2/3$  &  $-9$ & $1$ & $1/2$  &  $5$\\
$-8$ & $0$ & $2/3$  &  $-9$ & $1$ & $5/6$  &  $6$\\
$-8$ & $0$ & $0$  &  $-9$ & $2$ & $1/6$  &  $6$\\
$-8$ & $0$ & $0$  &  $-9$ & $2$ & $5/6$  &  $6$\\
$-8$ & $0$ & $1/3$  &  $-9$ & $2$ & $1/6$  &  $5$\\
$-8$ & $0$ & $1/3$  &  $-9$ & $2$ & $5/6$  &  $6$\\
$-8$ & $0$ & $2/3$  &  $-9$ & $2$ & $1/6$  &  $6$\\
$-8$ & $0$ & $2/3$  &  $-9$ & $2$ & $5/6$  &  $5$\\
$-8$ & $1$ & $1/3$  &  $-9$ & $0$ & $1/6$  &  $6$\\
$-8$ & $1$ & $1/3$  &  $-9$ & $0$ & $5/6$  &  $5$\\
$-8$ & $1$ & $2/3$  &  $-9$ & $0$ & $1/6$  &  $5$\\
$-8$ & $1$ & $2/3$  &  $-9$ & $0$ & $5/6$  &  $6$\\
$-8$ & $1$ & $1/3$  &  $-9$ & $1$ & $1/6$  &  $5$\\
$-8$ & $1$ & $1/3$  &  $-9$ & $1$ & $1/2$  &  $6$\\
$-8$ & $1$ & $1/3$  &  $-9$ & $1$ & $5/6$  &  $6$\\
$-8$ & $1$ & $2/3$  &  $-9$ & $1$ & $1/6$  &  $6$\\
$-8$ & $1$ & $2/3$  &  $-9$ & $1$ & $1/2$  &  $6$\\
$-8$ & $1$ & $2/3$  &  $-9$ & $1$ & $5/6$  &  $5$\\
$-8$ & $1$ & $1/3$  &  $-9$ & $2$ & $1/6$  &  $6$\\
$-8$ & $1$ & $1/3$  &  $-9$ & $2$ & $5/6$  &  $6$\\
$-8$ & $1$ & $2/3$  &  $-9$ & $2$ & $1/6$  &  $6$\\
$-8$ & $1$ & $2/3$  &  $-9$ & $2$ & $5/6$  &  $6$\\
$-8$ & $2$ & $0$  &  $-9$ & $0$ & $1/6$  &  $6$\\
$-8$ & $2$ & $0$  &  $-9$ & $0$ & $5/6$  &  $6$\\
$-8$ & $2$ & $0$  &  $-9$ & $1$ & $1/6$  &  $6$\\
$-8$ & $2$ & $0$  &  $-9$ & $1$ & $1/2$  &  $6$\\
$-8$ & $2$ & $0$  &  $-9$ & $1$ & $5/6$  &  $6$\\
$-8$ & $2$ & $0$  &  $-9$ & $2$ & $1/6$  &  $5$\\
$-8$ & $2$ & $0$  &  $-9$ & $2$ & $5/6$  &  $5$\\\hline
$-9$ & $0$ & $1/6$  &  $-9$ & $0$ & $1/6$  &  $6$\\
$-9$ & $0$ & $1/6$  &  $-9$ & $0$ & $5/6$  &  $7$\\
$-9$ & $0$ & $5/6$  &  $-9$ & $0$ & $5/6$  &  $6$\\
$-9$ & $0$ & $1/6$  &  $-9$ & $1$ & $1/6$  &  $6$\\
$-9$ & $0$ & $1/6$  &  $-9$ & $1$ & $1/2$  &  $6$\\
$-9$ & $0$ & $1/6$  &  $-9$ & $1$ & $5/6$  &  $5$\\
$-9$ & $0$ & $5/6$  &  $-9$ & $1$ & $1/6$  &  $5$\\
$-9$ & $0$ & $5/6$  &  $-9$ & $1$ & $1/2$  &  $6$\\
$-9$ & $0$ & $5/6$  &  $-9$ & $1$ & $5/6$  &  $6$\\
$-9$ & $0$ & $1/6$  &  $-9$ & $2$ & $1/6$  &  $6$\\
$-9$ & $0$ & $1/6$  &  $-9$ & $2$ & $5/6$  &  $6$\\
$-9$ & $0$ & $5/6$  &  $-9$ & $2$ & $1/6$  &  $6$\\
$-9$ & $0$ & $5/6$  &  $-9$ & $2$ & $5/6$  &  $6$\\
$-9$ & $1$ & $1/6$  &  $-9$ & $1$ & $1/6$  &  $6$\\
$-9$ & $1$ & $1/6$  &  $-9$ & $1$ & $1/2$  &  $6$\\
$-9$ & $1$ & $1/6$  &  $-9$ & $1$ & $5/6$  &  $6$\\
$-9$ & $1$ & $1/2$  &  $-9$ & $1$ & $1/2$  &  $6$\\
$-9$ & $1$ & $1/2$  &  $-9$ & $1$ & $5/6$  &  $6$\\
$-9$ & $1$ & $5/6$  &  $-9$ & $1$ & $5/6$  &  $6$\\
$-9$ & $1$ & $1/6$  &  $-9$ & $2$ & $1/6$  &  $6$\\
$-9$ & $1$ & $1/6$  &  $-9$ & $2$ & $5/6$  &  $7$\\
$-9$ & $1$ & $1/2$  &  $-9$ & $2$ & $1/6$  &  $6$\\
$-9$ & $1$ & $1/2$  &  $-9$ & $2$ & $5/6$  &  $6$\\
$-9$ & $1$ & $5/6$  &  $-9$ & $2$ & $1/6$  &  $7$\\
$-9$ & $1$ & $5/6$  &  $-9$ & $2$ & $5/6$  &  $6$\\
$-9$ & $2$ & $1/6$  &  $-9$ & $2$ & $1/6$  &  $6$\\
$-9$ & $2$ & $1/6$  &  $-9$ & $2$ & $5/6$  &  $5$\\
$-9$ & $2$ & $5/6$  &  $-9$ & $2$ & $5/6$  &  $6$\\\hline
$-6$ & $0$ & $0$  &  $-10$ & $0$ & $0$  &  $5$\\
$-6$ & $0$ & $0$  &  $-10$ & $1$ & $1/3$  &  $5$\\
$-6$ & $0$ & $0$  &  $-10$ & $1$ & $2/3$  &  $5$\\
$-6$ & $0$ & $0$  &  $-10$ & $2$ & $0$  &  $6$\\
$-6$ & $0$ & $0$  &  $-10$ & $2$ & $1/3$  &  $5$\\
$-6$ & $0$ & $0$  &  $-10$ & $2$ & $2/3$  &  $5$\\\hline
$-7$ & $0$ & $1/6$  &  $-10$ & $0$ & $0$  &  $6$\\
$-7$ & $0$ & $5/6$  &  $-10$ & $0$ & $0$  &  $6$\\
$-7$ & $0$ & $1/6$  &  $-10$ & $1$ & $1/3$  &  $6$\\
$-7$ & $0$ & $1/6$  &  $-10$ & $1$ & $2/3$  &  $5$\\
$-7$ & $0$ & $5/6$  &  $-10$ & $1$ & $1/3$  &  $5$\\
$-7$ & $0$ & $5/6$  &  $-10$ & $1$ & $2/3$  &  $6$\\
$-7$ & $0$ & $1/6$  &  $-10$ & $2$ & $0$  &  $6$\\
$-7$ & $0$ & $1/6$  &  $-10$ & $2$ & $1/3$  &  $5$\\
$-7$ & $0$ & $1/6$  &  $-10$ & $2$ & $2/3$  &  $6$\\
$-7$ & $0$ & $5/6$  &  $-10$ & $2$ & $0$  &  $6$\\
$-7$ & $0$ & $5/6$  &  $-10$ & $2$ & $1/3$  &  $6$\\
$-7$ & $0$ & $5/6$  &  $-10$ & $2$ & $2/3$  &  $5$\\
$-7$ & $1$ & $1/2$  &  $-10$ & $0$ & $0$  &  $6$\\
$-7$ & $1$ & $1/2$  &  $-10$ & $1$ & $1/3$  &  $6$\\
$-7$ & $1$ & $1/2$  &  $-10$ & $1$ & $2/3$  &  $6$\\
$-7$ & $1$ & $1/2$  &  $-10$ & $2$ & $0$  &  $6$\\
$-7$ & $1$ & $1/2$  &  $-10$ & $2$ & $1/3$  &  $6$\\
$-7$ & $1$ & $1/2$  &  $-10$ & $2$ & $2/3$  &  $6$\\\hline
$-8$ & $0$ & $0$  &  $-10$ & $0$ & $0$  &  $7$\\
$-8$ & $0$ & $1/3$  &  $-10$ & $0$ & $0$  &  $6$\\
$-8$ & $0$ & $2/3$  &  $-10$ & $0$ & $0$  &  $6$\\
$-8$ & $0$ & $0$  &  $-10$ & $1$ & $1/3$  &  $6$\\
$-8$ & $0$ & $0$  &  $-10$ & $1$ & $2/3$  &  $6$\\
$-8$ & $0$ & $1/3$  &  $-10$ & $1$ & $1/3$  &  $6$\\
$-8$ & $0$ & $1/3$  &  $-10$ & $1$ & $2/3$  &  $5$\\
$-8$ & $0$ & $2/3$  &  $-10$ & $1$ & $1/3$  &  $5$\\
$-8$ & $0$ & $2/3$  &  $-10$ & $1$ & $2/3$  &  $6$\\
$-8$ & $0$ & $0$  &  $-10$ & $2$ & $0$  &  $6$\\
$-8$ & $0$ & $0$  &  $-10$ & $2$ & $1/3$  &  $6$\\
$-8$ & $0$ & $0$  &  $-10$ & $2$ & $2/3$  &  $6$\\
$-8$ & $0$ & $1/3$  &  $-10$ & $2$ & $0$  &  $6$\\
$-8$ & $0$ & $1/3$  &  $-10$ & $2$ & $1/3$  &  $6$\\
$-8$ & $0$ & $1/3$  &  $-10$ & $2$ & $2/3$  &  $6$\\
$-8$ & $0$ & $2/3$  &  $-10$ & $2$ & $0$  &  $6$\\
$-8$ & $0$ & $2/3$  &  $-10$ & $2$ & $1/3$  &  $6$\\
$-8$ & $0$ & $2/3$  &  $-10$ & $2$ & $2/3$  &  $6$\\
$-8$ & $1$ & $1/3$  &  $-10$ & $0$ & $0$  &  $6$\\
$-8$ & $1$ & $2/3$  &  $-10$ & $0$ & $0$  &  $6$\\
$-8$ & $1$ & $1/3$  &  $-10$ & $1$ & $1/3$  &  $6$\\
$-8$ & $1$ & $1/3$  &  $-10$ & $1$ & $2/3$  &  $6$\\
$-8$ & $1$ & $2/3$  &  $-10$ & $1$ & $1/3$  &  $6$\\
$-8$ & $1$ & $2/3$  &  $-10$ & $1$ & $2/3$  &  $6$\\
$-8$ & $1$ & $1/3$  &  $-10$ & $2$ & $0$  &  $6$\\
$-8$ & $1$ & $1/3$  &  $-10$ & $2$ & $1/3$  &  $6$\\
$-8$ & $1$ & $1/3$  &  $-10$ & $2$ & $2/3$  &  $7$\\
$-8$ & $1$ & $2/3$  &  $-10$ & $2$ & $0$  &  $6$\\
$-8$ & $1$ & $2/3$  &  $-10$ & $2$ & $1/3$  &  $7$\\
$-8$ & $1$ & $2/3$  &  $-10$ & $2$ & $2/3$  &  $6$\\
$-8$ & $2$ & $0$  &  $-10$ & $0$ & $0$  &  $6$\\
$-8$ & $2$ & $0$  &  $-10$ & $1$ & $1/3$  &  $6$\\
$-8$ & $2$ & $0$  &  $-10$ & $1$ & $2/3$  &  $6$\\
$-8$ & $2$ & $0$  &  $-10$ & $2$ & $0$  &  $5$\\
$-8$ & $2$ & $0$  &  $-10$ & $2$ & $1/3$  &  $6$\\
$-8$ & $2$ & $0$  &  $-10$ & $2$ & $2/3$  &  $6$\\\hline
$-9$ & $0$ & $1/6$  &  $-10$ & $0$ & $0$  &  $7$\\
$-9$ & $0$ & $5/6$  &  $-10$ & $0$ & $0$  &  $7$\\
$-9$ & $0$ & $1/6$  &  $-10$ & $1$ & $1/3$  &  $6$\\
$-9$ & $0$ & $1/6$  &  $-10$ & $1$ & $2/3$  &  $6$\\
$-9$ & $0$ & $5/6$  &  $-10$ & $1$ & $1/3$  &  $6$\\
$-9$ & $0$ & $5/6$  &  $-10$ & $1$ & $2/3$  &  $6$\\
$-9$ & $0$ & $1/6$  &  $-10$ & $2$ & $0$  &  $6$\\
$-9$ & $0$ & $1/6$  &  $-10$ & $2$ & $1/3$  &  $7$\\
$-9$ & $0$ & $1/6$  &  $-10$ & $2$ & $2/3$  &  $6$\\
$-9$ & $0$ & $5/6$  &  $-10$ & $2$ & $0$  &  $6$\\
$-9$ & $0$ & $5/6$  &  $-10$ & $2$ & $1/3$  &  $6$\\
$-9$ & $0$ & $5/6$  &  $-10$ & $2$ & $2/3$  &  $7$\\
$-9$ & $1$ & $1/6$  &  $-10$ & $0$ & $0$  &  $6$\\
$-9$ & $1$ & $1/2$  &  $-10$ & $0$ & $0$  &  $6$\\
$-9$ & $1$ & $5/6$  &  $-10$ & $0$ & $0$  &  $6$\\
$-9$ & $1$ & $1/6$  &  $-10$ & $1$ & $1/3$  &  $7$\\
$-9$ & $1$ & $1/6$  &  $-10$ & $1$ & $2/3$  &  $6$\\
$-9$ & $1$ & $1/2$  &  $-10$ & $1$ & $1/3$  &  $6$\\
$-9$ & $1$ & $1/2$  &  $-10$ & $1$ & $2/3$  &  $6$\\
$-9$ & $1$ & $5/6$  &  $-10$ & $1$ & $1/3$  &  $6$\\
$-9$ & $1$ & $5/6$  &  $-10$ & $1$ & $2/3$  &  $7$\\
$-9$ & $1$ & $1/6$  &  $-10$ & $2$ & $0$  &  $7$\\
$-9$ & $1$ & $1/6$  &  $-10$ & $2$ & $1/3$  &  $6$\\
$-9$ & $1$ & $1/6$  &  $-10$ & $2$ & $2/3$  &  $7$\\
$-9$ & $1$ & $1/2$  &  $-10$ & $2$ & $0$  &  $6$\\
$-9$ & $1$ & $1/2$  &  $-10$ & $2$ & $1/3$  &  $7$\\
$-9$ & $1$ & $1/2$  &  $-10$ & $2$ & $2/3$  &  $7$\\
$-9$ & $1$ & $5/6$  &  $-10$ & $2$ & $0$  &  $7$\\
$-9$ & $1$ & $5/6$  &  $-10$ & $2$ & $1/3$  &  $7$\\
$-9$ & $1$ & $5/6$  &  $-10$ & $2$ & $2/3$  &  $6$\\
$-9$ & $2$ & $1/6$  &  $-10$ & $0$ & $0$  &  $6$\\
$-9$ & $2$ & $5/6$  &  $-10$ & $0$ & $0$  &  $6$\\
$-9$ & $2$ & $1/6$  &  $-10$ & $1$ & $1/3$  &  $6$\\
$-9$ & $2$ & $1/6$  &  $-10$ & $1$ & $2/3$  &  $7$\\
$-9$ & $2$ & $5/6$  &  $-10$ & $1$ & $1/3$  &  $7$\\
$-9$ & $2$ & $5/6$  &  $-10$ & $1$ & $2/3$  &  $6$\\
$-9$ & $2$ & $1/6$  &  $-10$ & $2$ & $0$  &  $6$\\
$-9$ & $2$ & $1/6$  &  $-10$ & $2$ & $1/3$  &  $6$\\
$-9$ & $2$ & $1/6$  &  $-10$ & $2$ & $2/3$  &  $6$\\
$-9$ & $2$ & $5/6$  &  $-10$ & $2$ & $0$  &  $6$\\
$-9$ & $2$ & $5/6$  &  $-10$ & $2$ & $1/3$  &  $6$\\
$-9$ & $2$ & $5/6$  &  $-10$ & $2$ & $2/3$  &  $6$\\\hline
$-10$ & $0$ & $0$  &  $-10$ & $0$ & $0$  &  $7$\\
$-10$ & $0$ & $0$  &  $-10$ & $1$ & $1/3$  &  $6$\\
$-10$ & $0$ & $0$  &  $-10$ & $1$ & $2/3$  &  $6$\\
$-10$ & $0$ & $0$  &  $-10$ & $2$ & $0$  &  $6$\\
$-10$ & $0$ & $0$  &  $-10$ & $2$ & $1/3$  &  $7$\\
$-10$ & $0$ & $0$  &  $-10$ & $2$ & $2/3$  &  $7$\\
$-10$ & $1$ & $1/3$  &  $-10$ & $1$ & $1/3$  &  $7$\\
$-10$ & $1$ & $1/3$  &  $-10$ & $1$ & $2/3$  &  $6$\\
$-10$ & $1$ & $2/3$  &  $-10$ & $1$ & $2/3$  &  $7$\\
$-10$ & $1$ & $1/3$  &  $-10$ & $2$ & $0$  &  $7$\\
$-10$ & $1$ & $1/3$  &  $-10$ & $2$ & $1/3$  &  $7$\\
$-10$ & $1$ & $1/3$  &  $-10$ & $2$ & $2/3$  &  $7$\\
$-10$ & $1$ & $2/3$  &  $-10$ & $2$ & $0$  &  $7$\\
$-10$ & $1$ & $2/3$  &  $-10$ & $2$ & $1/3$  &  $7$\\
$-10$ & $1$ & $2/3$  &  $-10$ & $2$ & $2/3$  &  $7$\\
$-10$ & $2$ & $0$  &  $-10$ & $2$ & $0$  &  $7$\\
$-10$ & $2$ & $0$  &  $-10$ & $2$ & $1/3$  &  $6$\\
$-10$ & $2$ & $0$  &  $-10$ & $2$ & $2/3$  &  $6$\\
$-10$ & $2$ & $1/3$  &  $-10$ & $2$ & $1/3$  &  $6$\\
$-10$ & $2$ & $1/3$  &  $-10$ & $2$ & $2/3$  &  $7$\\
$-10$ & $2$ & $2/3$  &  $-10$ & $2$ & $2/3$  &  $6$\\\hline
$-6$ & $0$ & $0$  &  $-11$ & $1$ & $1/2$  &  $6$\\
$-6$ & $0$ & $0$  &  $-11$ & $2$ & $1/6$  &  $6$\\
$-6$ & $0$ & $0$  &  $-11$ & $2$ & $5/6$  &  $6$\\\hline
$-7$ & $0$ & $1/6$  &  $-11$ & $1$ & $1/2$  &  $6$\\
$-7$ & $0$ & $5/6$  &  $-11$ & $1$ & $1/2$  &  $6$\\
$-7$ & $0$ & $1/6$  &  $-11$ & $2$ & $1/6$  &  $6$\\
$-7$ & $0$ & $1/6$  &  $-11$ & $2$ & $5/6$  &  $6$\\
$-7$ & $0$ & $5/6$  &  $-11$ & $2$ & $1/6$  &  $6$\\
$-7$ & $0$ & $5/6$  &  $-11$ & $2$ & $5/6$  &  $6$\\
$-7$ & $1$ & $1/2$  &  $-11$ & $1$ & $1/2$  &  $6$\\
$-7$ & $1$ & $1/2$  &  $-11$ & $2$ & $1/6$  &  $6$\\
$-7$ & $1$ & $1/2$  &  $-11$ & $2$ & $5/6$  &  $6$\\\hline
$-8$ & $0$ & $0$  &  $-11$ & $1$ & $1/2$  &  $6$\\
$-8$ & $0$ & $1/3$  &  $-11$ & $1$ & $1/2$  &  $6$\\
$-8$ & $0$ & $2/3$  &  $-11$ & $1$ & $1/2$  &  $6$\\
$-8$ & $0$ & $0$  &  $-11$ & $2$ & $1/6$  &  $6$\\
$-8$ & $0$ & $0$  &  $-11$ & $2$ & $5/6$  &  $6$\\
$-8$ & $0$ & $1/3$  &  $-11$ & $2$ & $1/6$  &  $7$\\
$-8$ & $0$ & $1/3$  &  $-11$ & $2$ & $5/6$  &  $6$\\
$-8$ & $0$ & $2/3$  &  $-11$ & $2$ & $1/6$  &  $6$\\
$-8$ & $0$ & $2/3$  &  $-11$ & $2$ & $5/6$  &  $7$\\
$-8$ & $1$ & $1/3$  &  $-11$ & $1$ & $1/2$  &  $6$\\
$-8$ & $1$ & $2/3$  &  $-11$ & $1$ & $1/2$  &  $6$\\
$-8$ & $1$ & $1/3$  &  $-11$ & $2$ & $1/6$  &  $6$\\
$-8$ & $1$ & $1/3$  &  $-11$ & $2$ & $5/6$  &  $7$\\
$-8$ & $1$ & $2/3$  &  $-11$ & $2$ & $1/6$  &  $7$\\
$-8$ & $1$ & $2/3$  &  $-11$ & $2$ & $5/6$  &  $6$\\
$-8$ & $2$ & $0$  &  $-11$ & $1$ & $1/2$  &  $6$\\
$-8$ & $2$ & $0$  &  $-11$ & $2$ & $1/6$  &  $6$\\
$-8$ & $2$ & $0$  &  $-11$ & $2$ & $5/6$  &  $6$\\\hline
$-9$ & $0$ & $1/6$  &  $-11$ & $1$ & $1/2$  &  $6$\\
$-9$ & $0$ & $5/6$  &  $-11$ & $1$ & $1/2$  &  $6$\\
$-9$ & $0$ & $1/6$  &  $-11$ & $2$ & $1/6$  &  $7$\\
$-9$ & $0$ & $1/6$  &  $-11$ & $2$ & $5/6$  &  $6$\\
$-9$ & $0$ & $5/6$  &  $-11$ & $2$ & $1/6$  &  $6$\\
$-9$ & $0$ & $5/6$  &  $-11$ & $2$ & $5/6$  &  $7$\\
$-9$ & $1$ & $1/6$  &  $-11$ & $1$ & $1/2$  &  $7$\\
$-9$ & $1$ & $1/2$  &  $-11$ & $1$ & $1/2$  &  $6$\\
$-9$ & $1$ & $5/6$  &  $-11$ & $1$ & $1/2$  &  $7$\\
$-9$ & $1$ & $1/6$  &  $-11$ & $2$ & $1/6$  &  $7$\\
$-9$ & $1$ & $1/6$  &  $-11$ & $2$ & $5/6$  &  $7$\\
$-9$ & $1$ & $1/2$  &  $-11$ & $2$ & $1/6$  &  $7$\\
$-9$ & $1$ & $1/2$  &  $-11$ & $2$ & $5/6$  &  $7$\\
$-9$ & $1$ & $5/6$  &  $-11$ & $2$ & $1/6$  &  $7$\\
$-9$ & $1$ & $5/6$  &  $-11$ & $2$ & $5/6$  &  $7$\\
$-9$ & $2$ & $1/6$  &  $-11$ & $1$ & $1/2$  &  $7$\\
$-9$ & $2$ & $5/6$  &  $-11$ & $1$ & $1/2$  &  $7$\\
$-9$ & $2$ & $1/6$  &  $-11$ & $2$ & $1/6$  &  $6$\\
$-9$ & $2$ & $1/6$  &  $-11$ & $2$ & $5/6$  &  $7$\\
$-9$ & $2$ & $5/6$  &  $-11$ & $2$ & $1/6$  &  $7$\\
$-9$ & $2$ & $5/6$  &  $-11$ & $2$ & $5/6$  &  $6$\\\hline
$-10$ & $0$ & $0$  &  $-11$ & $1$ & $1/2$  &  $6$\\
$-10$ & $0$ & $0$  &  $-11$ & $2$ & $1/6$  &  $7$\\
$-10$ & $0$ & $0$  &  $-11$ & $2$ & $5/6$  &  $7$\\
$-10$ & $1$ & $1/3$  &  $-11$ & $1$ & $1/2$  &  $7$\\
$-10$ & $1$ & $2/3$  &  $-11$ & $1$ & $1/2$  &  $7$\\
$-10$ & $1$ & $1/3$  &  $-11$ & $2$ & $1/6$  &  $8$\\
$-10$ & $1$ & $1/3$  &  $-11$ & $2$ & $5/6$  &  $7$\\
$-10$ & $1$ & $2/3$  &  $-11$ & $2$ & $1/6$  &  $7$\\
$-10$ & $1$ & $2/3$  &  $-11$ & $2$ & $5/6$  &  $8$\\
$-10$ & $2$ & $0$  &  $-11$ & $1$ & $1/2$  &  $8$\\
$-10$ & $2$ & $1/3$  &  $-11$ & $1$ & $1/2$  &  $7$\\
$-10$ & $2$ & $2/3$  &  $-11$ & $1$ & $1/2$  &  $7$\\
$-10$ & $2$ & $0$  &  $-11$ & $2$ & $1/6$  &  $7$\\
$-10$ & $2$ & $0$  &  $-11$ & $2$ & $5/6$  &  $7$\\
$-10$ & $2$ & $1/3$  &  $-11$ & $2$ & $1/6$  &  $6$\\
$-10$ & $2$ & $1/3$  &  $-11$ & $2$ & $5/6$  &  $7$\\
$-10$ & $2$ & $2/3$  &  $-11$ & $2$ & $1/6$  &  $7$\\
$-10$ & $2$ & $2/3$  &  $-11$ & $2$ & $5/6$  &  $6$\\\hline
$-11$ & $1$ & $1/2$  &  $-11$ & $1$ & $1/2$  &  $8$\\
$-11$ & $1$ & $1/2$  &  $-11$ & $2$ & $1/6$  &  $8$\\
$-11$ & $1$ & $1/2$  &  $-11$ & $2$ & $5/6$  &  $8$\\
$-11$ & $2$ & $1/6$  &  $-11$ & $2$ & $1/6$  &  $7$\\
$-11$ & $2$ & $1/6$  &  $-11$ & $2$ & $5/6$  &  $7$\\
$-11$ & $2$ & $5/6$  &  $-11$ & $2$ & $5/6$  &  $7$\\\hline
$-6$ & $0$ & $0$  &  $-12$ & $2$ & $0$  &  $6$\\\hline
$-7$ & $0$ & $1/6$  &  $-12$ & $2$ & $0$  &  $6$\\
$-7$ & $0$ & $5/6$  &  $-12$ & $2$ & $0$  &  $6$\\
$-7$ & $1$ & $1/2$  &  $-12$ & $2$ & $0$  &  $6$\\\hline
$-8$ & $0$ & $0$  &  $-12$ & $2$ & $0$  &  $6$\\
$-8$ & $0$ & $1/3$  &  $-12$ & $2$ & $0$  &  $7$\\
$-8$ & $0$ & $2/3$  &  $-12$ & $2$ & $0$  &  $7$\\
$-8$ & $1$ & $1/3$  &  $-12$ & $2$ & $0$  &  $7$\\
$-8$ & $1$ & $2/3$  &  $-12$ & $2$ & $0$  &  $7$\\
$-8$ & $2$ & $0$  &  $-12$ & $2$ & $0$  &  $7$\\\hline
$-9$ & $0$ & $1/6$  &  $-12$ & $2$ & $0$  &  $7$\\
$-9$ & $0$ & $5/6$  &  $-12$ & $2$ & $0$  &  $7$\\
$-9$ & $1$ & $1/6$  &  $-12$ & $2$ & $0$  &  $7$\\
$-9$ & $1$ & $1/2$  &  $-12$ & $2$ & $0$  &  $8$\\
$-9$ & $1$ & $5/6$  &  $-12$ & $2$ & $0$  &  $7$\\
$-9$ & $2$ & $1/6$  &  $-12$ & $2$ & $0$  &  $7$\\
$-9$ & $2$ & $5/6$  &  $-12$ & $2$ & $0$  &  $7$\\\hline
$-10$ & $0$ & $0$  &  $-12$ & $2$ & $0$  &  $8$\\
$-10$ & $1$ & $1/3$  &  $-12$ & $2$ & $0$  &  $8$\\
$-10$ & $1$ & $2/3$  &  $-12$ & $2$ & $0$  &  $8$\\
$-10$ & $2$ & $0$  &  $-12$ & $2$ & $0$  &  $7$\\
$-10$ & $2$ & $1/3$  &  $-12$ & $2$ & $0$  &  $7$\\
$-10$ & $2$ & $2/3$  &  $-12$ & $2$ & $0$  &  $7$\\\hline
$-11$ & $1$ & $1/2$  &  $-12$ & $2$ & $0$  &  $8$\\
$-11$ & $2$ & $1/6$  &  $-12$ & $2$ & $0$  &  $7$\\
$-11$ & $2$ & $5/6$  &  $-12$ & $2$ & $0$  &  $7$\\\hline
$-12$ & $2$ & $0$  &  $-12$ & $2$ & $0$  &  $7$\\\hline
\end{supertabular}
\end{center}

\newpage

\begin{center}
\tablehead{
 \hline
 \multicolumn{7}{|c|}{$T^2/Z_3$ with $M_{ab} < 0,\, M_{ca} > 0$} \\ \hline \hline
 \multicolumn{3}{|c||}{$ab$-sector} & \multicolumn{3}{c||}{$ca$-sector} & $bc$-sector \\ \hline
 $M_{ab}$ & $\omega^i_{ab}$ & $\alpha_{ab}$ &  $M_{ca}$ & $\omega^i_{ca}$ & $\alpha_{ca}$ &  \# of Higgs \\ \hline\hline
}
\tabletail{\hline}
\tablecaption{Possible parameter configurations on $T^2/Z_3$ with $M_{ab} < 0,\, M_{ca} > 0$.}
\label{tbl:T2Z3_positive}
\begin{supertabular}{|c|c|c||c|c|c||c|}
$-6$ & $0$ & $0$  &  $6$ & $0$ & $0$  &  $\fbox{1}$\\\hline
$-6$ & $0$ & $0$  &  $7$ & $0$ & $1/6$  &  $1$\\
$-6$ & $0$ & $0$  &  $7$ & $0$ & $5/6$  &  $1$\\\hline
$-7$ & $0$ & $1/6$  &  $7$ & $0$ & $5/6$  &  $\fbox{1}$\\
$-7$ & $0$ & $5/6$  &  $7$ & $0$ & $1/6$  &  $\fbox{1}$\\\hline
$-6$ & $0$ & $0$  &  $8$ & $0$ & $0$  &  $1$\\
$-6$ & $0$ & $0$  &  $8$ & $0$ & $1/3$  &  $1$\\
$-6$ & $0$ & $0$  &  $8$ & $0$ & $2/3$  &  $1$\\\hline
$-7$ & $0$ & $1/6$  &  $8$ & $0$ & $0$  &  $1$\\
$-7$ & $0$ & $1/6$  &  $8$ & $0$ & $2/3$  &  $1$\\
$-7$ & $0$ & $5/6$  &  $8$ & $0$ & $0$  &  $1$\\
$-7$ & $0$ & $5/6$  &  $8$ & $0$ & $1/3$  &  $1$\\\hline
$-8$ & $0$ & $0$  &  $8$ & $0$ & $0$  &  $\fbox{1}$\\
$-8$ & $0$ & $1/3$  &  $8$ & $0$ & $2/3$  &  $\fbox{1}$\\
$-8$ & $0$ & $2/3$  &  $8$ & $0$ & $1/3$  &  $\fbox{1}$\\\hline
$-6$ & $0$ & $0$  &  $9$ & $0$ & $1/6$  &  $1$\\
$-6$ & $0$ & $0$  &  $9$ & $0$ & $5/6$  &  $1$\\
$-6$ & $0$ & $0$  &  $9$ & $1$ & $1/6$  &  $1$\\
$-6$ & $0$ & $0$  &  $9$ & $1$ & $5/6$  &  $1$\\
$-6$ & $0$ & $0$  &  $9$ & $2$ & $1/6$  &  $1$\\
$-6$ & $0$ & $0$  &  $9$ & $2$ & $5/6$  &  $1$\\\hline
$-7$ & $0$ & $1/6$  &  $9$ & $0$ & $1/6$  &  $1$\\
$-7$ & $0$ & $1/6$  &  $9$ & $0$ & $5/6$  &  $1$\\
$-7$ & $0$ & $5/6$  &  $9$ & $0$ & $1/6$  &  $1$\\
$-7$ & $0$ & $5/6$  &  $9$ & $0$ & $5/6$  &  $1$\\
$-7$ & $0$ & $1/6$  &  $9$ & $1$ & $5/6$  &  $1$\\
$-7$ & $0$ & $5/6$  &  $9$ & $1$ & $1/6$  &  $1$\\
$-7$ & $0$ & $1/6$  &  $9$ & $2$ & $1/6$  &  $1$\\
$-7$ & $0$ & $5/6$  &  $9$ & $2$ & $5/6$  &  $1$\\
$-7$ & $1$ & $1/2$  &  $9$ & $1$ & $1/6$  &  $1$\\
$-7$ & $1$ & $1/2$  &  $9$ & $1$ & $5/6$  &  $1$\\
$-7$ & $1$ & $1/2$  &  $9$ & $2$ & $1/6$  &  $1$\\
$-7$ & $1$ & $1/2$  &  $9$ & $2$ & $5/6$  &  $1$\\\hline
$-8$ & $0$ & $0$  &  $9$ & $0$ & $1/6$  &  $1$\\
$-8$ & $0$ & $0$  &  $9$ & $0$ & $5/6$  &  $1$\\
$-8$ & $0$ & $1/3$  &  $9$ & $0$ & $5/6$  &  $1$\\
$-8$ & $0$ & $2/3$  &  $9$ & $0$ & $1/6$  &  $1$\\
$-8$ & $0$ & $1/3$  &  $9$ & $2$ & $1/6$  &  $1$\\
$-8$ & $0$ & $2/3$  &  $9$ & $2$ & $5/6$  &  $1$\\
$-8$ & $1$ & $1/3$  &  $9$ & $1$ & $1/6$  &  $1$\\
$-8$ & $1$ & $2/3$  &  $9$ & $1$ & $5/6$  &  $1$\\
$-8$ & $1$ & $1/3$  &  $9$ & $2$ & $5/6$  &  $1$\\
$-8$ & $1$ & $2/3$  &  $9$ & $2$ & $1/6$  &  $1$\\
$-8$ & $2$ & $0$  &  $9$ & $1$ & $1/6$  &  $1$\\
$-8$ & $2$ & $0$  &  $9$ & $1$ & $5/6$  &  $1$\\\hline
$-9$ & $0$ & $1/6$  &  $9$ & $0$ & $5/6$  &  $\fbox{1}$\\
$-9$ & $0$ & $5/6$  &  $9$ & $0$ & $1/6$  &  $\fbox{1}$\\
$-9$ & $1$ & $1/6$  &  $9$ & $2$ & $5/6$  &  $\fbox{1}$\\
$-9$ & $1$ & $5/6$  &  $9$ & $2$ & $1/6$  &  $\fbox{1}$\\
$-9$ & $2$ & $1/6$  &  $9$ & $1$ & $5/6$  &  $\fbox{1}$\\
$-9$ & $2$ & $5/6$  &  $9$ & $1$ & $1/6$  &  $\fbox{1}$\\\hline
$-6$ & $0$ & $0$  &  $10$ & $0$ & $0$  &  $1$\\
$-6$ & $0$ & $0$  &  $10$ & $1$ & $1/3$  &  $1$\\
$-6$ & $0$ & $0$  &  $10$ & $1$ & $2/3$  &  $1$\\
$-6$ & $0$ & $0$  &  $10$ & $2$ & $0$  &  $2$\\
$-6$ & $0$ & $0$  &  $10$ & $2$ & $1/3$  &  $1$\\
$-6$ & $0$ & $0$  &  $10$ & $2$ & $2/3$  &  $1$\\\hline
$-7$ & $0$ & $1/6$  &  $10$ & $0$ & $0$  &  $1$\\
$-7$ & $0$ & $5/6$  &  $10$ & $0$ & $0$  &  $1$\\
$-7$ & $0$ & $1/6$  &  $10$ & $1$ & $2/3$  &  $1$\\
$-7$ & $0$ & $5/6$  &  $10$ & $1$ & $1/3$  &  $1$\\
$-7$ & $0$ & $1/6$  &  $10$ & $2$ & $0$  &  $1$\\
$-7$ & $0$ & $1/6$  &  $10$ & $2$ & $1/3$  &  $1$\\
$-7$ & $0$ & $1/6$  &  $10$ & $2$ & $2/3$  &  $1$\\
$-7$ & $0$ & $5/6$  &  $10$ & $2$ & $0$  &  $1$\\
$-7$ & $0$ & $5/6$  &  $10$ & $2$ & $1/3$  &  $1$\\
$-7$ & $0$ & $5/6$  &  $10$ & $2$ & $2/3$  &  $1$\\
$-7$ & $1$ & $1/2$  &  $10$ & $1$ & $1/3$  &  $1$\\
$-7$ & $1$ & $1/2$  &  $10$ & $1$ & $2/3$  &  $1$\\
$-7$ & $1$ & $1/2$  &  $10$ & $2$ & $0$  &  $2$\\
$-7$ & $1$ & $1/2$  &  $10$ & $2$ & $1/3$  &  $1$\\
$-7$ & $1$ & $1/2$  &  $10$ & $2$ & $2/3$  &  $1$\\\hline
$-8$ & $0$ & $0$  &  $10$ & $0$ & $0$  &  $1$\\
$-8$ & $0$ & $1/3$  &  $10$ & $0$ & $0$  &  $1$\\
$-8$ & $0$ & $2/3$  &  $10$ & $0$ & $0$  &  $1$\\
$-8$ & $0$ & $1/3$  &  $10$ & $1$ & $2/3$  &  $1$\\
$-8$ & $0$ & $2/3$  &  $10$ & $1$ & $1/3$  &  $1$\\
$-8$ & $0$ & $0$  &  $10$ & $2$ & $1/3$  &  $1$\\
$-8$ & $0$ & $0$  &  $10$ & $2$ & $2/3$  &  $1$\\
$-8$ & $0$ & $1/3$  &  $10$ & $2$ & $0$  &  $1$\\
$-8$ & $0$ & $1/3$  &  $10$ & $2$ & $1/3$  &  $1$\\
$-8$ & $0$ & $2/3$  &  $10$ & $2$ & $0$  &  $1$\\
$-8$ & $0$ & $2/3$  &  $10$ & $2$ & $2/3$  &  $1$\\
$-8$ & $1$ & $1/3$  &  $10$ & $1$ & $1/3$  &  $1$\\
$-8$ & $1$ & $2/3$  &  $10$ & $1$ & $2/3$  &  $1$\\
$-8$ & $1$ & $1/3$  &  $10$ & $2$ & $0$  &  $1$\\
$-8$ & $1$ & $1/3$  &  $10$ & $2$ & $1/3$  &  $1$\\
$-8$ & $1$ & $1/3$  &  $10$ & $2$ & $2/3$  &  $1$\\
$-8$ & $1$ & $2/3$  &  $10$ & $2$ & $0$  &  $1$\\
$-8$ & $1$ & $2/3$  &  $10$ & $2$ & $1/3$  &  $1$\\
$-8$ & $1$ & $2/3$  &  $10$ & $2$ & $2/3$  &  $1$\\
$-8$ & $2$ & $0$  &  $10$ & $1$ & $1/3$  &  $1$\\
$-8$ & $2$ & $0$  &  $10$ & $1$ & $2/3$  &  $1$\\
$-8$ & $2$ & $0$  &  $10$ & $2$ & $0$  &  $1$\\\hline
$-9$ & $0$ & $1/6$  &  $10$ & $0$ & $0$  &  $1$\\
$-9$ & $0$ & $5/6$  &  $10$ & $0$ & $0$  &  $1$\\
$-9$ & $0$ & $1/6$  &  $10$ & $2$ & $1/3$  &  $1$\\
$-9$ & $0$ & $5/6$  &  $10$ & $2$ & $2/3$  &  $1$\\
$-9$ & $1$ & $1/6$  &  $10$ & $1$ & $1/3$  &  $1$\\
$-9$ & $1$ & $5/6$  &  $10$ & $1$ & $2/3$  &  $1$\\
$-9$ & $1$ & $1/6$  &  $10$ & $2$ & $0$  &  $1$\\
$-9$ & $1$ & $1/6$  &  $10$ & $2$ & $2/3$  &  $1$\\
$-9$ & $1$ & $1/2$  &  $10$ & $2$ & $1/3$  &  $1$\\
$-9$ & $1$ & $1/2$  &  $10$ & $2$ & $2/3$  &  $1$\\
$-9$ & $1$ & $5/6$  &  $10$ & $2$ & $0$  &  $1$\\
$-9$ & $1$ & $5/6$  &  $10$ & $2$ & $1/3$  &  $1$\\
$-9$ & $2$ & $1/6$  &  $10$ & $1$ & $2/3$  &  $1$\\
$-9$ & $2$ & $5/6$  &  $10$ & $1$ & $1/3$  &  $1$\\\hline
$-10$ & $0$ & $0$  &  $10$ & $0$ & $0$  &  $\fbox{1}$\\
$-10$ & $1$ & $1/3$  &  $10$ & $2$ & $2/3$  &  $\fbox{1}$\\
$-10$ & $1$ & $2/3$  &  $10$ & $2$ & $1/3$  &  $\fbox{1}$\\
$-10$ & $2$ & $1/3$  &  $10$ & $1$ & $2/3$  &  $\fbox{1}$\\
$-10$ & $2$ & $2/3$  &  $10$ & $1$ & $1/3$  &  $\fbox{1}$\\\hline
$-6$ & $0$ & $0$  &  $11$ & $1$ & $1/2$  &  $2$\\
$-6$ & $0$ & $0$  &  $11$ & $2$ & $1/6$  &  $2$\\
$-6$ & $0$ & $0$  &  $11$ & $2$ & $5/6$  &  $2$\\\hline
$-7$ & $0$ & $1/6$  &  $11$ & $1$ & $1/2$  &  $1$\\
$-7$ & $0$ & $5/6$  &  $11$ & $1$ & $1/2$  &  $1$\\
$-7$ & $0$ & $1/6$  &  $11$ & $2$ & $1/6$  &  $1$\\
$-7$ & $0$ & $1/6$  &  $11$ & $2$ & $5/6$  &  $2$\\
$-7$ & $0$ & $5/6$  &  $11$ & $2$ & $1/6$  &  $2$\\
$-7$ & $0$ & $5/6$  &  $11$ & $2$ & $5/6$  &  $1$\\
$-7$ & $1$ & $1/2$  &  $11$ & $1$ & $1/2$  &  $2$\\
$-7$ & $1$ & $1/2$  &  $11$ & $2$ & $1/6$  &  $2$\\
$-7$ & $1$ & $1/2$  &  $11$ & $2$ & $5/6$  &  $2$\\\hline
$-8$ & $0$ & $1/3$  &  $11$ & $1$ & $1/2$  &  $1$\\
$-8$ & $0$ & $2/3$  &  $11$ & $1$ & $1/2$  &  $1$\\
$-8$ & $0$ & $0$  &  $11$ & $2$ & $1/6$  &  $1$\\
$-8$ & $0$ & $0$  &  $11$ & $2$ & $5/6$  &  $1$\\
$-8$ & $0$ & $1/3$  &  $11$ & $2$ & $1/6$  &  $1$\\
$-8$ & $0$ & $1/3$  &  $11$ & $2$ & $5/6$  &  $1$\\
$-8$ & $0$ & $2/3$  &  $11$ & $2$ & $1/6$  &  $1$\\
$-8$ & $0$ & $2/3$  &  $11$ & $2$ & $5/6$  &  $1$\\
$-8$ & $1$ & $1/3$  &  $11$ & $1$ & $1/2$  &  $1$\\
$-8$ & $1$ & $2/3$  &  $11$ & $1$ & $1/2$  &  $1$\\
$-8$ & $1$ & $1/3$  &  $11$ & $2$ & $1/6$  &  $2$\\
$-8$ & $1$ & $1/3$  &  $11$ & $2$ & $5/6$  &  $1$\\
$-8$ & $1$ & $2/3$  &  $11$ & $2$ & $1/6$  &  $1$\\
$-8$ & $1$ & $2/3$  &  $11$ & $2$ & $5/6$  &  $2$\\
$-8$ & $2$ & $0$  &  $11$ & $1$ & $1/2$  &  $2$\\
$-8$ & $2$ & $0$  &  $11$ & $2$ & $1/6$  &  $1$\\
$-8$ & $2$ & $0$  &  $11$ & $2$ & $5/6$  &  $1$\\\hline
$-9$ & $0$ & $1/6$  &  $11$ & $2$ & $1/6$  &  $1$\\
$-9$ & $0$ & $5/6$  &  $11$ & $2$ & $5/6$  &  $1$\\
$-9$ & $1$ & $1/6$  &  $11$ & $1$ & $1/2$  &  $1$\\
$-9$ & $1$ & $5/6$  &  $11$ & $1$ & $1/2$  &  $1$\\
$-9$ & $1$ & $1/6$  &  $11$ & $2$ & $1/6$  &  $1$\\
$-9$ & $1$ & $1/6$  &  $11$ & $2$ & $5/6$  &  $1$\\
$-9$ & $1$ & $1/2$  &  $11$ & $2$ & $1/6$  &  $1$\\
$-9$ & $1$ & $1/2$  &  $11$ & $2$ & $5/6$  &  $1$\\
$-9$ & $1$ & $5/6$  &  $11$ & $2$ & $1/6$  &  $1$\\
$-9$ & $1$ & $5/6$  &  $11$ & $2$ & $5/6$  &  $1$\\
$-9$ & $2$ & $1/6$  &  $11$ & $1$ & $1/2$  &  $1$\\
$-9$ & $2$ & $5/6$  &  $11$ & $1$ & $1/2$  &  $1$\\
$-9$ & $2$ & $1/6$  &  $11$ & $2$ & $5/6$  &  $1$\\
$-9$ & $2$ & $5/6$  &  $11$ & $2$ & $1/6$  &  $1$\\\hline
$-10$ & $1$ & $1/3$  &  $11$ & $2$ & $5/6$  &  $1$\\
$-10$ & $1$ & $2/3$  &  $11$ & $2$ & $1/6$  &  $1$\\
$-10$ & $2$ & $1/3$  &  $11$ & $1$ & $1/2$  &  $1$\\
$-10$ & $2$ & $2/3$  &  $11$ & $1$ & $1/2$  &  $1$\\\hline
$-6$ & $0$ & $0$  &  $12$ & $2$ & $0$  &  $2$\\\hline
$-7$ & $0$ & $1/6$  &  $12$ & $2$ & $0$  &  $2$\\
$-7$ & $0$ & $5/6$  &  $12$ & $2$ & $0$  &  $2$\\
$-7$ & $1$ & $1/2$  &  $12$ & $2$ & $0$  &  $2$\\\hline
$-8$ & $0$ & $0$  &  $12$ & $2$ & $0$  &  $2$\\
$-8$ & $0$ & $1/3$  &  $12$ & $2$ & $0$  &  $1$\\
$-8$ & $0$ & $2/3$  &  $12$ & $2$ & $0$  &  $1$\\
$-8$ & $1$ & $1/3$  &  $12$ & $2$ & $0$  &  $2$\\
$-8$ & $1$ & $2/3$  &  $12$ & $2$ & $0$  &  $2$\\
$-8$ & $2$ & $0$  &  $12$ & $2$ & $0$  &  $1$\\\hline
$-9$ & $0$ & $1/6$  &  $12$ & $2$ & $0$  &  $1$\\
$-9$ & $0$ & $5/6$  &  $12$ & $2$ & $0$  &  $1$\\
$-9$ & $1$ & $1/6$  &  $12$ & $2$ & $0$  &  $1$\\
$-9$ & $1$ & $1/2$  &  $12$ & $2$ & $0$  &  $2$\\
$-9$ & $1$ & $5/6$  &  $12$ & $2$ & $0$  &  $1$\\
$-9$ & $2$ & $1/6$  &  $12$ & $2$ & $0$  &  $1$\\
$-9$ & $2$ & $5/6$  &  $12$ & $2$ & $0$  &  $1$\\\hline
$-10$ & $1$ & $1/3$  &  $12$ & $2$ & $0$  &  $1$\\
$-10$ & $1$ & $2/3$  &  $12$ & $2$ & $0$  &  $1$\\
$-10$ & $2$ & $0$  &  $12$ & $2$ & $0$  &  $1$\\\hline
\end{supertabular}
\end{center}

\newpage

\subsection{$T^2/Z_4$}

\begin{center}
\tablehead{
 \hline
 \multicolumn{7}{|c|}{$T^2/Z_4$ with $M_{ab} < 0,\, M_{ca} < 0$} \\ \hline \hline
 \multicolumn{3}{|c||}{$ab$-sector} & \multicolumn{3}{c||}{$ca$-sector} & $bc$-sector \\ \hline
 $M_{ab}$ & $\omega^i_{ab}$ & $\alpha_{ab}$  &  $M_{ca}$ & $\omega^i_{ca}$ & $\alpha_{ca}$  &  \# of Higgs \\ \hline\hline
}
\tabletail{\hline}
\tablecaption{Possible parameter configurations on $T^2/Z_4$ with $M_{ab} < 0,\, M_{ca} < 0$.}
\label{tbl:T2Z4_negative}
\begin{supertabular}{|c|c|c||c|c|c||c|}
$-8$ & $0$ & $0$  &  $-8$ & $0$ & $0$  &  $5$\\\hline
$-8$ & $0$ & $0$  &  $-9$ & $0$ & $0$  &  $5$\\
$-8$ & $0$ & $0$  &  $-9$ & $1$ & $1/2$  &  $4$\\\hline
$-9$ & $0$ & $0$  &  $-9$ & $0$ & $0$  &  $5$\\
$-9$ & $0$ & $0$  &  $-9$ & $1$ & $1/2$  &  $4$\\
$-9$ & $1$ & $1/2$  &  $-9$ & $1$ & $1/2$  &  $5$\\\hline
$-8$ & $0$ & $0$  &  $-10$ & $0$ & $0$  &  $5$\\
$-8$ & $0$ & $0$  &  $-10$ & $0$ & $1/2$  &  $5$\\
\end{supertabular}
\end{center}

\newpage

\begin{center}
\tablehead{
 \hline
 \multicolumn{7}{|c|}{$T^2/Z_4$ with $M_{ab} < 0,\, M_{ca} > 0$} \\ \hline \hline
 \multicolumn{3}{|c||}{$ab$-sector} & \multicolumn{3}{c||}{$ca$-sector} & $bc$-sector \\ \hline
 $M_{ab}$ & $\omega^i_{ab}$ & $\alpha_{ab}$  &  $M_{ca}$ & $\omega^i_{ca}$ & $\alpha_{ca}$  &  \# of Higgs \\ \hline\hline
}
\tabletail{\hline}
\tablecaption{Possible parameter configurations on $T^2/Z_4$ with $M_{ab} < 0,\, M_{ca} > 0$.}
\label{tbl:T2Z4_positive}
\begin{supertabular}{|c|c|c||c|c|c||c|}
$-8$ & $0$ & $0$  &  $8$ & $0$ & $0$  &  $\fbox{1}$\\\hline
$-8$ & $0$ & $0$  &  $9$ & $0$ & $0$  &  $1$\\\hline
$-9$ & $0$ & $0$  &  $9$ & $0$ & $0$  &  $\fbox{1}$\\\hline
$-8$ & $0$ & $0$  &  $10$ & $0$ & $0$  &  $1$\\
$-8$ & $0$ & $0$  &  $10$ & $0$ & $1/2$  &  $1$\\
$-8$ & $0$ & $0$  &  $10$ & $2$ & $0$  &  $1$\\\hline
$-9$ & $0$ & $0$  &  $10$ & $0$ & $0$  &  $1$\\
$-9$ & $1$ & $1/2$  &  $10$ & $2$ & $0$  &  $1$\\
\end{supertabular}
\end{center}

\newpage

\subsection{$T^2/Z_6$}

\begin{center}
\tablehead{
 \hline
 \multicolumn{7}{|c|}{$T^2/Z_6$ with $M_{ab} < 0,\, M_{ca} < 0$} \\ \hline \hline
 \multicolumn{3}{|c||}{$ab$-sector} & \multicolumn{3}{c||}{$ca$-sector} & $bc$-sector \\ \hline
 $M_{ab}$ & $\omega^i_{ab}$ & $\alpha_{ab}$  &  $M_{ca}$ & $\omega^i_{ca}$ & $\alpha_{ca}$  &  \# of Higgs \\ \hline\hline
}
\tabletail{\hline}
\tablecaption{Possible parameter configurations on $T^2/Z_6$ with $M_{ab} < 0,\, M_{ca} < 0$.}
\label{tbl:T2Z6_negative}
\begin{supertabular}{|c|c|c||c|c|c||c|}
$-12$ & $0$ & $0$  &  $-12$ & $0$ & $0$  &  $5$\\\hline
$-12$ & $0$ & $0$  &  $-13$ & $1$ & $1/2$  &  $4$\\\hline
$-13$ & $1$ & $1/2$  &  $-13$ & $1$ & $1/2$  &  $4$\\\hline
$-12$ & $0$ & $0$  &  $-14$ & $0$ & $0$  &  $5$\\
$-12$ & $0$ & $0$  &  $-14$ & $2$ & $0$  &  $4$\\\hline
$-13$ & $1$ & $1/2$  &  $-14$ & $0$ & $0$  &  $4$\\
$-13$ & $1$ & $1/2$  &  $-14$ & $2$ & $0$  &  $5$\\\hline
$-14$ & $0$ & $0$  &  $-14$ & $0$ & $0$  &  $5$\\
\end{supertabular}
\end{center}

\newpage

\begin{center}
\tablehead{
 \hline
 \multicolumn{7}{|c|}{$T^2/Z_6$ with $M_{ab} < 0,\, M_{ca} > 0$} \\ \hline \hline
 \multicolumn{3}{|c||}{$ab$-sector} & \multicolumn{3}{c||}{$ca$-sector} & $bc$-sector \\ \hline
 $M_{ab}$ & $\omega^i_{ab}$ & $\alpha_{ab}$  &  $M_{ca}$ & $\omega^i_{ca}$ & $\alpha_{ca}$  &  \# of Higgs \\ \hline\hline
}
\tabletail{\hline}
\tablecaption{Possible parameter configurations on $T^2/Z_6$ with $M_{ab} < 0,\, M_{ca} > 0$.}
\label{tbl:T2Z6_positive}
\begin{supertabular}{|c|c|c||c|c|c||c|}
$-12$ & $0$ & $0$  &  $12$ & $0$ & $0$  &  $\fbox{1}$\\\hline
$-12$ & $0$ & $0$  &  $14$ & $0$ & $0$  &  $1$\\\hline
$-14$ & $0$ & $0$  &  $14$ & $0$ & $0$  &  $\fbox{1}$\\\hline
$-12$ & $0$ & $0$  &  $15$ & $0$ & $1/2$  &  $1$\\
$-12$ & $0$ & $0$  &  $15$ & $3$ & $1/2$  &  $1$\\\hline
$-13$ & $1$ & $1/2$  &  $15$ & $3$ & $1/2$  &  $1$\\\hline
$-14$ & $2$ & $0$  &  $15$ & $3$ & $1/2$  &  $1$\\\hline
$-15$ & $0$ & $1/2$  &  $15$ & $0$ & $1/2$  &  $\fbox{1}$\\
\end{supertabular}
\end{center}


\bibliographystyle{utphys}
\bibliography{10D_classification}


\end{document}